\documentclass[aapm,mph,amsmath,amssymb,reprint,superscriptaddress,nofootinbib]{revtex4}
\usepackage[italian]{varioref}
\usepackage{amssymb,amsmath,physics,caption,booktabs,subfig,mathtools,xfrac}
\usepackage{gensymb,mathtools,amsfonts,amsthm,mathrsfs,slashed,relsize}
\usepackage{graphicx}
\usepackage{appendix}
\usepackage{dcolumn}
\usepackage{bm}
\usepackage{comment}
\usepackage{xcolor}
\usepackage[mathlines]{lineno}
\usepackage[colorlinks]{hyperref}
\modulolinenumbers[5]

\newcommand{\versor}[1]{\mathbf{\hat{#1}}}

\hypersetup{hidelinks}

\begin{document}
	
\preprint{AAPM/123-QED}
	
\title[]{Cosmic Birefringence: Cross-Spectra and Cross-Bispectra with CMB Anisotropies}
	
\author{Alessandro Greco}
\affiliation{Dipartimento di Fisica e Astronomia ``Galileo Galilei", Universit\`a degli Studi di Padova, via Marzolo 8, I-35131, Padova, Italy}
\affiliation{INFN, Sezione di Padova, via Marzolo 8, I-35131, Padova, Italy}

\author{Nicola Bartolo}
\affiliation{Dipartimento di Fisica e Astronomia ``Galileo Galilei", Universit\`a degli Studi di Padova, via Marzolo 8, I-35131, Padova, Italy}
\affiliation{INFN, Sezione di Padova, via Marzolo 8, I-35131, Padova, Italy}
\affiliation{INAF - Osservatorio Astronomico di Padova, vicolo dell' Osservatorio 5, I-35122 Padova, Italy}

\author{Alessandro Gruppuso}
\affiliation{Istituto Nazionale di Astrofisica - Osservatorio di Astrofisica e Scienza dello Spazio di Bologna, via Gobetti 101, I-40129 Bologna, Italy}
\affiliation{INFN, Sezione di Bologna, viale Berti Pichat 6/2, I-40127 Bologna, Italy}

\date{\today}

\begin{abstract}
Parity-violating extensions of Maxwell electromagnetism induce a rotation of the linear polarization plane of photons during propagation. This effect, known as cosmic birefringence, impacts on the Cosmic Microwave Background (CMB) observations producing a mixing of $E$ and $B$ polarization modes which is otherwise null in the standard scenario. Such an effect is naturally parametrized by a rotation angle which can be written as the sum of an isotropic component $\alpha_0$ and an anisotropic one $\delta\alpha(\versor{n})$. In this paper we compute angular power spectra and bispectra involving $\delta\alpha$ and the CMB  temperature and polarization maps. In particular, contrarily to what happens for the cross-spectra, we show that even in absence of primordial cross-correlations between the anisotropic birefringence angle and the CMB maps, there exist non-vanishing three-point correlation functions carrying signatures of parity-breaking physics. Furthermore, we find that such angular bispectra still survive in a regime of purely anisotropic cosmic birefringence, which corresponds to the conservative case of having $\alpha_0=0$. These bispectra represent an additional observable aimed at studying cosmic birefringence and its parity-violating nature beyond power spectrum analyses. They provide also a way to perform consistency checks for specific models of cosmic birefringence. Moreover, we estimate that among all the possible birefringent bispectra, $\expval{\delta\alpha\, TB}$ and $\expval{\delta\alpha\,EB}$ are the ones which contain the largest signal-to-noise ratio. Once the cosmic birefringence signal is taken to be at the level of current constraints, we show that these bispectra are within reach of future CMB experiments, as LiteBIRD.

\end{abstract}
	
\maketitle

\section{\label{sec:intro}Introduction}
The fact that nature exhibits a parity-violating behaviour has been mostly accepted since parity is maximally broken in the electroweak sector of the standard model of particle physics \cite{lee1956question}. Hence, a question naturally arises: is that possible that besides weak interactions also electromagnetism encode parity-breaking signatures? In order to address this intriguing  question, extensions of Maxwell electromagnetism have been proposed in the literature, e.g. in the form of a Chern-Simons coupling between photons and a scalar field $\chi$ \cite{carroll1990limits}:
\begin{equation}
\label{eqn:lagran}
\mathcal{L}=-\frac{1}{4}F_{\mu\nu}F^{\mu\nu}+h(\chi) F_{\mu\nu}\tilde{F}^{\mu\nu}\, , 
\end{equation}
where $F_{\mu\nu}\equiv\nabla_{\mu}A_{\nu}-\nabla_{\nu}A_{\mu}$ is the electromagnetic tensor, $\tilde{F}^{\mu\nu}=\epsilon^{\mu\nu\rho\sigma}F_{\rho\sigma}/2$ is its dual, and $h(\chi)$ is a dimensionless scalar coupling function of the field $\chi$.

Such a modification of standard electromagnetism affects the photons' propagation, by inducing a rotation of the Stokes parameters $Q$ and $U$:
\begin{equation}
\label{eqn:rotation}
Q\pm iU\mapsto\left(Q\pm iU\right)e^{\pm2i\alpha},
\end{equation}
where the function $\alpha$, called \emph{birefringence angle}, characterizes the amplitude of the deviation from the Maxwell theory, and therefore it is expected to be a small quantity.

In the literature several models of cosmic birefringence have been considered, in which the field $\chi$ is identified as an axion-like field \cite{sherwin2021cosmic,choi2021cosmic,fujita2021detection,nakagawa2021cosmic,li2008cosmological,caldwell2011cross,zhai2020effects,arvanitaki2010string,takahashi2021kilobyte}. Physically, the existence of such a field is predicted e.g. in the context of string theory \cite{marsh2016axion,kim2021cosmic}. Furthermore $\chi$ has been proposed as a candidate for dark matter (e.g. \cite{liu2017axion} and Refs. therein) or for early dark energy in the form of a pseudo Nambu-Goldstone boson to explain the Hubble tension \cite{capparelli2020cosmic}.

Since in the model described by Eq.~\eqref{eqn:lagran} the angle is proportional to the distance travelled by photons \cite{li2008cosmological}, a tiny coupling to the Chern-Simons term can become observable if the source of linearly polarized photons is the farthest possible. For this reason, Cosmic Microwave Background (CMB) radiation, being linearly polarized because of Thomson scattering at the last scattering surface (LSS), is a well suited observable for this kind of investigation.

Such an effect, known as \emph{cosmic birefringence}, could have left measurable imprints in the CMB angular power spectra $C_{\ell}$'s. Indeed, it is well known that the cross-correlations $C_{\ell}^{TB}$ and $C_{\ell}^{EB}$ are identically zero in a standard cosmological context, because the coefficients of the harmonic expansion $a_{T,\ell m}$ and $a_{E,\ell m}$ behave in an opposite way under parity transformation with respect to $a_{B,\ell m}$. Nevertheless, the presence of a Chern-Simons correction, like the one in Eq.~\eqref{eqn:lagran}, induces a rotation of the CMB angular power spectra \cite{lue1999cosmological,liu2006effect,feng2006searching}
\begin{align}
\label{eqn:Cte}
C^{TE}_{\ell,\text{obs}}&=C_{\ell}^{TE}\cos2\alpha,\\	
C^{TB}_{\ell,\text{obs}}&=C^{TE}_{\ell}\sin2\alpha,\\
C^{EE}_{\ell,\text{obs}}&=C_{\ell}^{EE}\cos^22\alpha+C_{\ell}^{BB}\sin^22\alpha,\\
C^{EB}_{\ell,\text{obs}}&=\left(C_{\ell}^{EE}-C_{\ell}^{BB}\right)\sin2\alpha\cos2\alpha,\\
\label{eqn:Cbb}
C^{BB}_{\ell,\text{obs}}&=C_{\ell}^{EE}\sin^22\alpha+C_{\ell}^{BB}\cos^22\alpha\, ,
\end{align}
where the subscript ``obs'' labels the ``observed'' angular power spectra, i.e. those that we observe by taking into account the rotation of the polarization plane induced by cosmic birefringence. Thus, thanks to the birefringent mechanism, parity-violating correlators $TB$ and $EB$ turn on, whereas by setting $\alpha=0$ we recover the standard results. Eqs.~\eqref{eqn:Cte}-\eqref{eqn:Cbb} are obtained under the assumption of isotropic cosmic birefringence, i.e. with the rotation angle being just a constant. Using \textit{Planck} data, such an angle has been estimated consistent with zero \cite{aghanim2016planck,gruppuso2020planck} within the total error budget, which is dominated by the uncertainty of the instrumental polarization angle. Interestingly, a recent reanalysis of \textit{Planck} 2018 data (PR3 release), which makes uses of the information contained also in the foreground emissions, is able to cut the total uncertainty by a factor of $2$ providing $\alpha=(0.35\pm0.14)^{\circ}$ at $68\%$ C.L. \cite{minami2020new} and a very similar result is confirmed with PR4 data\footnote{For more details on PR3 and PR4 {\it Planck} data release see \url{http://pla.esac.esa.int}.} $\alpha=(0.30\pm0.11)^{\circ}$ at $68\%$ C.L. \cite{diego2022cosmic}. However the cosmological significance of such measurements must be fully scrutinized in view of possible foreground polarization \cite{clark2021origin,diego2022cosmic,eskilt2022}. Other techniques based only on CMB data and aimed at disentangling the instrumental polarisation angle and the isotropic birefringence effect have been proposed \cite{sherwin2021cosmic} (see also~\cite{gruppuso2016note}), but current CMB data at low multipoles are not sufficiently signal-dominated to make them applicable.

In this paper we focus instead on the \emph{anisotropic} cosmic birefringence, for which the rotation angle also encodes a dependence on the observation direction:
\begin{equation}
\label{eqn:deltaAlpha}
\alpha=\alpha_0+\delta\alpha(\versor{n}), 
\end{equation}
so that $\delta\alpha$ can be regarded as a small perturbation around its isotropic counterpart. Such a spatial fluctuation in the rotation angle is currently constrained by CMB observations to be consistent with zero (see e.g. \cite{gruppuso2020planck} and the references therein). By including an anisotropic component, not only the expressions collected in Eqs.~\eqref{eqn:Cte}-\eqref{eqn:Cbb} are modified by the presence of further correction terms \cite{li2008cosmological}, but it is also possible to compute correlations between $\delta\alpha$ and the other CMB observables: $T$, $E$, $B$ \cite{caldwell2011cross,capparelli2020cosmic} (and eventually also with the weak gravitational lensing, as done in \cite{liu2017axion}).
	
In \cite{zhai2020effects} Zhai et al. calculate the CMB three-point angular correlation functions (in short the angular bispectra), for the temperature and rotated polarization fields, by taking into account the cosmic birefringence effects. The authors perform such calculations under the assumption that the temperature, the unrotated polarization fields of CMB and the anisotropic rotation angle are all Gaussian random fields. They find that rotated CMB bispectra would arise if two-point cross-correlations of $T$ with $\delta\alpha$, and $E$ with $\delta\alpha$ are non-vanishing. Instead, we calculate here the three-point angular correlations between $\delta\alpha$ and the CMB observables, showing that still keeping the Gaussian assumption for these fields, there exist non-zero rotated bispectra even in absence of an unrotated cross-correlation between the birefringence angle and $T$, $E$, $B$. Furthermore, we show in this paper that the  $\expval{\delta\alpha\,TB}$ and $\expval{\delta\alpha\, EB}$ bispectra are the three-point angular correlation functions with the largest signal-to-noise ratio.

The structure of the paper is organized as follows. In Sec.~\ref{sec:math} we briefly review how to deal with the anisotropic component of the rotation angle, and how cosmic birefringence affects the CMB harmonic coefficients. In Sec.~\ref{sec:power_spectrum} we compute the rotated two-point cross-correlations between $\delta\alpha$ and CMB observables, motivating our interest in going beyond power-spectra. In Sec.~\ref{sec:bispectra} we explicitly compute the three-point angular cross-correlation functions involving the same quantities, but this time with the assumption of no cross-correlation at the two-point level. In Sec.~\ref{sec:plots} we provide as an example some plots showing the behaviour of the angular bispectra obtained in Sec.~\ref{sec:bispectra} for a scale-invariant model of cosmic birefringence. In Sec.~\ref{sec:SNR} we estimate the signal-to-noise ratio for the angular bispectra we have computed in Sec.~\ref{sec:bispectra}. Sec.~\ref{sec:end} is dedicated to the conclusion and discussions. Some conceptual and mathematical details about the parity symmetry, the scale-invariant limit of the cosmic birefringence angular power spectrum, the Fisher forecast we made for the birefringent bispectra, and how to generalize our results by including reionization can be found in the Appendices~\ref{app:parRel},~\ref{app:PowerSpectrum},~\ref{app:FisherFormulas},~\ref{app:Reionization} respectively.

\section{\label{sec:math}Basic Equations and Formalism}
If cosmic birefringence is induced by a Chern-Simons coupling involving a scalar field, as in Eq.~\eqref{eqn:lagran}, then the full rotation angle of the polarization plane is related to the variation of such a field \cite{li2008cosmological}:
\begin{equation}
\label{eqn:angle}
\alpha=2\left[h(\chi_f)-h(\chi_i)\right],
\end{equation}
where $\chi_f$ and $\chi_i$ are the values of the field $\chi$ at the end (today) and at the beginning of the photon's travel (the time of hydrogen atom recombination, from which photons started to free-stream in the Universe according to the sudden recombination approximation). As customary in cosmological perturbations theory, we can decompose the field $\chi$ as the sum of a homogeneous background term plus a perturbative inhomogeneous contribution:
\begin{equation}
\chi(\mathbf{x},\tau)=\chi_0(\tau)+\delta\chi(\mathbf{x},\tau),
\end{equation}
where $\mathbf{x}$ is a vector of the comoving spatial coordinates and $\tau$ is the conformal time. Hence, by comparing side-by-side Eq.~\eqref{eqn:deltaAlpha} with Eq.~\eqref{eqn:angle}, we can infer that
\begin{equation}
\alpha_0=2h[\chi_0(\tau_0)]-2h[\chi_0(\tau_{\text{reco}})]
\end{equation}
and
\begin{equation}
\label{eqn:delta}
\delta\alpha(\versor{n})=-2\frac{\mathrm{d}h(\chi_0)}{\mathrm{d}\chi_0}\Bigg|_{\tau=\tau_{\text{reco}}}\delta\chi[(\tau_0-\tau_{\text{reco}})\versor{n},\tau_{\text{reco}}],
\end{equation}
where $\tau_0$ and $\tau_{\text{reco}}$ are the conformal time today and at the recombination epoch, respectively, while $\versor{n}$ denotes (minus) the coming direction of CMB photons in a spatially flat Universe, i.e. the line of sight direction. Let us notice that in writing down Eq.~\eqref{eqn:delta}, we have neglected the term proportional to the final value of the fluctuation  $\delta\chi$: this is due to the fact that such a term only gives rise to an unobservable monopole contribution due to the motion of the observer with respect to the CMB frame \cite{li2008cosmological}.

As usual, one can employ the standard multipole expansion to analyse CMB maps on the celestial sphere, expanding the temperature and polarization anisotropies by means of the proper harmonic coefficients and spin-weighted spherical-harmonics \cite{zaldarriaga55all}:
\begin{align}
T(\versor{n})&=\sum_{\ell m}a_{T,\ell m}Y_{\ell m}(\versor{n}),\\
\left[Q\pm iU\right](\versor{n})&=\sum_{\ell m}a_{\pm2,\ell m}\,_{\pm2}Y_{\ell m}(\versor{n}),
\end{align}
where
\begin{align}
\label{eqn:alpha}
a_{T,\ell m}&=\int\mathrm{d}\hat{n}\,Y^*_{\ell m}(\versor{n})T(\versor{n}),\\
a_{\pm 2,\ell m}&=\int\mathrm{d}\hat{n}\,_{\pm2}Y_{\ell m}^*(\versor{n})\left[Q\pm iU\right](\versor{n}).
\end{align}
Hence, we can extend such a treatment also to the anisotropic component of the birefringent angle, which will be written in terms of spin-zero spherical harmonics, being $\delta\alpha$ a scalar quantity:
\begin{equation}
\label{eqn:expansion}
\delta\alpha(\versor{n})=\sum_{\ell m}\alpha_{\ell m}Y_{\ell m}(\versor{n})\, ,
\end{equation}
with
\begin{equation}
\label{eqn:alpha_lm}
\alpha_{\ell m}=\int\mathrm{d}\hat{n}\,Y^*_{\ell m}(\versor{n})\,\delta\alpha(\versor{n}).
\end{equation}

Since cosmic birefringence induces a rotation of the Stokes parameters in the manner shown by Eq.~\eqref{eqn:rotation}, it is not difficult to see that the observed harmonic coefficients for the polarization modes rotates too as
\begin{equation}
a_{\pm2,\ell m}^{\text{obs}}=\int\mathrm{d}\hat{n}\,_{\pm2}Y_{\ell m}^*(\versor{n})\left[Q\pm iU\right](\versor{n})e^{\pm 2i\alpha(\versor{n})}.
\end{equation}
By introducing the standard linear combination \cite{hu1997cmb},
\begin{equation}
a_{\pm 2,\ell m}=-(a_{E,\ell m}\pm ia_{B,\ell m}),
\end{equation}
we can thus obtain a compact expression for the rotated harmonic coefficients of the $P_1\equiv E$ and $P_2\equiv B$ CMB polarization modes:
\begin{equation}
\label{eqn:harmonic}
\begin{split}
a_{P_j,\ell m}^{\text{obs}}=\sum_{s=\pm2}\frac{e^{is\alpha_0}}{2}&\sum_{LM}\int\mathrm{d}\hat{n}\,_sY_{\ell m}^*(\versor{n})\,_sY_{LM}(\versor{n})\mathscr{R}^{(s)}_{jk}\,a_{P_k,LM}e^{is\delta\alpha(\versor{n})}
\end{split}
\end{equation}
where for $j,k=1,2$ we have defined
\begin{equation}
\label{eqn:Greek_Matrix}
a_{P,\ell m}=\begin{pmatrix}
a_{E,\ell m} \\
a_{B,\ell m}
\end{pmatrix},
\qquad\mathscr{R}^{(s)}\equiv\begin{pmatrix}
1 & is/2 \\
-is/2 & 1
\end{pmatrix},
\end{equation}
and the sum over $k$ is understood. Of course it is trivial to see that $a_{T,\ell m}^{\text{obs}}=a_{T,\ell m}$ and  $\alpha_{\ell m}^{\text{obs}}=\alpha_{\ell m}$, since cosmic birefringence only affects the Stokes parameters $Q$ and $U$. 

Hence, we are now in the position to compute the correlation functions involving the rotated CMB fields and the anisotropic cosmic birefringence. For our treatment we consider the same (phenomenological) assumptions made in \cite{li2008cosmological}:
\begin{enumerate}
	\item the rotation field $\delta\alpha$ is uncorrelated with the primordial $T$, $E$, and $B$ modes;
	\item the rotation angle is small everywhere;
	\item the underlying inflationary model is parity-conserving;   $C_{\ell}^{TB}=C_{\ell}^{EB}=0$ for primordial modes;
	\item the unrotated anisotropy fields of CMB and $\delta\alpha$ are all Gaussian random fields.
\end{enumerate}
The last assumption allows us to evaluate the $n$-point correlation functions involving the harmonic coefficients of $\delta\alpha$, $T$, $E$ and $B$ modes by means of the Isserlis theorem \cite{isserlis1918formula}:
\begin{equation}
\label{eqn:gauss}
\begin{split}
\langle x_1&\prod_{j=2}^{n}x_j\rangle=\sum_{i=2}^n\langle x_1x_i\rangle\langle\partial_{x_i}\prod_{j=2}^{n}x_j\rangle,
	\end{split}
\end{equation}
when $n$ is an even integer. If instead $n$ is an odd integer the correlation function is identically vanishing.

\section{\label{sec:power_spectrum} Cross-Correlations of Cosmic Birefringence Angle with CMB Maps}

Under the assumption of statistical isotropy, the two-point angular correlation functions are simply given as the angular power spectra times Kronecker deltas: 
\begin{equation}
\label{eqn:definition}
\expval{a^*_{X,\ell_1 m_1}a_{Y,\ell_2m_2}}=C_{\ell_1}^{XY}\delta_{\ell_1\ell_2}\delta_{m_1m_2},
\end{equation}
where $X,Y=\delta\alpha, T, E, B$. Now, let us see how the observed cross-correlators are related to the unrotated ones: we want to compute the more general cross-correlation of the anisotropic component of birefringence angle with CMB polarization modes, which, by recalling Eq.~\eqref{eqn:harmonic}, is given as
\begin{equation}
\label{eqn:1p}
\begin{split}		\langle&\alpha^*_{\ell_1m_1}a_{P_j,\ell_2m_2}^{\text{obs}}\rangle=\sum_{s_2}\frac{e^{is_2\alpha_0}}{2}\sum_{L_2M_2}\int\mathrm{d}\hat{n}_2\,_{s_2}Y_{\ell_2 m_2}^*(\versor{n}_2)\,_{s_2}Y_{L_2M_2}(\versor{n}_2)\mathscr{R}^{(s_2)}_{jk}\expval{\alpha^*_{\ell_1m_1}a_{P_k,L_2M_2}e^{is_2\delta\alpha(\versor{n}_2)}}.
\end{split}
\end{equation}
The ensemble average on the right-hand side of the above equation can be easily evaluated by approximating the exponential to unity, since we want to work at the linear order in $\delta\alpha$. Thus, by recalling the definition of Eq.~\eqref{eqn:definition}, the term within angular brackets on the right-hand side of Eq.~\eqref{eqn:1p} simply reduces to
\begin{equation}
\label{eqn:expval}
\begin{split}
\langle\alpha_{\ell_1m_1}^*&a_{P_k,L_2M_2}e^{is_2\delta\alpha(\versor{n}_2)}\rangle \simeq C^{\alpha P_k}_{L_2}\delta_{\ell_1 L_2}\delta_{m_1M_2}.
\end{split}
\end{equation}
We then substitute the result above in Eq.~\eqref{eqn:1p} to obtain
\begin{equation}
\label{eqn:alphaP}
\begin{split}
\langle\alpha^*_{\ell_1m_1}a_{P_j,\ell_2m_2}^{\text{obs}}\rangle=\frac{1}{2}\sum_{s_2}&e^{is_2\alpha_0}\int\mathrm{d}\hat{n}_2\,_{s_2}Y_{\ell_2 m_2}^*(\versor{n}_2)\,_{s_2}Y_{\ell_1 m_1}(\versor{n}_2)\mathscr{R}^{(s_2)}_{jk}C^{\alpha P_k}_{\ell_1}.
\end{split}
\end{equation}
Therefore, by exploiting the orthogonality relation \cite{newman1966note} of spin-weighted spherical harmonics,
\begin{equation}
\label{eqn:orthogonality}
\int\mathrm{d}\hat{n}\,_{s}Y_{\ell_1m_2}(\versor{n})\,_{s}Y_{\ell_2,m_2}^*(\versor{n})=\delta_{\ell_1\ell_2}\delta_{m_1m_2},
\end{equation}
and, by performing the summation over $s_2=\pm 2$, we get the relation which associates $C^{X Y}_{\ell}$ with $C^{X Y}_{\ell,\text{obs}}$: 
\begin{align}
\label{eqn:AT}
C_{\ell,\text{obs}}^{\alpha T}&=C_{\ell}^{\alpha T}, \\
\label{eqn:AE}
C_{\ell,\text{obs}}^{\alpha E}&=C_{\ell}^{\alpha E}\cos2\alpha_0-C_{\ell}^{\alpha B}\sin2\alpha_0, \\
\label{eqn:AB}
C_{\ell,\text{obs}}^{\alpha B}&=C_{\ell}^{\alpha E}\sin2\alpha_0+C_{\ell}^{\alpha B}\cos2\alpha_0.
\end{align}
We can easily see that the observed cross-correlators are simply obtained from the ``primordial'' ones by performing a spatial rotation. This result is telling us that the observed angular power spectra would vanish if the unrotated primordial correlations are absent. This conclusion albeit trivial, is interesting also because it provides an additional motivation to investigate higher-order correlation functions. Indeed, as we will see very soon, we will reach a very different conclusion for the angular bispectra.

\section{\label{sec:bispectra} Birefringent Angular Bispectra}
Now we move on by evaluating all the three-point functions involving correlations between the anisotropic birefringence angle $\delta\alpha$ and the CMB maps. For this reason, we are going to calculate the ensemble averages for the combinations listed in Tab.~\ref{tab:bispectra}.
\begin{table}
\caption{\label{tab:bispectra}Bispectra involving the anisotropic birefringence angle $\delta\alpha$ and the CMB anisotropy maps, $T,E,B$, ordered according to the number of polarization fields $N_P$.}
\begin{ruledtabular}
\begin{tabular}{ccccc}
$N_P$ & & & & \\
\hline
$0$	& $\delta\alpha\,\delta\alpha\, T$ & $\delta\alpha\, TT$ & & \\
$1$	& $\delta\alpha\,\delta\alpha\, E$ & $\delta\alpha\,\delta\alpha\, B$ & $\delta\alpha\, T E$ & $\delta\alpha\, T B$  \\
$2$	& $\delta\alpha\, EE$  & $\delta\alpha\, EB$ & $\delta\alpha\, BB$ & \\
\end{tabular}
\end{ruledtabular}
\end{table}
Since cosmic birefringence only affects the Stokes parameters $Q$ and $U$, it is clear that the {observed correlators $\expval{\delta\alpha\,\delta\alpha\,T}$ and $\expval{\delta\alpha\,TT}$ correspond to the primordial ones, and so they are non-zero only in the presence of some intrinsic (primordial) underlying non-Gaussianity, a case that we are not considering in this paper, according to our previous assumptions, see the first line of Tab.~\ref{tab:bispectra}.  Anyway, as just mentioned before, differently from what occurs for the two-point correlation functions, vanishing primordial (three-point) correlation functions do not prevent the possibility to have non-vanishing observed three-point correlation functions for the other combinations. Therefore, let us focus on the three-point functions listed in the second and the third line of Tab.~\ref{tab:bispectra}.

\subsection{One Polarization Field}
In analogy with what we have done in Sec.~\ref{sec:power_spectrum}, the bispectra involving a single polarization field in Tab.~\ref{tab:bispectra} can be evaluated by firstly calculating the following general quantity:
\begin{equation}
	\label{eqn:below}
	\begin{split}
		\langle\alpha_{\ell_1m_1}\beta_{\ell_2m_2}a_{P_j,\ell_3m_3}^{\text{obs}}\rangle=\sum_{s_3}\frac{e^{is_3\alpha_0}}{2}\sum_{L_3M_3}\int\mathrm{d}\hat{n}_3\,_{s_3}Y_{\ell_3 m_3}^*(\versor{n}_3)&\,_{s_3}Y_{L_3M_3}(\versor{n}_3)\mathscr{R}^{(s_3)}_{jk}\times\\
		&\times\expval{\alpha_{\ell_1m_1}a_{c,\ell_2m_2}a_{P_k,L_3M_3}e^{is_3\delta\alpha(\versor{n}_3)}},
	\end{split}
\end{equation}
and then specializing for $\beta_{\ell m}=\alpha_{\ell m},a_{T,\ell m}$ and $P_{j}=E,B$. From Eq.~\eqref{eqn:below} is it easy to understand why we can have non-vanishing three-point correlation functions even under the Gaussian assumption: the ensemble average within the angular integral is effectively a trispectrum, i.e. a four-point correlation function in the harmonic space, which is in general non-zero for Gaussian random fields. In order to see this more clearly, we now expand the exponential as
\begin{equation}
e^{is_3\delta\alpha(\versor{n}_3)}\simeq1+is_3\sum_{L_3'M_3'}\alpha_{L_3'M_3'}Y_{L_3'M_3'}(\versor{n}_3).
\end{equation}
By means of Eq.~\eqref{eqn:gauss}, it is not difficult to show that the only non-vanishing contribution to the ensemble average on the right-hand side of Eq.~(\ref{eqn:below}) is given by
\begin{equation}
\label{eqn:above}
\begin{split}
\langle\alpha_{\ell_1m_1}&\beta_{\ell_2m_2}a_{P_k,L_3M_3}e^{is_3\delta\alpha(\versor{n}_3)}\rangle=is_3\delta_{\ell_2L_3}\delta_{m_2,-M_3}C_{\ell_1}^{\alpha\alpha}C_{\ell_2}^{\beta P_k}Y_{\ell_1m_1}^*(\versor{n}_3),
\end{split}
\end{equation}
where we have neglected all the terms proportional to the primordial cross-correlation between $\alpha$ and $P_k$, according to the assumptions listed in Sec~\ref{sec:math}. For the same reason, since our aim is to study what happens when the two-point cross-correlations between the anisotropic birefringence angle and the CMB anisotropies are absent, we can already infer that the only non-vanishing bispectra involving a single polarization field are those with $\beta_{\ell m}=a_{T,\ell m}$, and so from now on we replace the generic field $\beta$ with the CMB temperature anisotropies.

We now substitute Eq.~\eqref{eqn:above} in Eq.~\eqref{eqn:below}. Using the complex conjugate of spin-weighted spherical harmonics \cite{newman1966note},
\begin{equation}
_{s}Y_{\ell m}^*(\versor{n})=(-1)^s\,_{-s}Y_{\ell,-m}(\versor{n}),
\end{equation}
we perform the integration over the solid angle by means of the well-known formula of the triple integral~\cite{newman1966note}:
\begin{equation}
\label{eqn:triple}
\begin{split}
\int\mathrm{d}\hat{n}&\,_{s_1}Y_{\ell_1 ,m_1}(\versor{n})\,_{s_2}Y_{\ell_2,m_2}(\versor{n})\,_{s_3}Y_{\ell_3,m_3}(\versor{n})=I^{-s_1,-s_2,-s_3}_{\ell_1\ell_2\ell_3}\begin{pmatrix}
\ell_1 & \ell_2 & \ell_3 \\
m_1 & m_2 & m_3
\end{pmatrix},
\end{split}
\end{equation}
where we have defined
\begin{equation}
\begin{split}
I^{-s_1,-s_2,-s_3}_{\ell_1\ell_2\ell_3}&\equiv\sqrt{\frac{(2\ell_1+1)(2\ell_2+1)(2\ell_3+1)}{4\pi}}\begin{pmatrix}
\ell_1 & \ell_2 & \ell_3 \\
-s_1 & -s_2 & -s_3
\end{pmatrix},
\end{split}
\end{equation}
and where the ``matrix'' is a  Wigner $3j$-symbol, which obeys the symmetry \cite{varshalovich1988quantum}:
\begin{equation}
\label{eqn:flip}
\begin{pmatrix}
\ell_1 & \ell_2 & \ell_3 \\
-m_1 & -m_2 & -m_3
\end{pmatrix}=(-1)^{\ell_T}\begin{pmatrix}
\ell_1 & \ell_2 & \ell_3 \\
m_1 & m_2 & m_3
\end{pmatrix} \, , 
\end{equation}
with $\ell_T\equiv\ell_1+\ell_2+\ell_3$. We finally obtain that Eq.~\eqref{eqn:below} simply reduces to
\begin{equation}
\label{eqn:1P}
\begin{split}
\langle \alpha_{\ell_1m_1}&a_{T,\ell_2m_2}a_{P_j,\ell_3m_3}^{\text{obs}}\rangle=\frac{i}{2}\begin{pmatrix}
\ell_1 & \ell_2 & \ell_3\\
m_1    & m_2    & m_3
\end{pmatrix}\sum_{s_3}s_3e^{is_3\alpha_0}\mathscr{R}^{(s_3)}_{jk} C_{\ell_1}^{\alpha\alpha}C_{\ell_2}^{T P_k}I^{0,s_3,-s_3}_{\ell_1\ell_2\ell_3}\, .
\end{split}
\end{equation}
After performing the summation over $s_3=\pm2$, we find the following expressions for the observed $\delta\alpha\,TE$ and $\delta\alpha\,TB$ angular bispectra:
\begin{equation}
\label{eqn:ATE}
\begin{split}
\langle\alpha_{\ell_1m_1}&a_{T,\ell_2m_2}a_{E,\ell_3m_3}^{\text{obs}}\rangle=C_{\ell_1}^{\alpha\alpha}\Big\{i\left[1-(-1)^{\ell_{T}}\right]\cos2\alpha_0-\left[1+(-1)^{\ell_{T}}\right]\sin2\alpha_0\Big\}C_{\ell_2}^{TE}I^{0,2,-2}_{\ell_1\ell_2\ell_3}\begin{pmatrix}
\ell_1 & \ell_2 & \ell_3 \\
m_1 & m_2 & m_3
\end{pmatrix}
\end{split}
\end{equation}
and 
\begin{equation}
\label{eqn:ATB}
\begin{split}
\langle\alpha_{\ell_1m_1}&a_{T,\ell_2m_2}a_{B,\ell_3m_3}^{\text{obs}}\rangle=C_{\ell_1}^{\alpha\alpha}\Big\{\left[1+(-1)^{\ell_{T}}\right]\cos2\alpha_0+i\left[1-(-1)^{\ell_{T}}\right]\sin2\alpha_0\Big\}C_{\ell_2}^{TE}I^{0,2,-2}_{\ell_1\ell_2\ell_3}\begin{pmatrix}
	\ell_1 & \ell_2 & \ell_3 \\
	m_1 & m_2 & m_3
\end{pmatrix}.
\end{split}
\end{equation}
We have disregarded the terms proportional to $C_{\ell}^{TB}$ again, according to the assumptions listed in Sec.~\ref{sec:math}. Notice that in writing Eqs.~\eqref{eqn:AT}-\eqref{eqn:AB} and Eq.~(\ref{eqn:ATE}) and~(\ref{eqn:ATB}) (see also Eqs.~\eqref{eqn:AEE}-\eqref{eqn:AEB}) we are accounting only for the epoch of recombination as the time of polarization generation (see, e.g.~\cite{sherwin2021cosmic}). In App.~\ref{app:Reionization} we give the full expressions that account also for the reionization epoch.}

As we have anticipated previously, it is interesting to note that we obtain some non-zero bispectra between $\delta\alpha$ and CMB maps, even assuming no correlation between them at the two-(three) point level. Moreover, Eqs.~\eqref{eqn:ATE}-\eqref{eqn:ATB} contain terms of mixed parity for $\ell_{T}$, where the imaginary part of the bispectra is non-vanishing only if $\ell_{T}$ is an odd number, encoding parity-breaking signatures, see App.~\ref{app:parRel}. Moreover, let us notice that the two bispectra are non-vanishing even for $\alpha_0=0$, which corresponds to a regime of purely anisotropic cosmic birefringence. From Eqs.~\eqref{eqn:ATE}-\eqref{eqn:ATB} we see that setting $\alpha_0=0$ turns off one of the two terms associated to the parity properties of $\ell_{T}$, but the other term survives: this feature is suggesting that having a zero isotropic birefringence angle fixes the parity properties of the bispectrum itself, but it still produces a non-vanishing signal.

\subsection{Two Polarization Fields}
We now move on to consider the more complicated bispectra listed in the last line of Tab.~\ref{tab:bispectra}. Similarly to the previous case, we have
\begin{equation}
\label{eqn:2P}
\begin{split}
\langle&\alpha_{\ell_1m_1}a_{P_i,\ell_2m_2}^{\text{obs}}a^{\text{obs}}_{P_j,\ell_3m_3}\rangle=\sum_{s_2s_3}\frac{e^{i(s_2+s_3)\alpha_0}}{4}\sum_{L_2M_2}\sum_{L_3M_3}\int\mathrm{d}\hat{n}_2\,\int\mathrm{d}\hat{n}_3\,_{s_2}Y_{\ell_2 m_2}^*(\versor{n}_2)\,_{s_2}Y_{L_2M_2}(\versor{n}_2)\,\times\\
&\qquad\qquad\quad\times\,_{s_3}Y_{\ell_3 m_3}^*(\versor{n}_3)\,_{s_3}Y_{L_3M_3}(\versor{n}_3)\mathscr{R}^{(s_2)}_{ik}\mathscr{R}^{(s_3)}_{jl}\expval{\alpha_{\ell_1m_1}a_{P_k,L_2M_2}a_{P_{l},L_3M_3}e^{is_2\delta\alpha(\versor{n}_2)+is_3\delta\alpha(\versor{n}_3)}} \, , 
\end{split}
\end{equation}
where expanding again the exponential 
\begin{equation}
\begin{split}
e^{is_2\delta\alpha(\versor{n}_2)}&e^{is_3\delta\alpha(\versor{n}_3)}\simeq1+is_2\sum_{L_2'M_2'}\alpha_{L_2'M_2'}Y_{L_2'M_2'}(\versor{n}_2)+is_3\sum_{L_3'M_3'}\alpha_{L_3'M_3'}Y_{LM}(\versor{n}_3),
\end{split}
\end{equation}
and by exploiting Eq.~\eqref{eqn:gauss}, we work out the ensemble average within the integral as 
\begin{equation}
\label{eqn:sboldra}
\begin{split}
\langle\alpha_{\ell_1m_1}&a_{P_k,L_2M_2}a_{P_{l},L_3M_3}e^{is_2\delta\alpha(\versor{n}_2)+is_3\delta\alpha(\versor{n}_3)}\rangle\simeq iC^{\alpha\alpha}_{\ell_1}C_{L_2}^{P_kP_l}\delta_{L_2L_3}\delta_{M_2,-M_3}\left[s_2Y_{\ell_1m_1}^*(\versor{n}_2)+s_3Y_{\ell_1m_1}^*(\versor{n}_3)\right].
\end{split}
\end{equation}
We now replace Eq.~\eqref{eqn:sboldra} in Eq.~\eqref{eqn:2P} and, by performing the integration over the photons' direction with the help of Eq.~\eqref{eqn:orthogonality} and Eq.~\eqref{eqn:triple}, we finally get
\begin{equation}
\label{eqn:45}
\begin{split}
\langle\alpha_{\ell_1m_1}a_{P_i,\ell_2m_2}^{\text{obs}}a^{\text{obs}}_{P_j,\ell_3m_3}\rangle=
\frac{i}{4}\begin{pmatrix}
\ell_1 & \ell_2 & \ell_3 \\
m_1 & m_2 & m_3
\end{pmatrix}\sum_{s_2s_3}e^{i(s_2+s_3)\alpha_0}&R_{ik}^{(s_2)}R_{jl}^{(s_3)}C_{\ell_1}^{\alpha\alpha}\times\\
&\times\left[s_2C_{\ell_3}^{P_kP_l}I_{\ell_1\ell_2\ell_3}^{0,s_2,-s_2}+s_3C_{\ell_2}^{P_kP_l}I^{0,-s_3,s_3}_{\ell_1\ell_2\ell_3}\right].
\end{split}
\end{equation}

Thus, we can use the expression in Eq.~\eqref{eqn:45} to compute the bispectra listed in the last row of Tab.~\ref{tab:bispectra}. Let us notice that we get a relatively simple expression because, according to our assumptions, we have set equal to zero all the two-point cross-correlations of the anisotropic cosmic birefringent angle with CMB temperature and polarization modes, and because we are working at the leading order in $\delta\alpha$: differently from the case involving a single polarization field, this time no one of the configurations in the last line of Tab.~\ref{tab:bispectra} is vanishing. After some algebra we obtain
\begin{equation}
\label{eqn:AEE}
\begin{split}
\langle\alpha_{\ell_1m_1}&a_{E,\ell_2m_2}^{\text{obs}}a^{\text{obs}}_{E,\ell_3m_3}\rangle=\frac{1}{2}C_{\ell_1}^{\alpha\alpha}\Big\{[1+(-1)^{\ell_{T}}]\left[\left(C_{\ell_3}^{BB}+C_{\ell_2}^{BB}\right)-\left(C_{\ell_3}^{EE}+C_{\ell_2}^{EE}\right)\right]\sin4\alpha_0\,+\\
&+i\left[1-(-1)^{\ell_{T}}\right]\left[\left(C_{\ell_3}^{BB}-C_{\ell_2}^{BB}\right)\left(1-\cos4\alpha_0\right)+\left(C_{\ell_3}^{EE}-C_{\ell_2}^{EE}\right)\left(1+\cos4\alpha_0\right)\right]\Big\}I_{\ell_1\ell_2\ell_3}^{0,2,-2}\begin{pmatrix}
\ell_1 & \ell_2 & \ell_3 \\
m_1 & m_2 & m_3
\end{pmatrix}
\end{split}
\end{equation}

\begin{equation}
\label{eqn:ABB}
\begin{split}
\langle\alpha_{\ell_1m_1}&a_{B,\ell_2m_2}^{\text{obs}}a^{\text{obs}}_{B,\ell_3m_3}\rangle=\frac{1}{2}C_{\ell_1}^{\alpha\alpha}\Big\{[1+(-1)^{\ell_{T}}]\left[\left(C_{\ell_3}^{EE}+C_{\ell_2}^{EE}\right)-\left(C_{\ell_3}^{BB}+C_{\ell_2}^{BB}\right)\right]\sin4\alpha_0\,+\\
&+i\left[1-(-1)^{\ell_{T}}\right]\left[\left(C_{\ell_3}^{EE}-C_{\ell_2}^{EE}\right)\left(1-\cos4\alpha_0\right)+\left(C_{\ell_3}^{BB}-C_{\ell_2}^{BB}\right)\left(1+\cos4\alpha_0\right)\right]\Big\}I_{\ell_1\ell_2\ell_3}^{0,2,-2}\begin{pmatrix}
	\ell_1 & \ell_2 & \ell_3 \\
	m_1 & m_2 & m_3
\end{pmatrix}
\end{split}
\end{equation}

\begin{equation}
\label{eqn:AEB}
\begin{split}
\langle\alpha_{\ell_1m_1}&a_{E,\ell_2m_2}^{\text{obs}}a^{\text{obs}}_{B,\ell_3m_3}\rangle=\frac{1}{2}C_{\ell_1}^{\alpha\alpha}\Big\{i[1-(-1)^{\ell_{T}}]\left[\left(C_{\ell_3}^{EE}-C_{\ell_2}^{EE}\right)-\left(C_{\ell_3}^{BB}-C_{\ell_2}^{BB}\right)\right]\sin4\alpha_0\,+\\
&+\left[1+(-1)^{\ell_{T}}\right]\left[\left(C_{\ell_2}^{BB}-C_{\ell_3}^{EE}\right)\left(1-\cos4\alpha_0\right)+\left(C_{\ell_2}^{EE}-C_{\ell_3}^{BB}\right)\left(1+\cos4\alpha_0\right)\right]\Big\}I_{\ell_1\ell_2\ell_3}^{0,2,-2}\begin{pmatrix}
	\ell_1 & \ell_2 & \ell_3 \\
	m_1 & m_2 & m_3
\end{pmatrix}.
\end{split}
\end{equation}

Interestingly enough, all the bispectra collected in Eqs.~\eqref{eqn:AEE}-\eqref{eqn:AEB} contain terms of different parity, since they are proportional to $[1\mp(-1)^{\ell_{T}}]$ and thus vanish for $\ell_{T}=\text{even}\,(\text{odd})$. This is a manifest signature that these objects encode parity-violating features, that are due to the parity-breaking nature of cosmic birefringence. Moreover, as we could expect, all the bispectra that we have computed are proportional to the self-correlator $C_{\ell}^{\alpha\alpha}$, which obviously depends on the specific model which induces the birefringent mechanism.

It is interesting to see that, thanks to the symmetry properties of the Wigner-$3j$ symbols, the $\delta\alpha\,EE$ and $\delta\alpha\,BB$ angular bispectra are invariant under the index permutation $\ell_2\leftrightarrow\ell_3$. We point out this feature, because in general angular correlation functions involving different fields are not symmetric under the simultaneous interchange of their three multipole numbers $\ell_1$ $\ell_2$ $\ell_3$ (for example the $TTE$, $TET$, and $ETT$ combinations of a bispectrum of CMB temperature and polarization would correspond to three distinct bispectra \cite{bucher2016binned,yadav2007fast,babich2004primordial,shiraishi2019general}). We will use this property in Sec.~\ref{sec:SNR}, when we will estimate the signal-to-noise ratio of the cross-bispectra we have computed here.

\section{\label{sec:plots}Reduced Bispectra}
In order to explicitly evaluate some of the angular bispectra, we now adopt a phenomenological approach. We consider a scale-invariant model of cosmic birefringence, for which the self-correlator of $\delta\alpha$ can be parametrized as follows:
\begin{equation}
\label{eqn:Caa}
C_{\ell}^{\alpha\alpha}\equiv\frac{C_{\alpha\alpha}}{\ell(\ell+1)}, 
\end{equation}
where $C_{\alpha\alpha}$ is a model-dependent parameter which encodes the physics of the scalar field $\chi$ and that quantifies the amplitude of the anisotropic component of the birefringence angle. Indeed, the form of the power spectrum in Eq.~\eqref{eqn:Caa}  finds its motivation in the context of several models of cosmic birefringence \cite{li2008cosmological,caldwell2011cross,zhai2020effects} that approach this behaviour in the large-scale limit (see App.~\ref{app:PowerSpectrum} for more details).

Since we are working under the assumption of statistical isotropy, our bispectra should be invariant under spatial rotations, and this requires the angular bispectra should be proportional to the Wigner $3j$-symbol (see e.g. \cite{2001acoustic}):
\begin{equation}
\label{eqn:average}
\begin{split}
\langle a_{X,\ell_1m_1}a_{Y,\ell_2m_2}&a_{Z,\ell_3m_3}\rangle=\begin{pmatrix}
\ell_1 & \ell_2 & \ell_3 \\
m_1 & m_2 & m_3
\end{pmatrix}B_{\ell_1\ell_2\ell_3}^{XYZ},
\end{split}
\end{equation}
where $B_{\ell_1\ell_2\ell_3}^{XYZ}$ is the \emph{angular averaged bispectrum} and $X,Y,Z$ denote $\delta\alpha,T,E,B$.  However, in order to extract the physical information we display the so-called \emph{reduced bispectra} $b_{\ell_1\ell_2\ell_3}^{XYZ}$, that are related to the angular averaged one via
\begin{equation}
\label{eqn:reduced}
B_{\ell_1\ell_2\ell_3}^{XYZ}=G_{\ell_1\ell_2\ell_3}b_{\ell_1\ell_2\ell_3}^{XYZ},
\end{equation}
where the function $G_{\ell_1\ell_2\ell_3}$ is defined as:
\begin{equation}
\label{eqn:G}
\begin{split}
G_{\ell_1\ell_2\ell_3}\equiv-2\sqrt{\frac{(\ell_2+2)!(\ell_3+2)!}{(\ell_2-2)!(\ell_3-2)!}}\Big\{\ell_1&(\ell_1+1)\left[\ell_3(\ell_3+1)+\ell_2(\ell_2+1)-\ell_1(\ell_1+1)\right]+\\
&+\ell_2(\ell_2+1)\left[\ell_3(\ell_3+1)-\ell_2(\ell_2+1)+\ell_1(\ell_1+1)\right]+\\
&\quad+(\ell_3+2)(\ell_3-1)\left[\ell_1(\ell_1+1)-\ell_2(\ell_2+1)-\ell_3(\ell_3+1)\right]\Big\}^{-1}I_{\ell_1\ell_2\ell_3}^{0,2,-2}.
\end{split}
\end{equation}
Indeed such an expression matches the more common one for $\ell_{T}=$ even \cite{2001acoustic}, as can be shown via standard techniques in quantum theory of angular momentum,
\begin{equation}
\label{eqn:komu}
G_{\ell_1\ell_2\ell_3}\xrightarrow{\ell_{T}\,=\,\text{even}}I_{\ell_1\ell_2\ell_3}^{0,0,0},
\end{equation}
and can be found by exploiting the recursive formulas for the Wigner $3j$-symbols \cite{varshalovich1988quantum,kamionkowski2011odd}. The definition of Eq.~\eqref{eqn:G} is more general, since it remains non-zero for $\ell_{T}=$ odd. Differently, $I_{\ell_1\ell_2\ell_3}^{0,0,0}$ is vanishing for $\ell_{T}=$ odd, being it used e.g. for ensuring the parity-invariance of the primordial $TTT$ angular bispectrum in the context of parity-conserving inflationary models \cite{2001acoustic,kamionkowski2011odd,komatsu2010hunting}.

We have shown in Sec.~\ref{sec:bispectra} that our angular bispectra contain both parity-even and parity-odd components, so with the help of Eqs.~\eqref{eqn:average}-\eqref{eqn:reduced}, we can find the expression of the reduced bispectra associated with Eqs.~\eqref{eqn:ATE}-\eqref{eqn:ATB} and Eqs.~\eqref{eqn:AEE}-\eqref{eqn:AEB}. In particular we notice that the full reduced bispectrum can be regarded as a complex quantity, for which the real part corresponds to the $\ell_{T}$-even component, whereas the purely imaginary part corresponds to the $\ell_{T}$-odd component, that is (see App.~\ref{app:parRel}): 
\begin{equation}
\label{eqn:complex}
b_{\ell_1\ell_2\ell_3}^{XYZ}=b_{\ell_1\ell_2\ell_3}^{XYZ,\text{even}}+ib_{\ell_1\ell_2\ell_3}^{XYZ,\text{odd}}.
\end{equation}

\begin{figure}
	\subfloat[][\label{fig:ATEzero}Real part of the reduced bispectrum.]
	{\includegraphics[width=.45\textwidth]{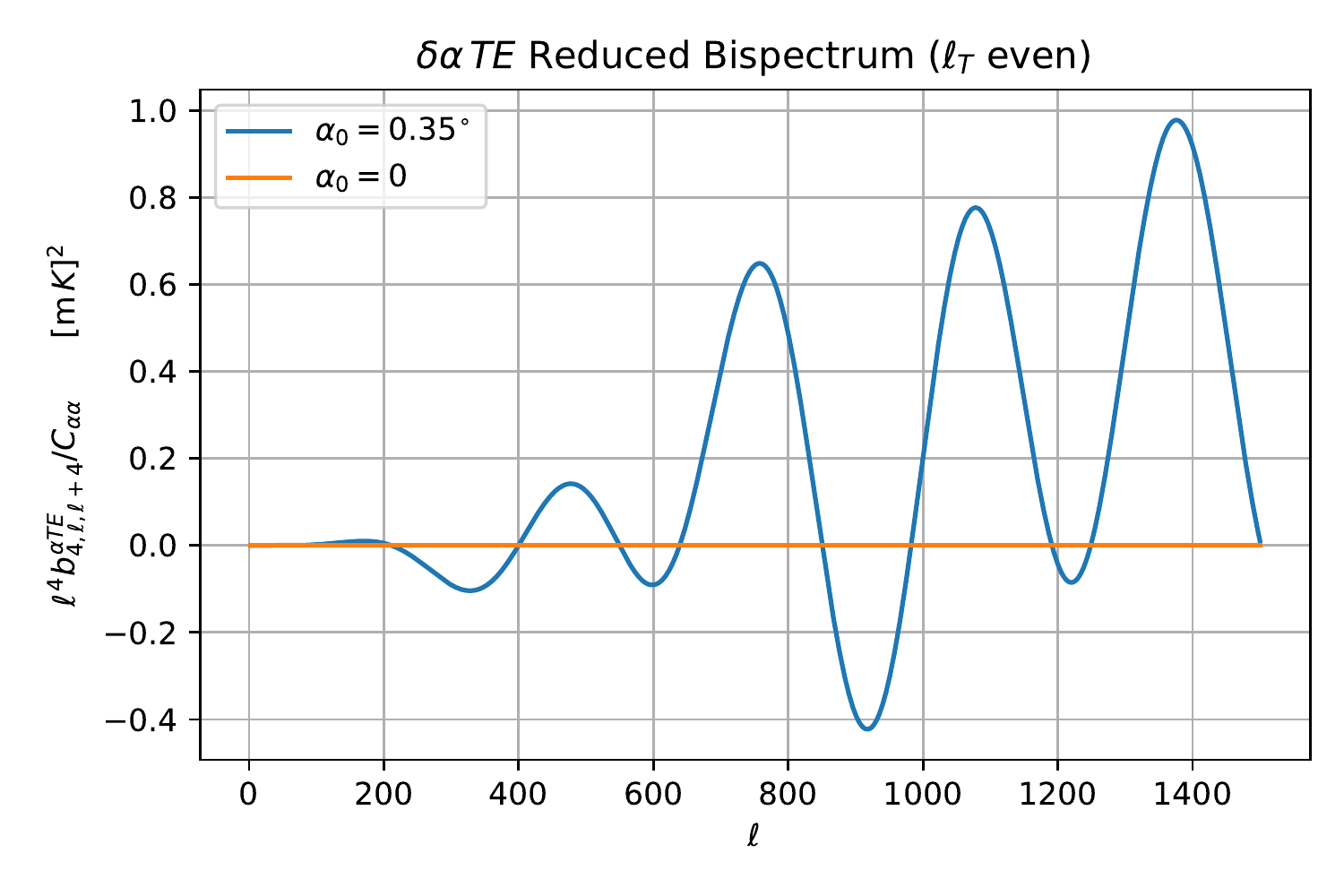}} 
	\quad
	\subfloat[][Imaginary part of the reduced bispectrum.]
	{\includegraphics[width=.45\textwidth]{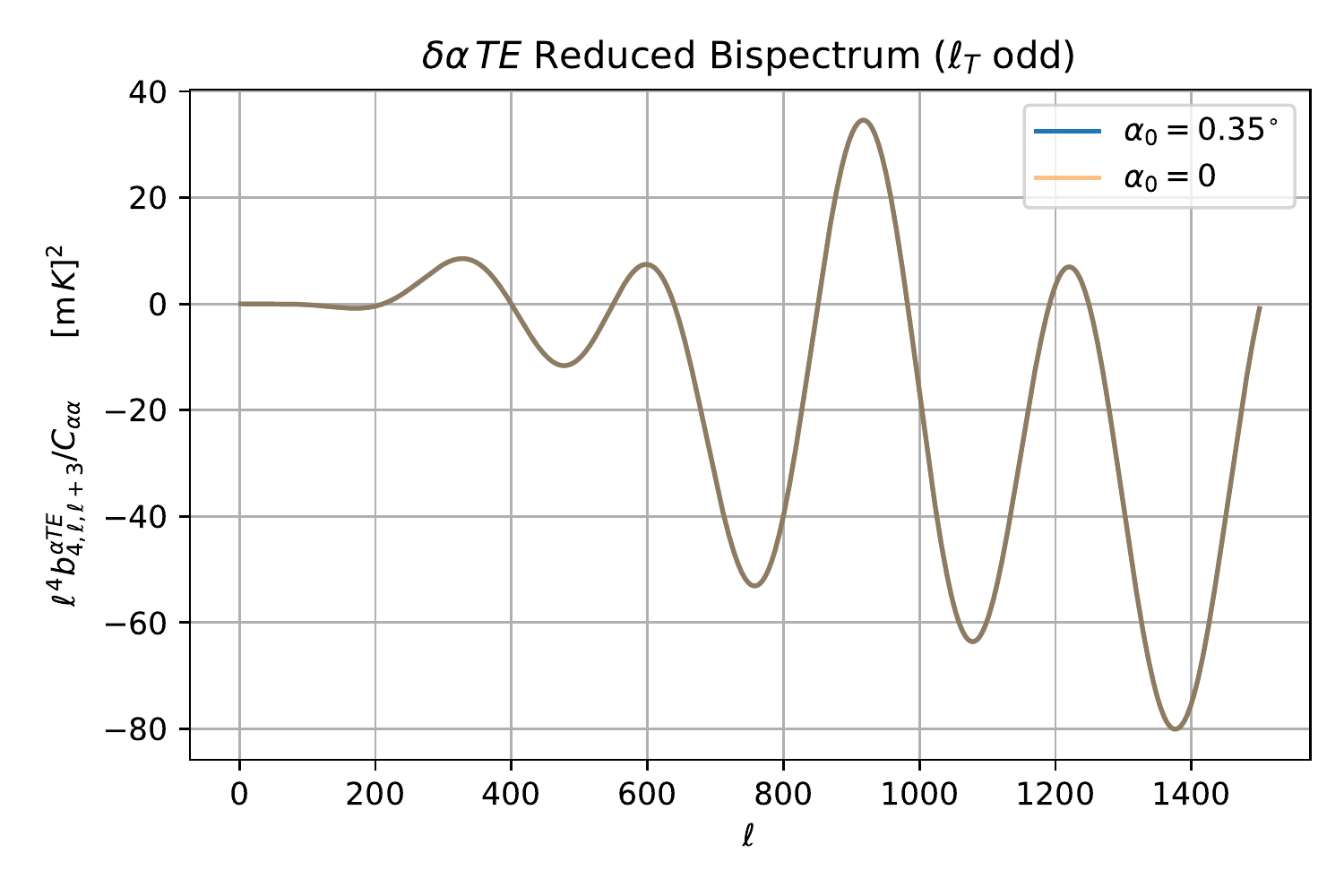}}
	\caption{Plot of $\ell^4_{2}b_{\ell_1\ell_2\ell_3}^{\alpha TE}/C_{\alpha\alpha}$ in units of $[\text{m\,K}]^2$.}
	\label{fig:ATE}
\end{figure}

\begin{figure}
	\subfloat[][Real part of the reduced bispectrum.]
	{\includegraphics[width=.45\textwidth]{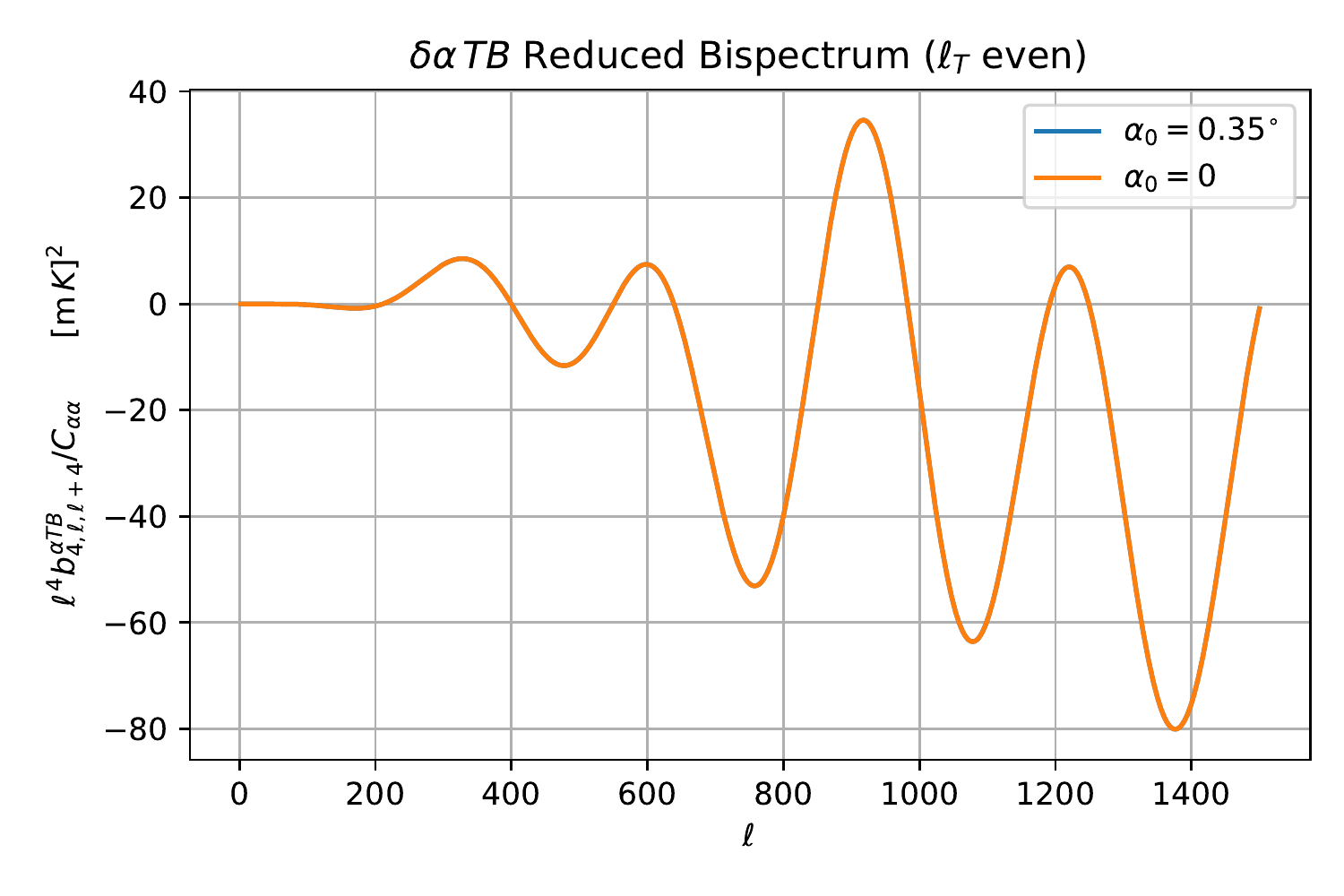}} \quad
	\subfloat[][\label{fig:ATBzero}Imaginary part of the reduced bispectrum.]
	{\includegraphics[width=.45\textwidth]{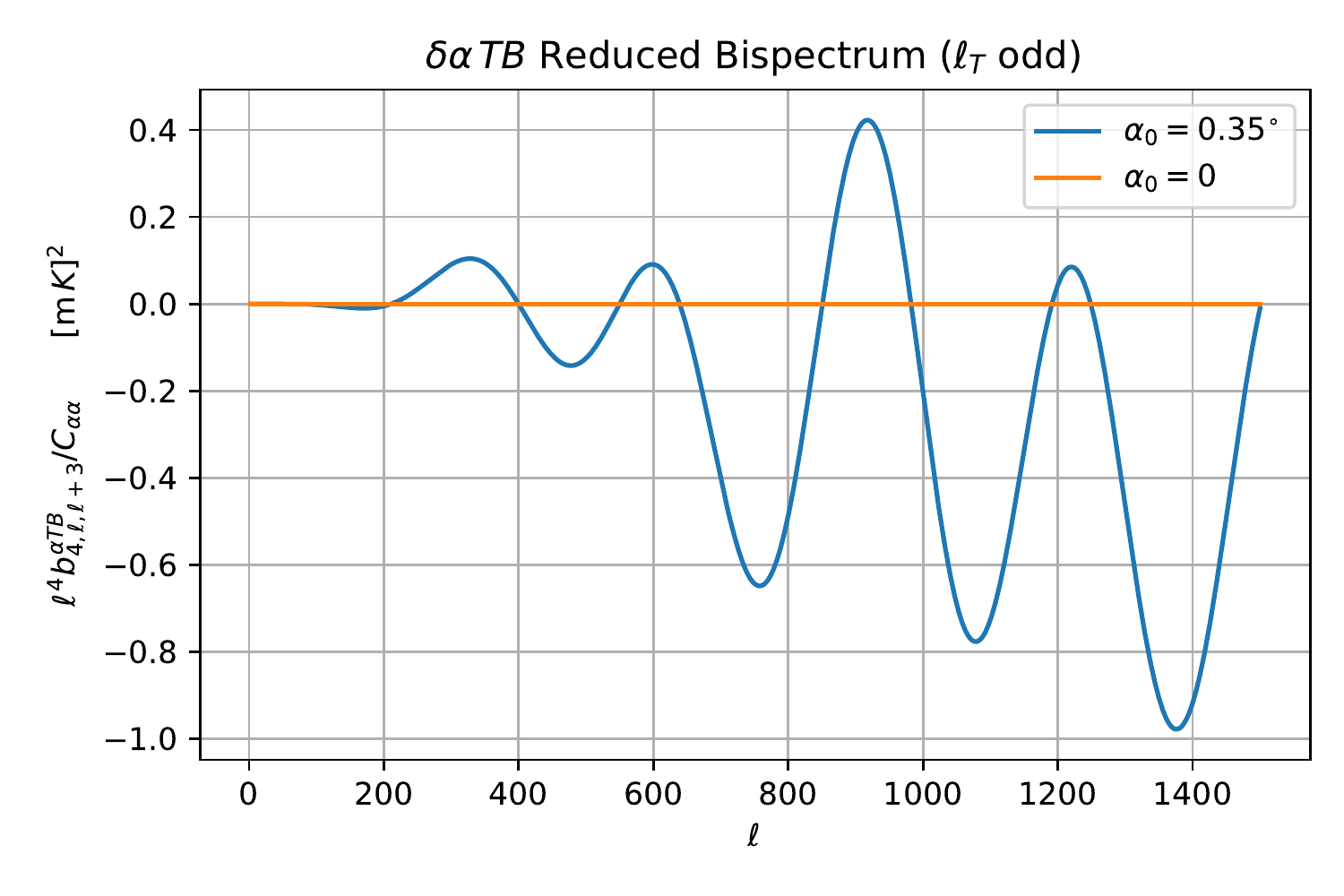}}
	\caption{Plot of $\ell^4_{2}b_{\ell_1\ell_2\ell_3}^{\alpha TB}/C_{\alpha\alpha}$ in units of $[\text{m\,K}]^2$.}
	\label{fig:ATB} 
\end{figure}

\begin{figure}
	\subfloat[][\label{fig:AEEzero}Real part of the reduced bispectrum.]
	{\includegraphics[width=.45\textwidth]{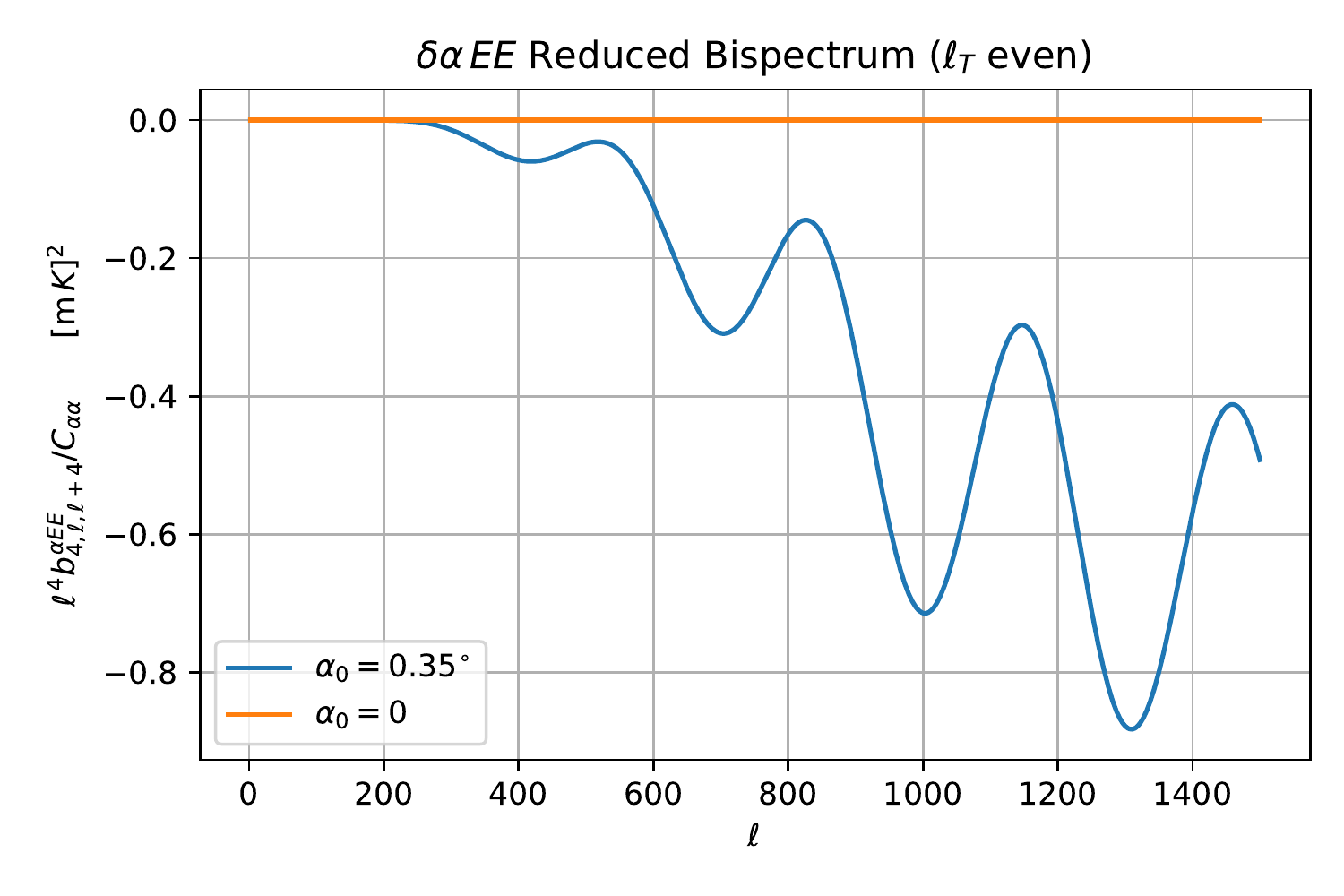}} \quad
	\subfloat[][Imaginary part of the reduced bispectrum.]
	{\includegraphics[width=.45\textwidth]{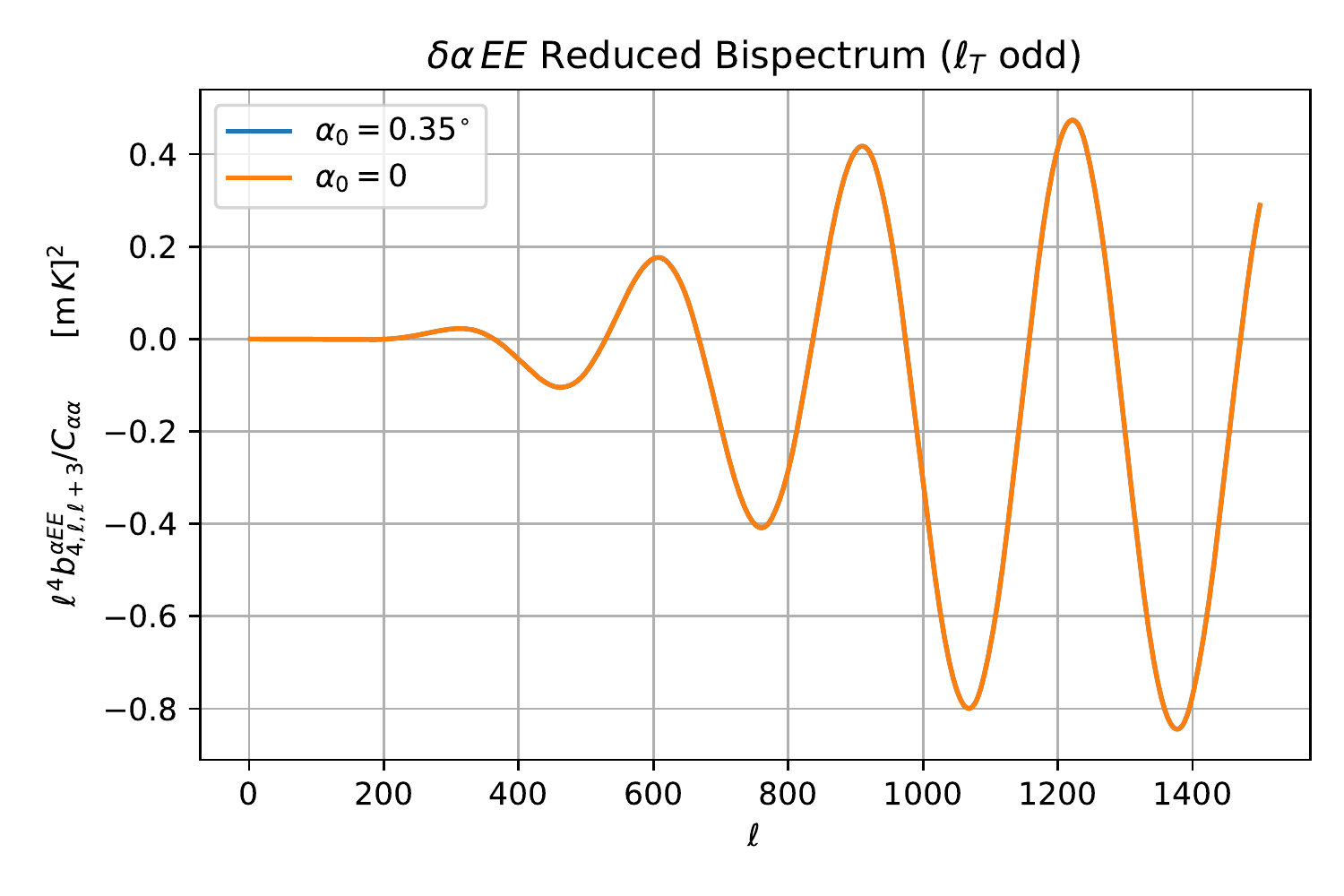}}
	\caption{Plot of $\ell^4_{2}b_{\ell_1\ell_2\ell_3}^{\alpha EE}/C_{\alpha\alpha}$ in units of $[\text{m\,K}]^2$.}
	\label{fig:AEE}
\end{figure}

\begin{figure}
	\subfloat[][\label{fig:ABBzero}Real part of the reduced bispectrum.]
	{\includegraphics[width=.45\textwidth]{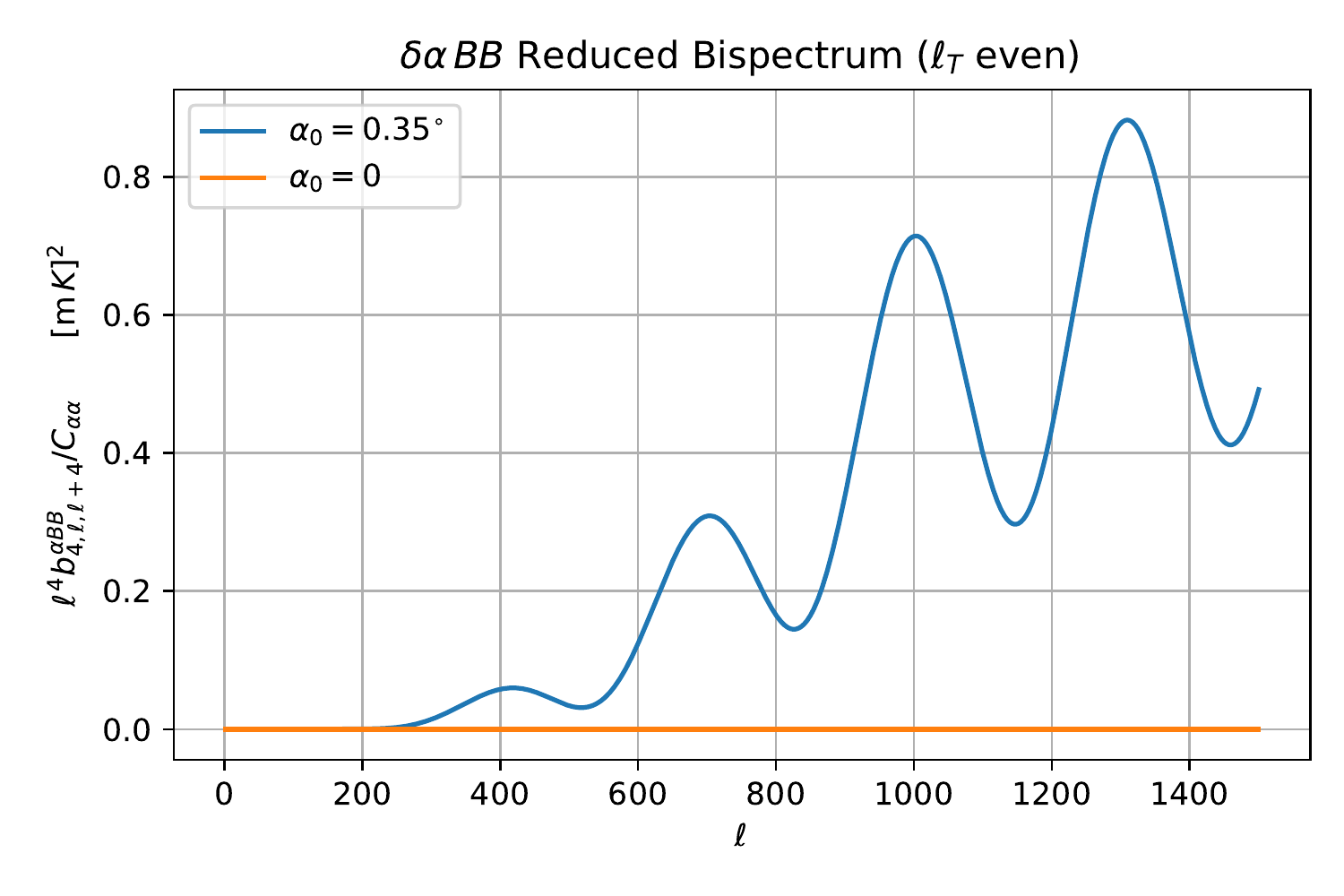}} \quad
	\subfloat[][Imaginary part of the reduced bispectrum.]
	{\includegraphics[width=.45\textwidth]{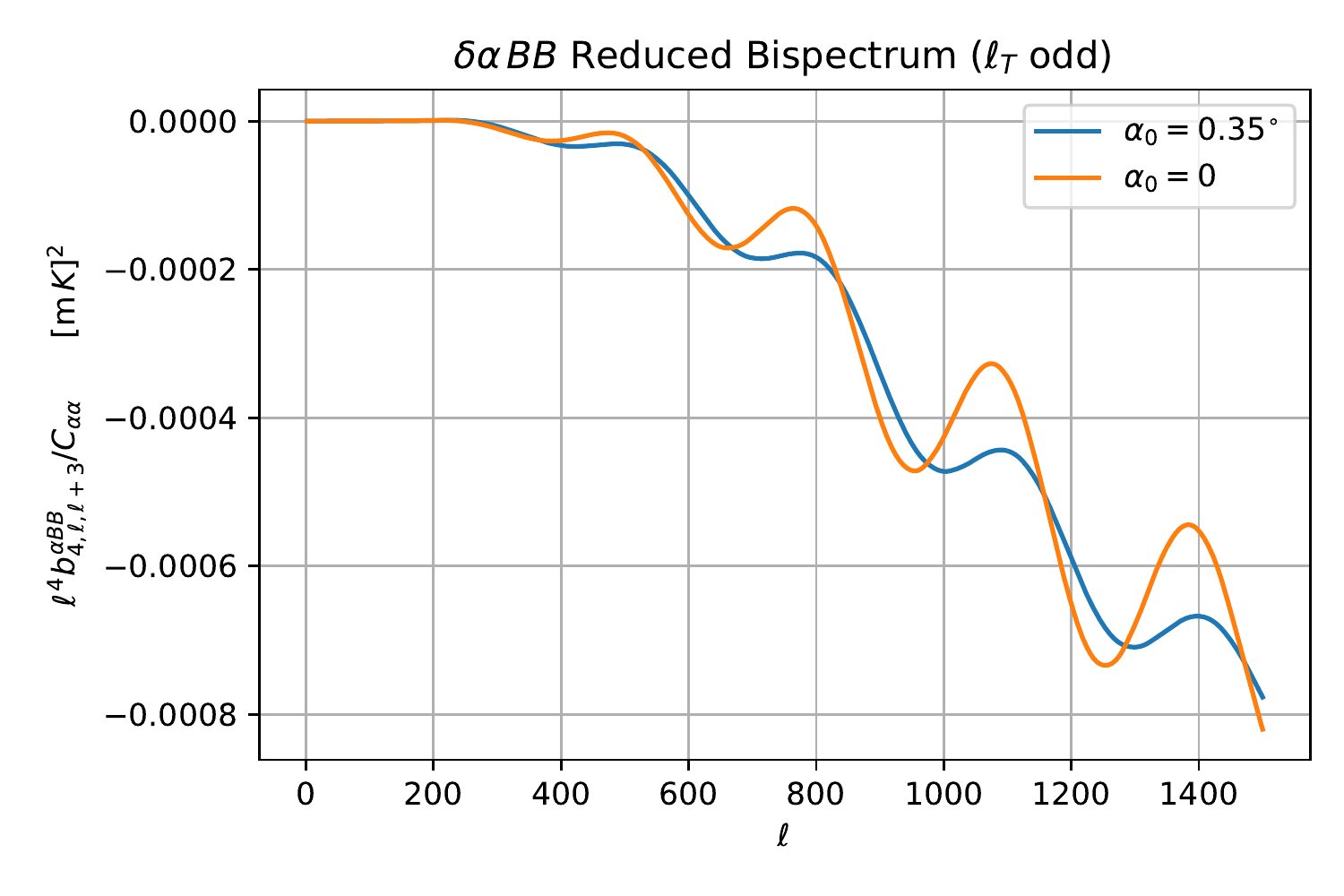}}
	\caption{Plot of $\ell^4_{2}b_{\ell_1\ell_2\ell_3}^{\alpha BB}/C_{\alpha\alpha}$ in units of $[\text{m\,K}]^2$.}
	\label{fig:ABB}
\end{figure}

\begin{figure}
	\subfloat[][Real part of the reduced bispectrum.]
	{\includegraphics[width=.45\textwidth]{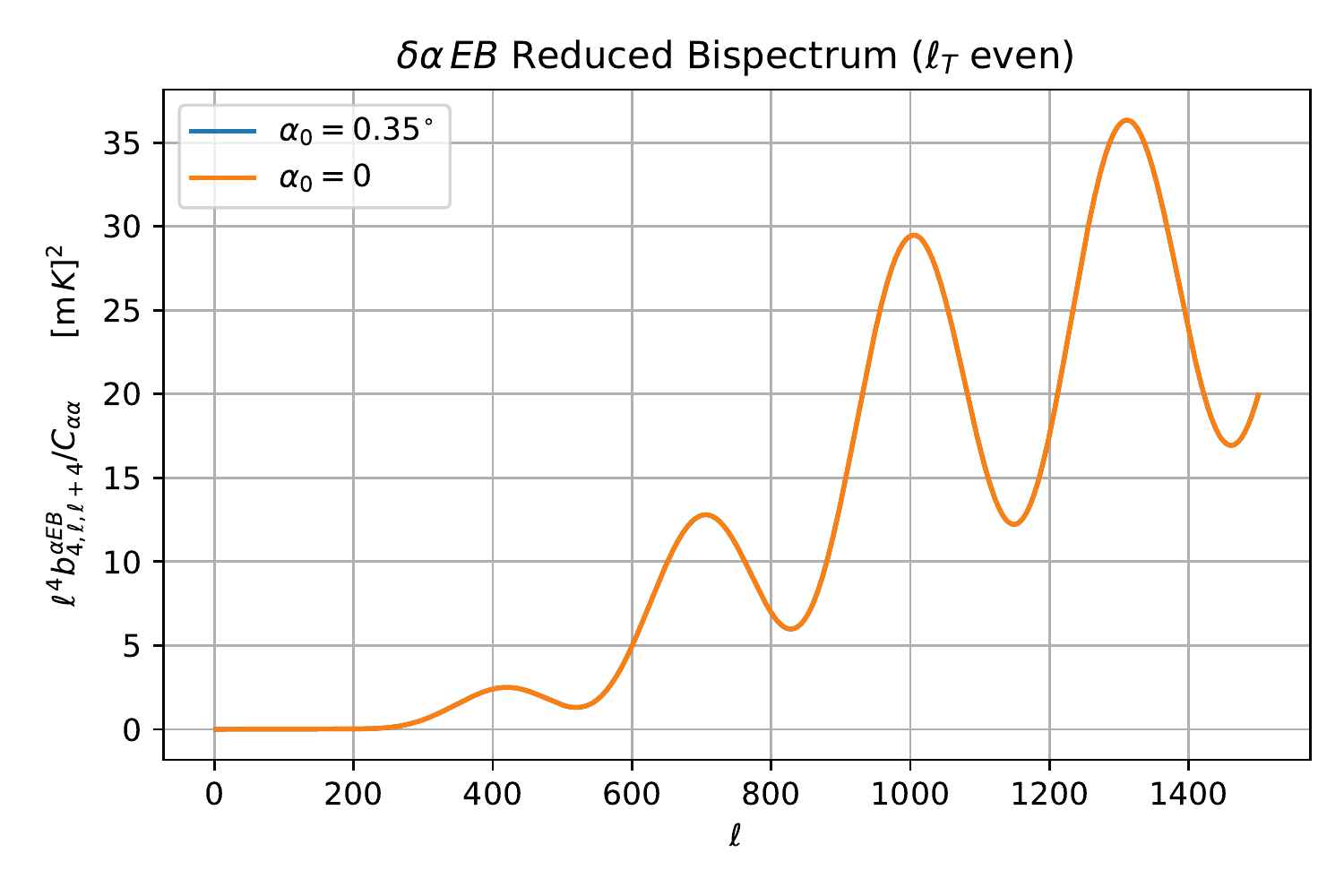}} \quad
	\subfloat[][\label{fig:AEBzero}Imaginary part of the reduced bispectrum.]
	{\includegraphics[width=.45\textwidth]{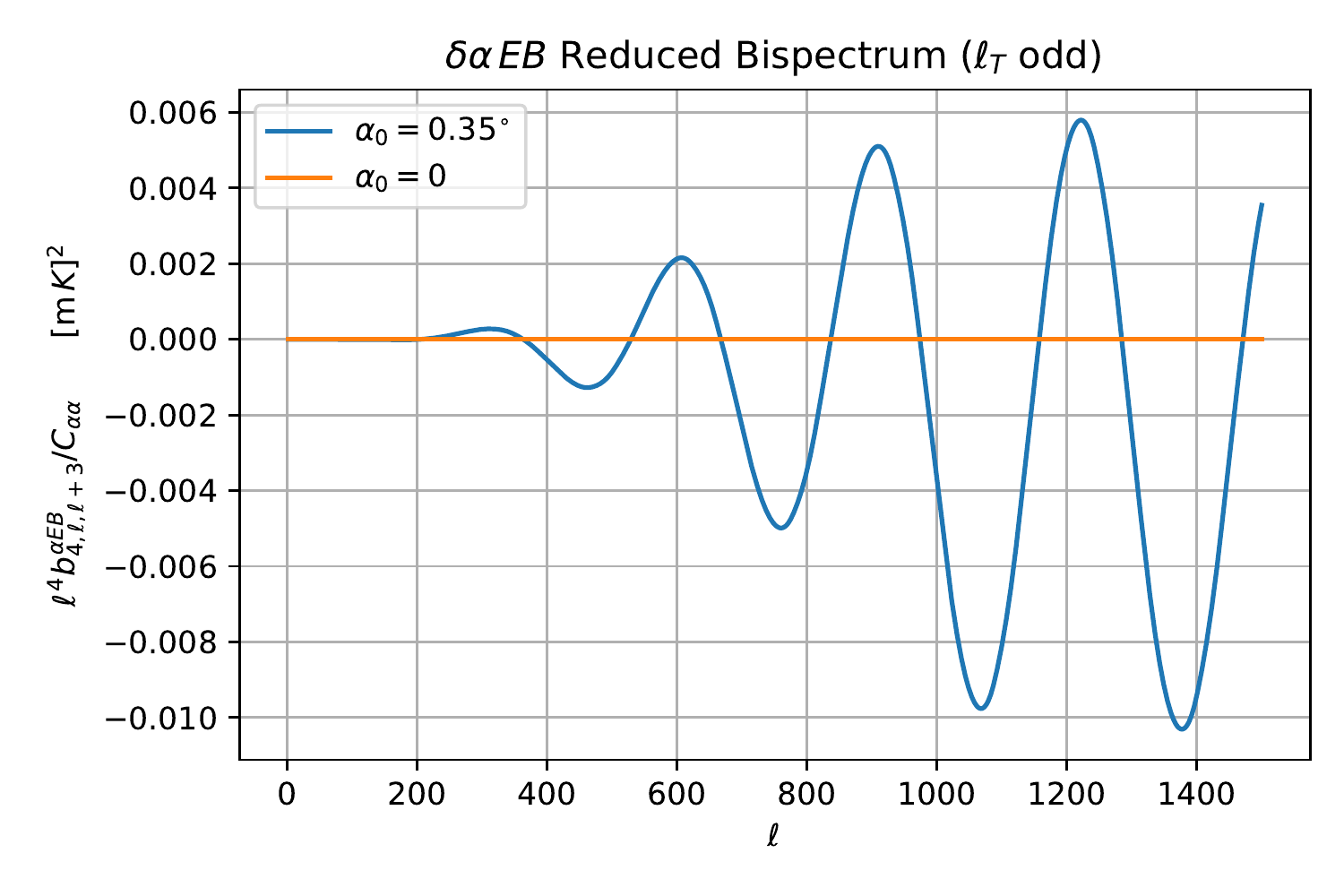}}
	\caption{Plot of $\ell^4_{2}b_{\ell_1\ell_2\ell_3}^{\alpha EB}/C_{\alpha\alpha}$ in units of $[\text{m\,K}]^2$.}
	\label{fig:AEB}
\end{figure}

Therefore, we can adopt such a decomposition and plot the reduced bispectra starting from the angular three-point correlation functions we have computed in Sec.~\ref{sec:bispectra}. In order to display our bispectra we fix two of the three different $\ell$'s by using the following configurations \cite{zhai2020effects}:
\begin{align}
	\ell_{T} &=\text{even}: & \left\{\ell_1,\ell_2,\ell_3\right\} &=\left\{4,\ell,\ell+4\right\}\\
	\ell_{T} &=\text{odd}: & \left\{\ell_1,\ell_2,\ell_3\right\} &=\left\{4,\ell,\ell+3\right\}
\end{align}
that automatically determine the overall parity properties, and ensure the triangular selection rule guaranteed by the Wigner $3j$-symbol, i.e.
\begin{equation}
|\ell_1-\ell_2|\le\ell_3\le\ell_1+\ell_2.
\end{equation}
We numerically compute the reduced angular bispectra by using Eq.~\eqref{eqn:Caa}, and the Boltzmann code CLASS \cite{lesgourgues2011cosmic} to evaluate the CMB angular power spectra: the results are shown in Figs.~\ref{fig:ATE}-\ref{fig:AEB}, where we have assumed a zero tensor-to-scalar ratio and taken into account the contributions from the weak gravitational lensing.\footnote{We have checked that introducing a non-zero tensor-to-scalar ratio $r$, consistent with present constraints~\cite{Planck:2018jri,BICEP:2021xfz,Tristram:2021tvh}, does not affect our plots and main conclusions in a significant way.}
	
Let us notice that the oscillating behaviour exhibited by the bispectra involving a single polarization field in Figs.~\ref{fig:ATE}-\ref{fig:ATB} is due to the the fact that these objects are proportional to the CMB cross-correlator $C_{\ell}^{TE}$ \cite{aghanim2020planck}. As mentioned before, an interesting result worth to be noticed is that even by assuming no isotropic cosmic birefringence, anyhow a non-vanishing reduced bispectra with a certain parity is generated (see Figs.~\ref{fig:ATEzero},~\ref{fig:ATBzero},~\ref{fig:AEEzero},~\ref{fig:ABBzero},~\ref{fig:AEBzero}). 

\section{\label{sec:SNR}Estimation of the Signal-to-noise ratio}
In this section, we compute the \emph{signal-to-noise ratio} (SNR) by for the birefringent bispectra of Eqs.~\eqref{eqn:ATE}-\eqref{eqn:ATB} and~\eqref{eqn:AEE}-\eqref{eqn:AEB}. According to its definition, the SNR is the ratio of the signal power to the noise power, and so, in order to estimate the uncertainty in the measurement of the bispectra, we invoke the \emph{Cramer-Rao inequality}, which states that the variance of an unbiased estimator for a given theoretical parameter cannot be less than the diagonal element of the inverse Fisher matrix \cite{kendall1969advanced}. By working in analogy with what is usually done in the context of primordial non-Gaussianity \cite{babich2004primordial,2001acoustic,komatsu2002pursuit,bartolo2004non,akrami2020planckNG,liguori2010primordial}, we consider here the simplest scenario where the form of a given bispectrum is considered known and the only parameter of interest is its overall amplitude. Since we are considering a single parameter, the Fisher matrix reduces to being just a number (see e.g. \cite{cooray2000imprint}):
\begin{equation}
\begin{split}
\label{eqn:F}
\mathsf{F}_{(XYZ)}=&\sum_{\ell_1\le\ell_2\le\ell_3}\sum_{\ell_1'\le\ell_2'\le\ell_3'}\sum_{ii'}B_{\ell_1\ell_2\ell_3}^{i}\left[\mathsf{Cov}(\hat{B}^{i}_{\ell_1\ell_2\ell_3},\hat{B}_{\ell_1'\ell_2'\ell_3'}^{i'})\right]^{-1}B_{\ell_1'\ell_2'\ell_3'}^{i'}.
\end{split}
\end{equation}
where $X,Y,Z=\delta\alpha,T,E,B$ and the index $i$
\begin{equation}
i=(X,Y,Z),(Y,Z,X),(Z,X,Y),\,\dots 
\end{equation}
labels all the possible non-redundant permutations of a fixed triplet of fields (i.e. they are $6$ when $X\ne Y\ne Z$, $1$ when $X=Y=Z$ and $3$ otherwise). $\hat{B}^{XYZ}_{\ell_1\ell_2\ell_3}$ is an unbiased estimator for the observed angular averaged bispectrum \cite{gangui2000cosmic},
\begin{equation}
\begin{split}
&\hat{B}_{\ell_1\ell_2\ell_3}^{XYZ}\equiv\sum_{m_1m_2m_3}\begin{pmatrix} \ell_1 & \ell_2 & \ell_3 \\
m_1 & m_2 & m_3
\end{pmatrix}a_{\ell_1m_1}^{X,\text{obs}}a_{\ell_2m_2}^{Y,\text{obs}}a_{\ell_3m_3}^{Z,\text{obs}},\end{split}
\end{equation}
and $\mathsf{Cov}(\hat{B}^{i}_{\ell_1\ell_2\ell_3},\hat{B}_{\ell_1'\ell_2'\ell_3'}^{i'})$ is the covariance matrix element. Differently from what is done e.g. in \cite{yadav2007fast}, we are not summing over all the possible combinations of different fields, but we are treating separately each contribution from the five non-vanishing bispectra that we have found in Sec.~\ref{sec:bispectra} (a similar approach is adopted in the context of CMB bispectra induced by weak gravitational lensing \cite{hu2000weak}). The covariance matrix element is defined as
\begin{equation}
\label{eqn:cov}
\begin{split}
&\mathsf{Cov}(\hat{B}^{XYZ}_{\ell_1\ell_2\ell_3},\hat{B}_{\ell_1'\ell_2'\ell_3'}^{X'Y'Z'})\equiv\langle\hat{B}_{\ell_1\ell_2\ell_3}^{XYZ}\hat{B}_{\ell_1'\ell_2'\ell_3'}^{X'Y'Z'}\rangle-\langle\hat{B}_{\ell_1\ell_2\ell_3}^{XYZ}\rangle\langle\hat{B}_{\ell_1'\ell_2'\ell_3'}^{X'Y'Z'}\rangle.
\end{split}
\end{equation}
As we are going to show, the first term encodes several contributions, and some of them are at least quadratic in $\delta\alpha$, whereas the second term is at least quartic in $\delta\alpha$ because, as can be seen from Eqs.~\eqref{eqn:ATE}-\eqref{eqn:ATB} and~\eqref{eqn:AEE}-\eqref{eqn:AEB}, all our bispectra are proportional to $C_{\ell_1}^{\alpha\alpha}$. For this reason we can disregard the second term on the right-hand side of Eq.~\eqref{eqn:cov}, and approximate the covariance matrix element as 
\begin{equation}
\label{62}
\begin{split}
\mathsf{Cov}(\hat{B}^{XYZ}_{\ell_1\ell_2\ell_3},\hat{B}_{\ell_1'\ell_2'\ell_3'}^{X'Y'Z'})\simeq\sum_{m_1m_2m_3}\sum_{m_1'm_2'm_3'}\begin{pmatrix} 
\ell_1 & \ell_2 & \ell_3 \\		
m_1 & m_2 & m_3\end{pmatrix}&\begin{pmatrix} 
\ell_1' & \ell_2' & \ell_3' \\		
m_1' & m_2' & m_3'\end{pmatrix}\times\\
&\times\expval{a_{\ell_1m_1}^Xa_{\ell_2m_2}^Ya_{\ell_3m_3}^Za_{\ell_1'm_1'}^{X'}a_{\ell_2'm_2'}^{Y'}a_{\ell_3'm_3'}^{Z'}}_{\text{obs}},
\end{split}
\end{equation}
where the underscript ``obs'' is just a shorthand notation to denote that all the harmonic coefficients within the ensemble average are the observed ones. In our case, one of three fields $X,Y,Z$ is always set to be $\delta\alpha$, and since we are working at the leading perturbative order, here we can neglect the factor $\exp(\pm2i\delta\alpha)$ appearing in Eq.~\eqref{eqn:harmonic}. As a consequence, according to our assumptions, there are no connected terms arising from non-Gaussian contributions in the six-point correlation function above, and so we can again exploit Eq.~\eqref{eqn:gauss}. Furthermore, as shown in Eqs.~\eqref{eqn:AT}-\eqref{eqn:AB}, the observed two-point cross-correlations between $\delta\alpha$ and the CMB maps are simply a rotation of the primordial ones, and so, according to our phenomenological assumptions, they vanish too. Thanks to all these approximations, the covariance matrix element reduces to
\begin{equation}	
\label{eqn:new_cov}
\begin{split}
&\mathsf{Cov}(\hat{B}^{XYZ}_{\ell_1\ell_2\ell_3},\hat{B}_{\ell_1'\ell_2'\ell_3'}^{X'Y'Z'})\simeq\sum_{m_1m_2m_3}\sum_{m_1'm_2'm_3'}\begin{pmatrix}
			\ell_1 & \ell_2 & \ell_3 \\
			m_1 & m_2 & m_3
		\end{pmatrix}\begin{pmatrix}
			\ell_1' & \ell_2' & \ell_3' \\
			m_1' & m_2' & m_3'
		\end{pmatrix}\times\\
		&\times\Big[\langle a_{X,\ell_1m_1}a_{Y,\ell_2m_2}\rangle\langle a_{Z,\ell_3m_3}a_{X',\ell_1'm_1'}\rangle\langle a_{Y',\ell_2'm_2'}a_{Z',\ell_3'm_3'}\rangle+\langle a_{X,\ell_1m_1}a_{Z,\ell_3m_3}\rangle\langle a_{Y,\ell_2m_2}a_{X',\ell_1'm_1'}\rangle\langle a_{Y',\ell_2'm_2'}a_{Z',\ell_3'm_3'}\rangle\,+\\
		&+\langle a_{X,\ell_1m_1}a_{X',\ell_1'm_1'}\rangle\langle a_{Y,\ell_2m_2}a_{Z,\ell_3m_3}\rangle\langle a_{Y',\ell_2'm_2'}a_{Z',\ell_3'm_3'}\rangle+\langle a_{X,\ell_1m_1}a_{Y',\ell_2'm_2'}\rangle\langle a_{Y,\ell_2m_2}a_{Z,\ell_3m_3}\rangle\langle a_{X',\ell_1'm_1'}a_{Z',\ell_3'm_3'}\rangle\,+\\
		&+\langle a_{X,\ell_1m_1}a_{Z',\ell_3'm_3'}\rangle\langle a_{Y,\ell_2m_2}a_{Z,\ell_3m_3}\rangle\langle a_{X',\ell_1'm_1'}a_{Y',\ell_2'm_2'}\rangle+\langle a_{X,\ell_1m_1}a_{Y,\ell_2m_2}\rangle\langle a_{Z,\ell_3m_3}a_{Y',\ell_2'm_2'}\rangle\langle a_{X',\ell_1'm_1'}a_{Z',\ell_3'm_3'}\rangle\,+\\
		&+\langle a_{X,\ell_1m_1}a_{Y,\ell_2m_2}\rangle\langle a_{Z,\ell_3m_3}a_{Z',\ell_3'm_3'}\rangle\langle a_{X',\ell_1'm_1'}a_{Y',\ell_2'm_2'}\rangle+\langle a_{X,\ell_1m_1}a_{Z,\ell_3m_3}\rangle\langle a_{Y,\ell_2m_2}a_{Y',\ell_2'm_2'}\rangle\langle a_{X',\ell_1'm_1'}a_{Z',\ell_3'm_3'}\rangle\,+\\
		&+\langle a_{X,\ell_1m_1}a_{Z,\ell_3m_3}\rangle\langle a_{Y,\ell_2m_2}a_{Z',\ell_3'm_3'}\rangle\langle a_{X',\ell_1'm_1'}a_{Y',\ell_2'm_2'}\rangle+\langle a_{X,\ell_1m_1}a_{X',\ell_1'm_1'}\rangle\langle a_{Y,\ell_2m_2}a_{Y',\ell_2'm_2'}\rangle\langle a_{Z,\ell_3m_3}a_{Z',\ell_3'm_3'}\rangle\,+\\
		&+\langle a_{X,\ell_1m_1}a_{X',\ell_1'm_1'}\rangle\langle a_{Y,\ell_2m_2}a_{Z',\ell_3'm_3'}\rangle\langle a_{Z,\ell_3m_3}a_{Y',\ell_2'm_2'}\rangle+\langle a_{X,\ell_1m_1}a_{Y',\ell_2'm_2'}\rangle\langle a_{Y,\ell_2m_2}a_{X',\ell_1'm_1'}\rangle\langle a_{Z,\ell_3m_3}a_{Z',\ell_3'm_3'}\rangle\,+\\
		&+\langle a_{X,\ell_1m_1}a_{Y',\ell_2'm_2'}\rangle\langle a_{Y,\ell_2m_2}a_{Z',\ell_3'm_3'}\rangle\langle a_{Z,\ell_3m_3}a_{X',\ell_1'm_1'}\rangle+\langle a_{X,\ell_1m_1}a_{Z',\ell_3'm_3'}\rangle\langle a_{Y,\ell_2m_2}a_{X',\ell_1'm_1'}\rangle\langle a_{Z,\ell_3m_3}a_{Y',\ell_2'm_2'}\rangle\,+\\
		&\hspace{239pt}+\langle a_{X,\ell_1m_1}a_{Z',\ell_3'm_3'}\rangle\langle a_{Y,\ell_2m_2}a_{Y',\ell_2'm_2'}\rangle\langle a_{Z,\ell_3m_3}a_{X',\ell_1'm_1'}\rangle\Big]_{\text{obs}}.
\end{split}
\end{equation}
Because of statistical isotropy, we can rewrite the two-point correlation functions in terms of the angular power spectra with Eq.~\eqref{eqn:definition}: by doing this, the first nine terms in the square brackets on the right-hand side of Eq.~\eqref{eqn:new_cov} become proportional to \cite{rotenberg19593}
\begin{equation}
\begin{split}
\sum_{mm'}&\begin{pmatrix}
	L & \ell & \ell \\
	0 & m & -m
\end{pmatrix}\begin{pmatrix}
	L & \ell' & \ell' \\
	0 & m' & -m'
\end{pmatrix}=\sum_{mm'}\frac{(-1)^{\ell+\ell'-m-m'}}{\sqrt{(2\ell+1)(2\ell'+1)}}\delta_{L,0}.
\end{split}
\end{equation}
However, the observable multipoles start from $L\ge2$, and so this means that the term above gives no contribution in the estimation of the signal-to-noise ratio. Therefore we are only left with 
\begin{equation}
\label{eqn:non-diagonal}
\begin{split}
\mathsf{Cov}(\hat{B}^{XYZ}_{\ell_1\ell_2\ell_3},\hat{B}_{\ell_1'\ell_2'\ell_3'}^{X'Y'Z'})&=(-1)^{\ell_T}C_{\ell_1,\text{obs}}^{XX'}C_{\ell_2,\text{obs}}^{YY'}C_{\ell_3,\text{obs}}^{ZZ'}\delta_{\ell_1\ell_1'}\delta_{\ell_2\ell_2'}\delta_{\ell_3\ell_3'}+C_{\ell_1,\text{obs}}^{XX'}C_{\ell_2,\text{obs}}^{YZ'}C_{\ell_3,\text{obs}}^{ZY'}\delta_{\ell_1\ell_1'}\delta_{\ell_2\ell_3'}\delta_{\ell_3\ell_2'}\,+\\
&\quad+C_{\ell_1,\text{obs}}^{XY'}C_{\ell_2,\text{obs}}^{YX'}C_{\ell_3,\text{obs}}^{ZZ'}\delta_{\ell_1\ell_2'}\delta_{\ell_2\ell_1'}\delta_{\ell_3\ell_3'}+(-1)^{\ell_{T}}C_{\ell_1,\text{obs}}^{XY'}C_{\ell_2,\text{obs}}^{YZ'}C_{\ell_3,\text{obs}}^{ZX'}\delta_{\ell_1\ell_2'}\delta_{\ell_2\ell_3'}\delta_{\ell_3\ell_1'}\,+\\
&\quad+(-1)^{\ell_{T}}C_{\ell_1,\text{obs}}^{XZ'}C_{\ell_2,\text{obs}}^{YX'}C_{\ell_3,\text{obs}}^{ZY'}\delta_{\ell_1\ell_3'}\delta_{\ell_2\ell_1'}\delta_{\ell_3\ell_2'}+C_{\ell_1,\text{obs}}^{XZ'}C_{\ell_2,\text{obs}}^{YY'}C_{\ell_3,\text{obs}}^{ZX'}\delta_{\ell_1\ell_3'}\delta_{\ell_2\ell_2'}\delta_{\ell_3\ell_1'}\, ,
\end{split}
\end{equation}
where we have exploited the relation of orthonormality for the Wigner-$3j$ symbols \cite{varshalovich1988quantum}:
\begin{equation}
\sum_{m_1m_2m_3}\begin{pmatrix}
\ell_1 & \ell_2 & \ell_3 \\
m_1 & m_2 & m_3
\end{pmatrix}\begin{pmatrix}
\ell_1 & \ell_2 & \ell_3 \\
m_1 & m_2 & m_3
\end{pmatrix}=1.
\end{equation}

It can seem not formally trivial to obtain the inverse covariance matrix starting from Eq.~\eqref{eqn:non-diagonal}. Anyway, we know that the covariance matrix element is non-vanishing only when connecting the same triplets, i.e. when $(\ell_1,\ell_2,\ell_3)$ is equal to $(\ell_1',\ell_2',\ell_3')$ or to a permutation of it. Thus, since we restrict the summation in the $(\ell_1\le\ell_2\le\ell_3)$ and $(\ell_1'\le\ell_2'\le\ell_3')$ domains, we can observe that the covariance matrix is already diagonal in the triplets space, so that we can rewrite Eq.~\eqref{eqn:F} as
\begin{equation}
\label{eqn:fisher}
\begin{split}
\mathsf{F}_{(XYZ)}=&\sum_{\ell_1\le\ell_2\le\ell_3}\sum_{ii'}B_{\ell_1\ell_2\ell_3}^{i}\left[\mathsf{Cov}(\hat{B}^{i}_{\ell_1\ell_2\ell_3},\hat{B}_{\ell_1\ell_2\ell_3}^{i'})\right]^{-1}B_{\ell_1\ell_2\ell_3}^{i'}.
\end{split}
\end{equation}
Let us just mention that the procedure of the domain restriction in the triplet space we have adopted from the beginning of this section is physically correct: this is due to the fact that any angular averaged bispectrum is symmetric under the simultaneous interchange of its three multipole numbers $\ell_1,\ell_2,\ell_3$ and its three field indices $X,Y,Z$ \cite{bucher2016binned}. Thus, in order to extract the information content, it is enough to study just the subspace $\ell_1\le\ell_2\le\ell_3$, since we are already summing over all the possible field permutations. 

We have now to specify the general formula of Eq.~\eqref{eqn:fisher} for the five bispectra collected in Eqs.~\eqref{eqn:ATE}-\eqref{eqn:ATB} and~\eqref{eqn:AEE}-\eqref{eqn:AEB}. Before doing this, we make a further approximation: since the total angular averaged bispectrum is the sum of different terms, it is reasonable to expect that the dominant contribution in the signal would come from those terms that are non-vanishing even for $\alpha_0=0$. Hence, in this section we consider a regime of purely anisotropic cosmic birefringence, which allows us to replace $C_{\ell}^{\text{obs}}\simeq C_{\ell}$ within the covariance in Eq.~\eqref{eqn:fisher}. 

Moreover, it is convenient to express the signal-to-noise ratio by means of a matrix formalism by defining a proper quadratic form, which should involve data vectors containing all the permutations of the given bispectrum and a suitable expression for the covariance matrix element:
\begin{equation}
\label{eqn:covariance}
\begin{split}
\mathsf{Cov}^{XYZ}_{\ell_1\ell_2\ell_3}&=C_{\ell_1}^{XX'}C_{\ell_2}^{YZ'}C_{\ell_3}^{ZY'}\delta_{\ell_2\ell_3}+C_{\ell_1}^{XY'}C_{\ell_2}^{YX'}C_{\ell_3}^{ZZ'}\delta_{\ell_1\ell_2}+C_{\ell_1}^{XZ'}C_{\ell_2}^{YY'}C_{\ell_3}^{ZX'}\delta_{\ell_1\ell_3}\delta_{\ell_2\ell_3}+\\
&\quad+(-1)^{\ell_{T}}\left[C_{\ell_1}^{XX'}C_{\ell_2}^{YY'}C_{\ell_3}^{ZZ'}+C_{\ell_1}^{XY'}C_{\ell_2}^{YZ'}C_{\ell_3}^{ZX'}\delta_{\ell_1\ell_2}\delta_{\ell_2\ell_3}+C_{\ell_1}^{XZ'}C_{\ell_2}^{YX'}C_{\ell_3}^{ZY'}\delta_{\ell_1\ell_3}\delta_{\ell_2\ell_3}\right].
\end{split}
\end{equation}
Therefore, we can substitute the general expression of Eq.~\eqref{eqn:covariance} in the definition of the Fisher matrix, so that we obtain five formulas for the squared signal-to-noise ratios (one for each birefringent bispectrum), whose explicit forms are listed in App.~\ref{app:FisherFormulas}. We report the results of the numerical evaluation of Eqs.~\eqref{eqn:SNR_ATE}-\eqref{eqn:SNR_ABB} in Tab.~\ref{tab:snr}: they have been obtained by summing up to $\ell_{\text{max}}=200$ and by considering an ideal regime with zero instrumental noise. Our choice for $\ell_{\text{max}}=200$ is dictated by two reasons: in some realistic models for birefringence with a Chern-Simons term these are the typical multipole values up to which the power-spectrum of the anisotropic birefringence angle $C_{\ell}^{\alpha\alpha}$ is approximately scale-invariant, which is the kind of spectrum we are using here as a toy-model; secondly we are going to specialize our Fisher forecast to a typical LiteBIRD-like satellite mission. 
\begin{table}
\caption{\label{tab:snr} Numerical estimation of the SNR for the birefringent bispectra in the ideal case (zero instrumental noise) in the purely anisotropic regime.}
\begin{ruledtabular}
\begin{tabular}{cll}
\textbf{Bispectrum} & \textbf{SNR} (in units of $C_{\alpha\alpha}^{1/2}$) & \textbf{SNR} (if $C_{\alpha\alpha}\sim6\times10^{-5}$) \\
\hline
$\delta\alpha\, TE$	& $\approx80$ & $\approx 0.62$ \\
$\delta\alpha\, TB$	& $\approx1926$ & $\approx 14.92$ \\
$\delta\alpha\, EB$	& $\approx3680$  & $\approx28.51$\\
$\delta\alpha\, EE$	& $\approx83$   & $\approx0.64$\\
$\delta\alpha\, BB$	& $\approx4$   & $\approx0.03$\\
\end{tabular}
\end{ruledtabular}
\end{table}
In Tab.~\ref{tab:snr} we have reported both the signal-to-noise ratio in units of $\sqrt{C_{\alpha\alpha}}$ and according to the current tightest upper observational constraints on the amplitude of a scale-invariant angular power spectrum of anisotropic cosmic birefringence from ACTPol \cite{namikawa2020atacama} and SPTpol \cite{bianchini2020searching}:
\begin{align}
C_{\alpha\alpha}&<6.3\times10^{-5}\,\left[\text{rad}^2\right]\qquad\text{($95\%$ C.L., ACTPol, SPTpol)}.
\end{align}
Present constraints on anisotropic birefringence, provided as amplitude $C_{\alpha\alpha}$ of the scale-invariant spectrum of $\delta\alpha(\versor{n})$, are also given by \textit{Planck} 2015 \cite{contreras2017constraints}, by \textit{Planck} 2018 \cite{gruppuso2020planck}, and Bicep-Keck data \cite{ade2017bicep2}. Other compatible, even though weaker, constraints on this parameter are provided by Polarbear \cite{ade2015polarbear} and WMAP \cite{gluscevic2012first} observations. Future CMB observations are
expected to improve the current bounds on cosmic birefringence by orders of magnitude \cite{pogosian2019future}.

Similarly, we can now examine the detection possibility of our bispectra for a future CMB experiment, like the LiteBIRD satellite (see e.g. \cite{matsumura2014mission}), as we have mentioned before. Here, we analyse an idealized experimental configuration where foregrounds are neglected. Thus, the signal-to-noise ratio is evaluated using again Eqs.~\eqref{eqn:SNR_ATE}-\eqref{eqn:SNR_ABB}, but this time we have to add the instrumental noise to the power spectra appearing in Eq.~\eqref{eqn:covariance}. By assuming a Gaussian form for the experimental window function of beam $\theta$, and by considering a white instrumental noise, we can use the Knox formulae \cite{knox1995determination} for the CMB correlators:
\begin{align}
\label{eqn:noise_CMB}
C_{\ell}^{XY}&\mapsto C_{\ell}^{XY} + \mu\,w^{-1}\exp\left(\frac{\ell^2\theta^2}{8\ln 2}\right)
\end{align}
where $\mu$ is a numerical factor defined in Tab.~\ref{tab:mu}, and $\sqrt{1/w}$ is the power noise. \\

\begin{table}
\caption{\label{tab:mu} Value of the factor $\mu$ for all the CMB maps.}
\begin{ruledtabular}
	\begin{tabular}{lccc}
		$\sfrac{X}{Y}$ & $T$ & $E$ & $B$ \\
		$T$	& $1$ & $0$  & $0$ \\
		$E$	& $0$ & $2$ &  $0$ \\
		$B$	& $0$  & $0$ &  $2$\\
	\end{tabular}
\end{ruledtabular}
\end{table}

From Tab.~\ref{tab:mu} we can see that the CMB cross-correlations have no noise contribution, since the noises from different maps are not correlated (see e.g. \cite{ng1999correlation}). A more complicated expression has to be considered instead for the auto-spectrum of anisotropic cosmic birefringence \cite{caldwell2011cross}:
\begin{equation}
\label{eqn:noise_AA}
C_{\ell}^{\alpha\alpha}\mapsto C_{\ell}^{\alpha\alpha}+\left\{\sum_{L_1L_2}\frac{\pi (2L_1+1)(2L_2+1)(C_{L_1}^{EE})^2e^{-\left(L_1^2+L_2^2\right)\theta^2/(8\ln 2)}}{\left[C_{L_1}^{BB}e^{-L_1^2\theta^2/(8\ln2)} + 2\,w^{-1}\right]\left[C_{L_2}^{EE}e^{-L_2^2\theta^2/(8\ln2)} + 2\,w^{-1}\right]}\begin{pmatrix}
L_1 & \ell & L_2 \\
2 & 0 & -2
\end{pmatrix}^2\right\}^{-1}.
\end{equation}
By substituting Eqs.~\eqref{eqn:noise_CMB}-\eqref{eqn:noise_AA} in Eqs.~\eqref{eqn:SNR_ATE}-\eqref{eqn:SNR_ABB} and by multiplying the overall result by the fraction of the sky $f_{\text{sky}}$ to which the experiment is sensitive, we can estimate the SNR according for a LiteBIRD-like experiment to the following instrumental parameters \cite{hazumi2020litebird,capparelli2020cosmic}:
\begin{equation}
\theta = 30^{\prime},\qquad w^{-1/2}=4.5\text{ $\mu$K-arcmin},\qquad f_{\text{sky}}=0.7.
\end{equation}
Our results are reported in Tab.~\ref{tab:litebird}.
\begin{table}
\caption{\label{tab:litebird} Numerical estimation of the SNR for the birefringent bispectra (including the LiteBIRD satellite instrumental noise) in the purely anisotropic regime.}
\begin{ruledtabular}
	\begin{tabular}{cl}
		\textbf{Bispectrum} & \textbf{SNR} (if $C_{\alpha\alpha}\sim6\times10^{-5}$) \\
		\hline
		$\delta\alpha\, TE$	 & $\approx 0.0661$ \\
		$\delta\alpha\, TB$ & $\approx 4.0635$ \\
		$\delta\alpha\, EB$	 & $\approx 7.5658$\\
		$\delta\alpha\, EE$	 & $\approx 0.0543$\\
		$\delta\alpha\, BB$	 & $\approx  0.0004$\\
	\end{tabular}
\end{ruledtabular}
\end{table}
From Tab.~\ref{tab:litebird} we can see that the bispectra involving a single $B$-mode in the polarization pattern, i.e. $\expval{\delta\alpha\,TB}$ and $\expval{\delta\alpha\,EB}$, are the more promising for what concerns a possible future detection. This is due to the form assumed by the covariance matrix in Eq.~\eqref{eqn:covariance} for these two specific cases, which, strictly speaking, once inverted results in a matrix of fractions with denominators that are smaller than in the case of the other bispectra. The reason for that is the dependence of the covariance matrix elements on quantities like $C_{\ell}^{TB}$ or $C_{\ell}^{EB}$ that are null by hypothesis (and also the fact that the covariance matrix will contain terms proportional to the power spectrum of the $B$ modes). Moreover, it is not surprising that SNR for the $\expval{\delta\alpha\,EB}$ bispectrum is larger than that for the $\expval{\delta\alpha\,TB}$, since in the former case the covariance matrix elements depend on the CMB temperature power spectrum, whose amplitude is estimated to be larger than that of $C_{\ell}^{EE}$ \cite{aghanim2020planck}. These results and considerations further motivate our choice of performing a Fisher forecast for a LiteBIRD-like experiment, that is a $B$-mode devoted satellite mission. They also justify our choice of analysing the SNR for specific combinations of the various fields involved in the observations, since, according to our results, we do expect that the bispectra involving a single  $B$-mode would provide the dominant contribution to the total SNR. The results shown in Tab.~\ref{tab:litebird} are indeed quite promising, showing in principle that the constraints they could provide are comparable to the present limits we have on anisotropic birefringence. \\
A few further comments are in order here. We have checked that the SNR remains very small either if we start from $l_{\rm min}=10$ up to $l_{\rm max}=200$, or in the case where we stop at $l_{\rm max}=10$, which is indeed telling us that the main contribution to the SNR comes from squeezed configurations where, e.g. $\ell_1 \ll \ell_2 \sim \ell_3$. 
Also, as mentioned after Eq.~\eqref{eqn:ATB}, we accounted only for the recombination epoch as the time of polarization generation. We do expect that adding the reionization epoch as well would not dramatically modify the SNR. Indeed we have verified that this is the case, in the simplifying assumption that the power spectrum of anisotropic birefringence from the reionization epoch is scale invariant and with the same amplitude as that adopted in~\eqref{eqn:Caa}. For example we find that for the $\expval{\alpha EB}$ bispectrum the SNR slightly increases to $\text{SNR}_{\alpha EB}^{\alpha_0=0}=8.0334$.

\section{\label{sec:end}Conclusions}
Cosmic Birefringence represents a fascinating phenomenon which is useful to investigate parity-breaking effects in physics. In this paper we have considered an extension of Maxwell electromagnetism consisting in a Chern-Simons coupling between the electromagnetic field and a scalar field. In particular we have shown what is the relation between the observed angular correlation functions involving the anisotropic birefringence angle and the CMB maps, and their unrotated counterparts. The observed angular power spectra are simply obtained by a rotation of the primordial ones, but this simple relation cannot be extended to higher-order correlators. Indeed we have computed the angular three-point functions and the corresponding reduced bispectra: we have shown that even by assuming that $\delta\alpha$, $T$ and the (unrotated primordial) $E$ and $B$ fields are all Gaussian random fields, and although any two-point cross-correlation $C_{\ell}^{\alpha X}$ (with $X=T,E,B$) is taken to be zero, we obtain non-vanishing parity-breaking bispectra. Moreover, from the results shown in Figs.~\ref{fig:ATE}-\ref{fig:AEB}, it is possible to see that there are non-vanishing contributions also in a purely anisotropic regime. We have estimated the signal-to-noise ratio for the birefringent bispectra, showing that a future LiteBIRD-like  experiment could be eventually able to detect the signals encoded in the $\delta\alpha TB$ and $\delta\alpha EB$ bispectra. 

Our analysis does not account for the presence of foregrounds, which goes beyond the scope of this paper. In computing the signal-to-noise ratio we have not accounted for possible secondary effects, e.g. due to lensing, that could in principle switch on some connected terms in Eq.~\eqref{62}. On the other hand, notice that we have not accounted  for the possibility of de-lensing (see, e.g.~\cite{Knox:2002pe,Kesden:2002ku,Hirata:2002jy,Seljak:2003pn} and in particular for our purposes, the discussion in~\cite{Kalaja:2020mkq,Coulton:2019odk}), in which case we would expect an improvement of the SNR, especially for the case of bispectra involving two $B$ modes. A detailed investigation about the validity of such approximations is left to future investigations.

These results have been obtained by using a scale-invariant model of cosmic birefringence, so that a future development of this work should be to extend our treatment to other models, like the ones described in, e.g., \cite{caldwell2011cross,capparelli2020cosmic}. Other intriguing possibilities arise if we ask ourselves what would happen if we relax some of the assumptions listed in Sec.~\ref{sec:math}. Indeed, an interesting development of this work would be including the possibility of primordial non-Gaussianity effects, and/or taking into account also eventual primordial two-point correlation functions between $\delta\alpha$ and the CMB maps (as done for instance in the context of CMB bispectra  in \cite{zhai2020effects}). Moreover we have assumed an underlying parity-invariant inflationary model, so that the only source of parity-violation in our bispectra comes from cosmic birefringence, but it would be interesting to include also primordial parity-violation from inflation in our analysis. Finally an interesting aspect of the results we found is the possibility to perform a tomographic analysis of the effects of anisotropic birefringence from both the recombination and the reionization epochs, following our results in App.~\ref{app:Reionization} \cite{workinprogress}. However we leave these discussions for future works.

\begin{acknowledgments}
The authors would like to thank Michele Liguori for many useful discussions about some technical aspects during the late stages of this work. N. Bartolo and A. Gruppuso acknowledge support from the COSMOS network (www.cosmosnet.it) 
through the ASI (Italian Space Agency) Grants 2016-24-H.0, 2016-24-H.1-2018 and 2020-9-HH.0. 
\end{acknowledgments}

\appendix

\section{Real and Imaginary Part of the Angular Bispectra and their Parity Properties}
\label{app:parRel}
In this appendix we are going to motivate the form of Eq.~\eqref{eqn:complex} (similar discussions can be found in \cite{duivenvoorden2020cmb,meerburg2016cmb}).

Let us consider the most general CMB angular bispectrum, expressed in terms of its related angular averaged bispectrum, which is given according to Eq.~\eqref{eqn:average} as
\begin{equation}
\label{eqn:normal}
\begin{split}
\langle a_{X,\ell_1m_1}a_{Y,\ell_2m_2}&a_{Z,\ell_3m_3}\rangle=\begin{pmatrix}
		\ell_1 & \ell_2 & \ell_3 \\
		m_1 & m_2 & m_3
	\end{pmatrix}B_{\ell_1\ell_2\ell_3}^{XYZ},
\end{split}
\end{equation}
being $X,Y,Z=\delta\alpha,T,E,B$. Let us see what happens by taking the complex conjugate:
\begin{equation}
\begin{split}
\langle a_{X,\ell_1m_1}^*&a_{Y,\ell_2m_2}^*a_{Z,\ell_3m_3}^*\rangle=\expval{a_{X,\ell_1,-m_1}a_{Y,\ell_2,-m_2}a_{Z,\ell_3,-m_3}},
\end{split}
\end{equation}
where we have exploited the reality condition:
\begin{equation}
a_{X,\ell_1m_1}^*=(-1)^{m_1}a_{X,\ell_1,-m_1}.
\end{equation}
The Wigner-$3j$ symbol automatically ensures the following selection rule on the $m$'s:
\begin{equation}
m_1+m_2+m_3=0,
\end{equation}
so that we can express the complex conjugate of the angular bispectrum by means of Eq.~\eqref{eqn:normal} as before, but with all the signs in front of the $m$'s flipped:
\begin{equation}
\begin{split}
\Big[\langle a_{X,\ell_1m_1}a_{Y,\ell_2m_2}a_{Z,\ell_3m_3}\rangle\Big]^*&=\begin{pmatrix}
\ell_1 & \ell_2 & \ell_3 \\
-m_1 & -m_2 & -m_3
\end{pmatrix}B_{\ell_1\ell_2\ell_3}^{XYZ}\\
&=(-1)^{\ell_T}\begin{pmatrix}
\ell_1 & \ell_2 & \ell_3 \\
m_1 & m_2 & m_3
\end{pmatrix}B_{\ell_1\ell_2\ell_3}^{XYZ}\\
&=(-1)^{\ell_T}\expval{a_{X,\ell_1m_1}a_{Y,\ell_2m_2}a_{Z,\ell_3m_3}}, 
\end{split}
\end{equation}
where we have used Eq.~\eqref{eqn:flip}. This means that we can write the transformation properties of the angular bispectrum under complex-conjugation as
\begin{equation}
\text{ang. bispectrum}\xrightarrow{\text{c.c}}\begin{cases}
+\,\text{itself}\quad&\text{if $\ell_{T}=\,$even},\\
-\,\text{itself}\quad&\text{if $\ell_{T}=\,$odd}.\\
\end{cases}
\end{equation} 
This is equivalent to say that: if $\ell_{T}$ is equal to an even number, then the angular bispectrum has to be a purely real quantity, whereas if $\ell_{T}$ is equal to an odd number, then the angular bispectrum has to be a purely imaginary quantity. Therefore, the most general bispectrum, without any assumption about the parity of $\ell_{T}$ can be written as
\begin{equation}
\label{eqn:CC}
B_{\ell_1\ell_2\ell_3}^{XYZ}=B_{\ell_1\ell_2\ell_3}^{XYZ,\text{even}}+iB_{\ell_1\ell_2\ell_3}^{XYZ,\text{odd}},
\end{equation}
where the subscripts ``even'' and ``odd'' refers to the parity of $\ell_{T}$. Moreover, this decomposition also plays a role in defining the overall parity of the angular bispectrum, since
\begin{equation}
\begin{split}
\langle &a_{X,\ell_1m_1}a_{Y,\ell_2m_2}a_{Z,\ell_3m_3}\rangle\xrightarrow{\text{parity}}(-1)^{\ell_{T}+N_B}\expval{a_{X,\ell_1m_1}a_{Y,\ell_2m_2}a_{Z,\ell_3m_3}},
\end{split}
\end{equation}
being $N_B$ the number of $B$-mode polarization CMB fields involved in the ensemble average. Hence, it is clear that parity is violated if $\ell_{T}+N_{\text{B}}$ is equal to an odd number. Since $N_{\text{B}}$ is fixed from the type of bispectrum one wants to evaluate, it follows that parity is automatically broken if the correlation function can be written as the sum of terms of mixed parity of $\ell_{T}$. Therefore, from Eq.~\eqref{eqn:CC}, we can now infer that the most general parity-violating CMB bispectrum is a complex quantity, since it involves components of different parity for $\ell_{T}$. 

For instance, for the case of the $TTB$ bispectrum, parity is conserved if $\ell_{T}$ is equal to an odd number, and it is broken if $\ell_{T}$ is equal to an even number, since $N_B=1$. In such an example, the term responsible for parity-violation is the real part of the angular bispectrum. The opposite occurs for the $TTE$ bispectrum, for which $N_B=0$, and so $\ell_{T}$ must be odd to break parity (so that the parity-violating term is the imaginary part of the correlation function). To sum up, parity-violation in angular bispectra can manifest itself with the presence of both real and imaginary parts of the angular averaged bispectrum. 

\section{\label{app:PowerSpectrum} Large-Scale Limit of the Angular Power Spectrum of Anisotropic Cosmic Birefringence}
It is possible to find a general expression for the power spectrum of anisotropic cosmic birefringence, by substituting Eq.~\eqref{eqn:delta} into Eq.~\eqref{eqn:alpha_lm}:
\begin{equation}
	\alpha_{\ell m}=-2\frac{\mathrm{d}h(\chi_0)}{\mathrm{d}\chi_0}\Bigg|_{\tau_{\text{reco}}}\int\mathrm{d}\hat{n}\,Y^*_{\ell m}(\versor{n})\,\delta\chi(\Delta\tau\versor{n})\, ,
\end{equation}
being $\Delta\tau\equiv\tau_0-\tau_{\text{reco}}$. If we move to the Fourier space we get the following expression:
\begin{equation}
\begin{split}
\alpha_{\ell m}=-8\pi &i^{\ell}\frac{\mathrm{d}h(\chi_0)}{\mathrm{d}\chi_0}\Bigg|_{\tau_{\text{reco}}}\int\frac{\mathrm{d}^3k}{(2\pi)^3}j_{\ell}(k\Delta\tau)\,Y_{\ell m}^*(\versor{k})\delta\chi(\mathbf{k},\tau_{\text{reco}})\, ,
\end{split}
\end{equation}
where, in order to express the complex exponential in terms of the $\ell$-th spherical Bessel function $j_{\ell}(x)$, we have adopted the plane-wave expansion,
\begin{equation}
	e^{i\mathbf{k}\cdot\mathbf{r}}=4\pi\sum_{\ell m}i^{\ell}j_{\ell}(kr)Y_{\ell m}^*(\versor{k})Y_{\ell m}(\versor{r}),
\end{equation}
and we have performed the integral over the solid angle with Eq.~\eqref{eqn:orthogonality}. By recalling Eq.~\eqref{eqn:definition}, the angular power spectrum of anisotropic cosmic birefringence is easily computed as
\begin{equation}
\label{eqn:aa}
\begin{split}
C_{\ell}^{\alpha\alpha}=16\pi&\left[\frac{\mathrm{d}h(\chi_0)}{\mathrm{d}\chi_0}\right]^2\Bigg|_{\tau_{\text{reco}}}\int_0^{+\infty}\frac{\mathrm{d}k}{k}j_{\ell}^2(k\Delta\tau)\Delta^2_{\delta\chi}(k,\tau_{\text{reco}}),
\end{split}
\end{equation}
where we have defined the Fourier transform of the two-point correlation function for the field fluctuation as
\begin{equation}
\expval{\delta\chi^*(\mathbf{k},\tau)\delta\chi(\mathbf{k}',\tau)}\equiv\frac{16\pi^5}{k^3}\delta^{(3)}(\mathbf{k}-\mathbf{k}')\Delta_{\delta\chi}^2(k,\tau)
\end{equation}
being $\Delta_{\delta\chi}^2(k,\tau)$ the dimensionless power spectrum of $\delta\chi$.

As customary, we can decompose the power spectrum as the product of a squared transfer function $T_{\delta\chi}^2(k,\tau)$ times the primordial power spectrum predicted by inflation:
\begin{equation}
\Delta_{\delta\chi}^2(k,\tau)=T_{\delta\chi}^2(k,\tau)\Delta^2_{\delta\chi}(k).
\end{equation}
The transfer function has the role of evolving the perturbation from primordial to the given time. For a ``standard'' model of cosmic birefringence like that depicted in Eq.~\eqref{eqn:lagran}, it is possible to relate the power spectrum of the field fluctuation to that of more common cosmological scalar perturbations. For instance, as shown in \cite{li2008cosmological,caldwell2011cross}, let us take $\chi$ as a
\emph{quintessence} field with a non-zero potential whose evolution is governed by the following Lagrangian density 
\begin{equation}
\label{eqn:chi_lagr}
\mathcal{L}_{\chi}=-\frac{1}{2}g^{\mu\nu}\partial_{\mu}\chi\partial_{\nu}\chi-V(\chi),
\end{equation}
with a perturbed metric that in the Newtonian conformal gauge reads
\begin{align}
g_{00}&=-a^2(\tau)\left[1+2\Psi(\mathbf{x},\tau)\right]\\ g_{0i}&=g_{i0}=0\\ g_{ij}&=a^2(\tau)\left[1+2\Phi(\mathbf{x},\tau)\right]\delta_{ij},\end{align}
where $\Psi$ and $\Phi$ are the two gauge-invariant Bardeen's potentials. By neglecting ultra-relativistic neutrinos we can approximate $\Phi\simeq-\Psi$ (which is reasonable at the recombination time, since it occurs during the matter-dominated epoch), and by linearly perturbing Eq.~\eqref{eqn:chi_lagr}, it is possible to solve this equation for adiabatic initial conditions and find \cite{li2008cosmological}
\begin{equation}
\delta\chi(\mathbf{k},\tau_{\text{reco}})=\frac{2\dot{\chi}_0(\tau_{\text{reco}})}{3H(\tau_{\text{reco}})}\Psi(\mathbf{k},\tau_{\text{reco}})\, , 
\end{equation}
$H$ being the Hubble parameter. Then, we can infer 
\begin{equation}
\Delta_{\delta\chi}^2(k,\tau_{\text{reco}})=\frac{4\dot{\chi}_0^2(\tau_{\text{reco}})}{9H^2(\tau_{\text{reco}})}T_{\Psi}^2(k,\tau_{\text{reco}})\Delta_{\Psi}^2(k),
\end{equation}
so that by substituting the expression above in Eq.~\eqref{eqn:aa}, we finally obtain
\begin{equation}
\begin{split}
C_{\ell}^{\alpha\alpha}=&\frac{64\pi}{9}\left[\frac{\dot{\chi}_0}{H}\frac{\mathrm{d}h(\chi_0)}{\mathrm{d}\chi_0}\right]^2\Bigg|_{\tau_{\text{reco}}}\int_0^{+\infty}\frac{\mathrm{d}k}{k}j_{\ell}^2(k\Delta\tau)T_{\Psi}^2(k,\tau_{\text{reco}})\Delta_{\Psi}^2(k).
\end{split}
\end{equation}
If we assume a standard inflationary scenario, we know that the primordial power spectrum is approximately scale-invariant, $\Delta^2_{\Psi}(k)=(9/25)A_s$, where $A_s$ is the amplitude of scalar perturbations (the numerical factor is just due to the fact that $\Psi(\mathbf{k})=-(3/5)\mathcal{R}(\mathbf{k})$, being $\mathcal{R}$ the comoving curvature perturbation). Now, the gravitational potential $\Psi$ is constant for those scales that enter the horizon during the matter-dominated epoch. This means that we can approximate $T_{\Psi}(k,\tau_{\text{reco}})\simeq1$ for all those modes that satisfy $k\tau_{\text{eq}}\le1$, which guarantees that the mode is outside the horizon at matter-radiation equivalence and hence that it entered during matter domination. Since for the $\ell$-th spherical Bessel function $j_{\ell}(x)$ the maximum value is attained roughly when
$x\simeq\ell$, we can deduce that we can set the transfer function equal to $1$ up to
\begin{equation}
\ell\simeq k\Delta\tau\simeq k\tau_0\le\frac{\tau_0}{\tau_{\text{eq}}}\simeq10^2,
\end{equation}
where we have used CLASS to estimate the $\tau/\tau_{\text{eq}}$ ratio according to the $\Lambda$CDM model \cite{lesgourgues2011cosmic}. Therefore, we can approximate the power spectrum of anisotropic cosmic birefringence for $\ell\le100$ as \cite{bloomfield2017indefinite}
\begin{equation}
\begin{split}
C_{\ell}^{\alpha\alpha}&=A_s\frac{64\pi}{25}\left[\frac{\dot{\chi}_0}{H}\frac{\mathrm{d}h(\chi_0)}{\mathrm{d}\chi_0}\right]^2\Bigg|_{\tau_{\text{reco}}}\int_0^{+\infty}\frac{\mathrm{d}k}{k}j_{\ell}^2(k\Delta\tau)=\frac{32\pi}{25}\left[\frac{\dot{\chi}_0}{H}\frac{\mathrm{d}h(\chi_0)}{\mathrm{d}\chi_0}\right]^2\Bigg|_{\tau_{\text{reco}}}\frac{A_s}{\ell(\ell+1)}.
\end{split}
\end{equation}
By defining 
\begin{equation}
C_{\alpha\alpha}\equiv\frac{32\pi A_s}{25}\left[\frac{\dot{\chi}_0}{H}\frac{\mathrm{d}h(\chi_0)}{\mathrm{d}\chi_0}\right]^2\Bigg|_{\tau_{\text{reco}}},
\end{equation}
we finally get the simple expression of Eq.~\eqref{eqn:Caa}.

\section{\label{app:FisherFormulas} Fisher Forecast for Birefringent Bispectra}
By following the procedure described in Sec.~\ref{sec:SNR}, we can obtain the following formulas for squared SNRs of the five birefringent bispectra we have computed in this paper. We report here their explicit expressions:
\begin{align}
	\label{eqn:SNR_ATE}
	\left[\mathsf{SNR}^{(\alpha_0=0)}_{(\alpha TE)}\right]^2&=\sum_{2\le\ell_1<\ell_2<\ell_3}^{\ell_{\text{max}}}\begin{pmatrix}
		B^{\alpha TE}_{\ell_1\ell_2\ell_3} & B^{TE\alpha}_{\ell_1\ell_2\ell_3} & B^{E\alpha T}_{\ell_1\ell_2\ell_3} & B^{\alpha ET}_{\ell_1\ell_2\ell_3} & B^{ET\alpha}_{\ell_1\ell_2\ell_3} & B^{T\alpha E}_{\ell_1\ell_2\ell_3}
	\end{pmatrix}\left[\mathsf{Cov}^{}_{\ell_1\ell_2\ell_3}\right]^{-1}\begin{pmatrix}
		B^{\alpha TE}_{\ell_1\ell_2\ell_3} \\
		\\
		B^{TE\alpha}_{\ell_1\ell_2\ell_3} \\
		\\
		B^{E\alpha T}_{\ell_1\ell_2\ell_3} \\
		\\
		B^{\alpha ET}_{\ell_1\ell_2\ell_3} \\
		\\
		B^{ET\alpha}_{\ell_1\ell_2\ell_3} \\
		\\
		B^{T\alpha E}_{\ell_1\ell_2\ell_3}
	\end{pmatrix},\\
	\label{eqn:SNR_ATB}
	\left[\mathsf{SNR}^{(\alpha_0=0)}_{(\alpha TB)}\right]^2&=\sum_{2\le\ell_1<\ell_2<\ell_3}^{\ell_{\text{max}}}\begin{pmatrix}
		B^{\alpha TB}_{\ell_1\ell_2\ell_3} & B^{TB\alpha}_{\ell_1\ell_2\ell_3} & B^{B\alpha T}_{\ell_1\ell_2\ell_3} & B^{\alpha BT}_{\ell_1\ell_2\ell_3}  & B^{BT\alpha}_{\ell_1\ell_2\ell_3} & B^{T\alpha B}_{\ell_1\ell_2\ell_3}
	\end{pmatrix}\left[\mathsf{Cov}^{}_{\ell_1\ell_2\ell_3}\right]^{-1}\begin{pmatrix}
		B^{\alpha TB}_{\ell_1\ell_2\ell_3} \\
		\\
		B^{TB\alpha}_{\ell_1\ell_2\ell_3} \\
		\\
		B^{B\alpha T}_{\ell_1\ell_2\ell_3} \\
		\\
		B^{\alpha BT}_{\ell_1\ell_2\ell_3} \\
		\\
		B^{BT\alpha}_{\ell_1\ell_2\ell_3} \\
		\\
		B^{T\alpha B}_{\ell_1\ell_2\ell_3}
	\end{pmatrix},\\
	\label{eqn:SNR_AEB}
	\left[\mathsf{SNR}^{(\alpha_0=0)}_{(\alpha EB)}\right]^2&=\sum_{2\le\ell_1<\ell_2<\ell_3}^{\ell_{\text{max}}}\begin{pmatrix}
		B^{\alpha EB}_{\ell_1\ell_2\ell_3} & B^{EB\alpha}_{\ell_1\ell_2\ell_3} & B^{B\alpha E}_{\ell_1\ell_2\ell_3} & B^{\alpha BE}_{\ell_1\ell_2\ell_3} & B^{BE\alpha}_{\ell_1\ell_2\ell_3} & B^{E\alpha B}_{\ell_1\ell_2\ell_3}
	\end{pmatrix}\left[\mathsf{Cov}^{}_{\ell_1\ell_2\ell_3}\right]^{-1}\begin{pmatrix}
		B^{\alpha EB}_{\ell_1\ell_2\ell_3} \\
		\\
		B^{EB\alpha}_{\ell_1\ell_2\ell_3} \\
		\\
		B^{B\alpha E}_{\ell_1\ell_2\ell_3} \\
		\\
		B^{\alpha BE}_{\ell_1\ell_2\ell_3} \\
		\\
		B^{EB\alpha}_{\ell_1\ell_2\ell_3} \\
		\\
		B^{BE\alpha}_{\ell_1\ell_2\ell_3} \\
	\end{pmatrix}\\
	\label{eqn:SNR_AEE}
	\left[\mathsf{SNR}^{(\alpha_0=0)}_{(\alpha EE)}\right]^2&=\sum_{2\le\ell_1<\ell_2<\ell_3}^{\ell_{\text{max}}}\begin{pmatrix}
		B^{\alpha EE}_{\ell_1\ell_2\ell_3} & B^{EE\alpha}_{\ell_1\ell_2\ell_3} & B^{E\alpha E}_{\ell_1\ell_2\ell_3}
	\end{pmatrix}\left[\mathsf{Cov}^{}_{\ell_1\ell_2\ell_3}\right]^{-1}\begin{pmatrix}
		B^{\alpha EE}_{\ell_1\ell_2\ell_3} \\
		\\
		B^{EE\alpha}_{\ell_1\ell_2\ell_3} \\
		\\
		B^{E\alpha E}_{\ell_1\ell_2\ell_3} \\
	\end{pmatrix},\\
	\label{eqn:SNR_ABB}
	\left[\mathsf{SNR}^{(\alpha_0=0)}_{(\alpha BB)}\right]^2&=\sum_{2\le\ell_1<\ell_2<\ell_3}^{\ell_{\text{max}}}\begin{pmatrix}
		B^{\alpha BB}_{\ell_1\ell_2\ell_3} & B^{BB\alpha}_{\ell_1\ell_2\ell_3} & B^{B\alpha B}_{\ell_1\ell_2\ell_3}
	\end{pmatrix}\left[\mathsf{Cov}^{}_{\ell_1\ell_2\ell_3}\right]^{-1}\begin{pmatrix}
		B^{\alpha BB}_{\ell_1\ell_2\ell_3} \\
		\\
		B^{BB\alpha}_{\ell_1\ell_2\ell_3} \\
		\\
		B^{B\alpha B}_{\ell_1\ell_2\ell_3}
	\end{pmatrix}.
\end{align}
In the evaluation of Eqs.~\eqref{eqn:SNR_ATE}-\eqref{eqn:SNR_ABB} we have considered only the configurations with $\ell_1\ne\ell_2\ne\ell_3$ in order to simplify the numerical code. This is a reasonable approximation, since the huge majority of the triplets satisfies this requirement. 

\section{\label{app:Reionization} Cosmic Birefringence from Reionization}
In our treatment we have considered  the CMB polarization as generated just at recombination. However, a more accurate analysis should also include the contribution coming from the reionization epoch. According to this, the observed CMB Stokes parameters measured after the rotation due to cosmic birefringence are given by \cite{sherwin2021cosmic}:
\begin{equation}
\label{eqn:reioreco}
\left[Q_{\text{obs}}\pm iU_{\text{obs}}\right](\versor{n})=\sum_{x=\text{reco, reio}}\left[Q_x\pm iU_x\right](\versor{n})e^{\pm2i[\alpha_{0,x}+\delta\alpha_x(\versor{n})]}
\, ,
\end{equation}
where we have defined a different birefringence angle for the recombination epoch and for the reionization one, respectively. Indeed, Eqs.~\eqref{eqn:angle} and~\eqref{eqn:delta} simply generalize to
\begin{align}
\alpha_{0,x}&=2h[\chi_0(\tau_0)]-2h[\chi_0(\tau_x)]\, ,\\
\delta\alpha_x(\versor{n})&=-2\frac{\mathrm{d}h(\chi_0)}{\mathrm{d}\chi_0}\Big|_{\tau=\tau_x}\delta\chi[(\tau_0-\tau_x)\versor{n},\tau_x],
\end{align}
being $x=\text{recombination, reionizaton}$, so that the harmonic coefficients can be written as
\begin{equation}
\alpha_{\ell m}^x=\int\mathrm{d}^2\hat{n}\,Y_{\ell m}^{*}(\versor{n})\,\delta\alpha_{x}(\versor{n})\qquad\qquad\begin{pmatrix}a_{E,\ell m} \\
a_{B,\ell m} \end{pmatrix}=\sum_{x=\text{reco, reio}}\begin{pmatrix}a_{E,\ell m}^{x} \\ 
a_{B,\ell m}^{x}
\end{pmatrix}
\, .
\end{equation}
Because of this linear correction due to the inclusion of the contributions from the reionization epoch, then it is straightforward to show that the observed harmonic coefficients of the CMB polarization are just given as
\begin{equation}
\begin{pmatrix}
a_{E,\ell m}^{\text{obs}} \\
a_{B,\ell m}^{\text{obs}}
\end{pmatrix}=\sum_{x=\text{reco, reio}}\begin{pmatrix}
a_{E,\ell m}^{\text{obs},x} \\
a_{B,\ell m}^{\text{obs},x}
\end{pmatrix}
\, ,
\end{equation}
where
\begin{equation}
\begin{pmatrix}
	a_{E,\ell m}^{\text{obs},x} \\
	a_{B,\ell m}^{\text{obs},x}
\end{pmatrix}=
\sum_{s=\pm2}\frac{e^{is\alpha_{0,x}}}{2}\sum_{LM}\int\mathrm{d}\hat{n}\,_sY_{\ell m}^*(\versor{n})\,_sY_{LM}(\versor{n})\begin{pmatrix}
1 & is/2 \\
-is/2 & 1
\end{pmatrix}
\begin{pmatrix}
	a_{E,LM}^{x} \\
	a_{B,LM}^{x}
\end{pmatrix}e^{is\delta\alpha_x(\versor{n})}.
\end{equation}
Thus, it becomes trivial to generalize our formulae accounting also for reionization. Indeed, Eqs.~\eqref{eqn:AT}-\eqref{eqn:AB} become
\begin{align}
C_{\ell,\text{obs}}^{\alpha_x T}&=C_{\ell}^{\alpha_x T}, \\
C_{\ell,\text{obs}}^{\alpha_x E}&=\sum_{y=\text{reco, reio}}\left[C_{\ell}^{\alpha_x E_y}\cos2\alpha_{0,y}-C_{\ell}^{\alpha_x B_y}\sin2\alpha_{0,y}\right], \\
C_{\ell,\text{obs}}^{\alpha_x B}&=\sum_{y=\text{reco, reio}}\left[C_{\ell}^{\alpha_x E_y}\sin2\alpha_{0,y}+C_{\ell}^{\alpha_x B_y}\cos2\alpha_{0,y}\right]\, , 
\end{align}
whereas Eqs.~\eqref{eqn:ATE}-\eqref{eqn:ATB} become
\begin{align}
B_{\ell_1\ell_2\ell_3}^{\alpha_x TE}&=I^{0,2,-2}_{\ell_1\ell_2\ell_3}\sum_{y}C_{\ell_1}^{\alpha_x\alpha_y}C_{\ell_2}^{TE_y}\Big\{i\left[1-(-1)^{\ell_{T}}\right]\cos2\alpha_{0,y}-\left[1+(-1)^{\ell_{T}}\right]\sin2\alpha_{0,y}\Big\}\, ,\\
B^{\alpha_x TB}_{\ell_1\ell_2\ell_3}&=I^{0,2,-2}_{\ell_1\ell_2\ell_3}\sum_{y}C_{\ell_1}^{\alpha_x\alpha_y}C_{\ell_2}^{TE_y}\Big\{\left[1+(-1)^{\ell_{T}}\right]\cos2\alpha_{0,y}+i\left[1-(-1)^{\ell_{T}}\right]\sin2\alpha_{0,y}\Big\}\, ,
\end{align}
and similarly Eqs.~\eqref{eqn:AEE}-\eqref{eqn:AEB} generalize to
\begin{equation}
\begin{split}
B_{\ell_1\ell_2\ell_3}^{\alpha_x EE}=\frac{i}{4}\sum_{y,z}\sum_{s_2,s_3}e^{is_2\alpha_{0,y}+is_3\alpha_{0,z}}\Big[C_{\ell_1}^{\alpha_x\alpha_y}I_{\ell_1\ell_2\ell_3}^{0,s_2,-s_2}&\left(s_2C_{\ell_3}^{E_yE_z}-s_3C_{\ell_3}^{B_yB_z}\right)\,+\\
&+C_{\ell_1}^{\alpha_x\alpha_z}I_{\ell_1\ell_2\ell_3}^{0,-s_3,s_3}\left(s_3C_{\ell_2}^{E_yE_z}-s_2C_{\ell_2}^{B_yB_z}\right)\Big]\, ,
\end{split}
\end{equation}
\begin{equation}
	\begin{split}
		B_{\ell_1\ell_2\ell_3}^{\alpha_x BB}=\frac{i}{4}\sum_{y,z}\sum_{s_2,s_3}e^{is_2\alpha_{0,y}+is_3\alpha_{0,z}}\Big[C_{\ell_1}^{\alpha_x\alpha_y}I_{\ell_1\ell_2\ell_3}^{0,s_2,-s_2}&\left(s_2C_{\ell_3}^{B_yB_z}-s_3C_{\ell_3}^{E_yE_z}\right)\,+\\
		&+C_{\ell_1}^{\alpha_x\alpha_z}I_{\ell_1\ell_2\ell_3}^{0,-s_3,s_3}\left(s_3C_{\ell_2}^{B_yB_z}-s_2C_{\ell_2}^{E_yE_z}\right)\Big]\, ,
	\end{split}
\end{equation}
\begin{equation}
	\begin{split}
		B_{\ell_1\ell_2\ell_3}^{\alpha_x EB}=\frac{1}{4}\sum_{y,z}\sum_{s_2,s_3}e^{is_2\alpha_{0,y}+is_3\alpha_{0,z}}\Big[C_{\ell_1}^{\alpha_x\alpha_y}I_{\ell_1\ell_2\ell_3}^{0,s_2,-s_2}&\left(\frac{s_2s_3}{2}C_{\ell_3}^{E_yE_z}-2C_{\ell_3}^{B_yB_z}\right)\,+\\
		&+C_{\ell_1}^{\alpha_x\alpha_z}I_{\ell_1\ell_2\ell_3}^{0,-s_3,s_3}\left(\frac{s_2s_3}{2}C_{\ell_2}^{B_yB_z}-s_2C_{\ell_2}^{E_yE_z}\right)\Big]\, ,
	\end{split}
\end{equation}
being $x,y,z=\text{recombination, reionization}$ and $s_2,s_3=\pm2$.

By the way, since we expect the transfer functions at the recombination and at the reionization to peak at very different redshifts, it is not unreasonable to neglect a cross-correlation evaluated at the two epochs when they are not the same, i.e.
\begin{equation}
C_{\ell}^{X_xY_y}\approx C_{\ell}^{X_xY_x}\delta_{x,y}\qquad\qquad\qquad(X,Y=\delta\alpha,E,B).
\end{equation}
This approximation strongly simplifies the general expressions provided in this appendix as follows: for the cross-spectra we just have
\begin{align}
\label{eqn:a_xT}
C_{\ell,\text{obs}}^{\alpha_x T}&=C_{\ell}^{\alpha_x T}, \\
C_{\ell,\text{obs}}^{\alpha_x E}&=C_{\ell}^{\alpha_x E_x}\cos2\alpha_{0,x}-C_{\ell}^{\alpha_x B_x}\sin2\alpha_{0,x}, \\
C_{\ell,\text{obs}}^{\alpha_x B}&=C_{\ell}^{\alpha_x E_x}\sin2\alpha_{0,x}+C_{\ell}^{\alpha_x B_x}\cos2\alpha_{0,x},
\end{align}
and analogously for the cross-bispectra we have
\begin{align}
B_{\ell_1\ell_2\ell_3}^{\alpha_x TE}&=C_{\ell_1}^{\alpha_x\alpha_x}\Big\{i\left[1-(-1)^{\ell_{T}}\right]\cos2\alpha_{0,x}-\left[1+(-1)^{\ell_{T}}\right]\sin2\alpha_{0,x}\Big\}C_{\ell_2}^{TE_x}I^{0,2,-2}_{\ell_1\ell_2\ell_3},\\
B^{\alpha_x TB}_{\ell_1\ell_2\ell_3}&=C_{\ell_1}^{\alpha_x\alpha_x}\Big\{\left[1+(-1)^{\ell_{T}}\right]\cos2\alpha_{0,x}+i\left[1-(-1)^{\ell_{T}}\right]\sin2\alpha_{0,x}\Big\}C_{\ell_2}^{TE_x}I^{0,2,-2}_{\ell_1\ell_2\ell_3},
\end{align}
\begin{equation}
\begin{split}
B_{\ell_1\ell_2\ell_3}^{\alpha_x EE}&=\frac{1}{2}C_{\ell_1}^{\alpha_x\alpha_x}\Big\{[1+(-1)^{\ell_{T}}]\left[\left(C_{\ell_3}^{B_xB_x}+C_{\ell_2}^{B_xB_x}\right)-\left(C_{\ell_3}^{E_xE_x}+C_{\ell_2}^{E_xE_x}\right)\right]\sin4\alpha_{0,x}\,+\\
&+i\left[1-(-1)^{\ell_{T}}\right]\left[\left(C_{\ell_3}^{B_xB_x}-C_{\ell_2}^{B_xB_x}\right)\left(1-\cos4\alpha_{0,x}\right)+\left(C_{\ell_3}^{E_xE_x}-C_{\ell_2}^{E_xE_x}\right)\left(1+\cos4\alpha_{0,x}\right)\right]\Big\}I_{\ell_1\ell_2\ell_3}^{0,2,-2},
\end{split}
\end{equation}
\begin{equation}
\begin{split}
B_{\ell_1\ell_2\ell_3}^{\alpha_x BB}&=\frac{1}{2}C_{\ell_1}^{\alpha_x\alpha_x}\Big\{[1+(-1)^{\ell_{T}}]\left[\left(C_{\ell_3}^{E_xE_x}+C_{\ell_2}^{E_xE_x}\right)-\left(C_{\ell_3}^{B_xB_x}+C_{\ell_2}^{B_xB_x}\right)\right]\sin4\alpha_{0,x}\,+\\
&+i\left[1-(-1)^{\ell_{T}}\right]\left[\left(C_{\ell_3}^{E_xE_x}-C_{\ell_2}^{E_xE_x}\right)\left(1-\cos4\alpha_{0,x}\right)+\left(C_{\ell_3}^{B_xB_x}-C_{\ell_2}^{B_xB_x}\right)\left(1+\cos4\alpha_{0,x}\right)\right]\Big\}I_{\ell_1\ell_2\ell_3}^{0,2,-2},
\end{split}
\end{equation}
\begin{equation}
\label{eqn:a_xEE}
\begin{split}
B_{\ell_1\ell_2\ell_3}^{\alpha_x EB}&=\frac{1}{2}C_{\ell_1}^{\alpha_x\alpha_x}\Big\{i[1-(-1)^{\ell_{T}}]\left[\left(C_{\ell_3}^{E_xE_x}-C_{\ell_2}^{E_xE_x}\right)-\left(C_{\ell_3}^{B_xB_x}-C_{\ell_2}^{B_xB_x}\right)\right]\sin4\alpha_{0,x}\,+\\
&+\left[1+(-1)^{\ell_{T}}\right]\left[\left(C_{\ell_2}^{B_xB_x}-C_{\ell_3}^{E_xE_x}\right)\left(1-\cos4\alpha_{0,x}\right)+\left(C_{\ell_2}^{E_xE_x}-C_{\ell_3}^{B_xB_x}\right)\left(1+\cos4\alpha_{0,x}\right)\right]\Big\}I_{\ell_1\ell_2\ell_3}^{0,2,-2}.
\end{split}
\end{equation}
Let us then notice that Eqs.~\eqref{eqn:a_xT}-\eqref{eqn:a_xEE} are nothing but the same formulae we derived in Secs.~\ref{sec:power_spectrum}-\ref{sec:bispectra} but modified by the presence of an extra label associated with the source of CMB polarization, i.e. recombination or reionization.

\nocite{*}
\bibliography{CBdraft}

\begin{thebibliography}{73}
\expandafter\ifx\csname natexlab\endcsname\relax\def\natexlab#1{#1}\fi
\expandafter\ifx\csname bibnamefont\endcsname\relax
  \def\bibnamefont#1{#1}\fi
\expandafter\ifx\csname bibfnamefont\endcsname\relax
  \def\bibfnamefont#1{#1}\fi
\expandafter\ifx\csname citenamefont\endcsname\relax
  \def\citenamefont#1{#1}\fi
\expandafter\ifx\csname url\endcsname\relax
  \def\url#1{\texttt{#1}}\fi
\expandafter\ifx\csname urlprefix\endcsname\relax\def\urlprefix{URL }\fi
\providecommand{\bibinfo}[2]{#2}
\providecommand{\eprint}[2][]{\url{#2}}

\bibitem[{\citenamefont{Lee and Yang}(1956)}]{lee1956question}
\bibinfo{author}{\bibfnamefont{T.}~\bibnamefont{Lee}} \bibnamefont{and}
  \bibinfo{author}{\bibfnamefont{C.}~\bibnamefont{Yang}},
  \bibinfo{journal}{Physical Review} \textbf{\bibinfo{volume}{104}},
  \bibinfo{pages}{254} (\bibinfo{year}{1956}).

\bibitem[{\citenamefont{Carroll et~al.}(1990)\citenamefont{Carroll, Field, and
  Jackiw}}]{carroll1990limits}
\bibinfo{author}{\bibfnamefont{S.~M.} \bibnamefont{Carroll}},
  \bibinfo{author}{\bibfnamefont{G.~B.} \bibnamefont{Field}}, \bibnamefont{and}
  \bibinfo{author}{\bibfnamefont{R.}~\bibnamefont{Jackiw}},
  \bibinfo{journal}{Physical Review D} \textbf{\bibinfo{volume}{41}},
  \bibinfo{pages}{1231} (\bibinfo{year}{1990}).

\bibitem[{\citenamefont{Sherwin and Namikawa}(2021)}]{sherwin2021cosmic}
\bibinfo{author}{\bibfnamefont{B.~D.} \bibnamefont{Sherwin}} \bibnamefont{and}
  \bibinfo{author}{\bibfnamefont{T.}~\bibnamefont{Namikawa}},
  \bibinfo{journal}{arXiv preprint arXiv:2108.09287}  (\bibinfo{year}{2021}).

\bibitem[{\citenamefont{Choi et~al.}(2021)\citenamefont{Choi, Lin, Visinelli,
  and Yanagida}}]{choi2021cosmic}
\bibinfo{author}{\bibfnamefont{G.}~\bibnamefont{Choi}},
  \bibinfo{author}{\bibfnamefont{W.}~\bibnamefont{Lin}},
  \bibinfo{author}{\bibfnamefont{L.}~\bibnamefont{Visinelli}},
  \bibnamefont{and} \bibinfo{author}{\bibfnamefont{T.~T.}
  \bibnamefont{Yanagida}}, \bibinfo{journal}{Physical Review D}
  \textbf{\bibinfo{volume}{104}}, \bibinfo{pages}{L101302}
  (\bibinfo{year}{2021}).

\bibitem[{\citenamefont{Fujita et~al.}(2021)\citenamefont{Fujita, Murai,
  Nakatsuka, and Tsujikawa}}]{fujita2021detection}
\bibinfo{author}{\bibfnamefont{T.}~\bibnamefont{Fujita}},
  \bibinfo{author}{\bibfnamefont{K.}~\bibnamefont{Murai}},
  \bibinfo{author}{\bibfnamefont{H.}~\bibnamefont{Nakatsuka}},
  \bibnamefont{and}
  \bibinfo{author}{\bibfnamefont{S.}~\bibnamefont{Tsujikawa}},
  \bibinfo{journal}{Physical Review D} \textbf{\bibinfo{volume}{103}},
  \bibinfo{pages}{043509} (\bibinfo{year}{2021}).

\bibitem[{\citenamefont{Nakagawa et~al.}(2021)\citenamefont{Nakagawa,
  Takahashi, and Yamada}}]{nakagawa2021cosmic}
\bibinfo{author}{\bibfnamefont{S.}~\bibnamefont{Nakagawa}},
  \bibinfo{author}{\bibfnamefont{F.}~\bibnamefont{Takahashi}},
  \bibnamefont{and} \bibinfo{author}{\bibfnamefont{M.}~\bibnamefont{Yamada}},
  \bibinfo{journal}{Physical review letters} \textbf{\bibinfo{volume}{127}},
  \bibinfo{pages}{181103} (\bibinfo{year}{2021}).

\bibitem[{\citenamefont{Li and Zhang}(2008)}]{li2008cosmological}
\bibinfo{author}{\bibfnamefont{M.}~\bibnamefont{Li}} \bibnamefont{and}
  \bibinfo{author}{\bibfnamefont{X.}~\bibnamefont{Zhang}},
  \bibinfo{journal}{Physical Review D} \textbf{\bibinfo{volume}{78}},
  \bibinfo{pages}{103516} (\bibinfo{year}{2008}).

\bibitem[{\citenamefont{Caldwell et~al.}(2011)\citenamefont{Caldwell,
  Gluscevic, and Kamionkowski}}]{caldwell2011cross}
\bibinfo{author}{\bibfnamefont{R.~R.} \bibnamefont{Caldwell}},
  \bibinfo{author}{\bibfnamefont{V.}~\bibnamefont{Gluscevic}},
  \bibnamefont{and}
  \bibinfo{author}{\bibfnamefont{M.}~\bibnamefont{Kamionkowski}},
  \bibinfo{journal}{Physical Review D} \textbf{\bibinfo{volume}{84}},
  \bibinfo{pages}{043504} (\bibinfo{year}{2011}).

\bibitem[{\citenamefont{Zhai et~al.}(2020)\citenamefont{Zhai, Li, Li, Li, and
  Zhang}}]{zhai2020effects}
\bibinfo{author}{\bibfnamefont{H.}~\bibnamefont{Zhai}},
  \bibinfo{author}{\bibfnamefont{S.}~\bibnamefont{Li}},
  \bibinfo{author}{\bibfnamefont{M.}~\bibnamefont{Li}},
  \bibinfo{author}{\bibfnamefont{H.}~\bibnamefont{Li}}, \bibnamefont{and}
  \bibinfo{author}{\bibfnamefont{X.}~\bibnamefont{Zhang}},
  \bibinfo{journal}{Journal of Cosmology and Astroparticle Physics}
  \textbf{\bibinfo{volume}{2020}}, \bibinfo{pages}{051} (\bibinfo{year}{2020}).

\bibitem[{\citenamefont{Arvanitaki et~al.}(2010)\citenamefont{Arvanitaki,
  Dimopoulos, Dubovsky, Kaloper, and March-Russell}}]{arvanitaki2010string}
\bibinfo{author}{\bibfnamefont{A.}~\bibnamefont{Arvanitaki}},
  \bibinfo{author}{\bibfnamefont{S.}~\bibnamefont{Dimopoulos}},
  \bibinfo{author}{\bibfnamefont{S.}~\bibnamefont{Dubovsky}},
  \bibinfo{author}{\bibfnamefont{N.}~\bibnamefont{Kaloper}}, \bibnamefont{and}
  \bibinfo{author}{\bibfnamefont{J.}~\bibnamefont{March-Russell}},
  \bibinfo{journal}{Physical Review D} \textbf{\bibinfo{volume}{81}},
  \bibinfo{pages}{123530} (\bibinfo{year}{2010}).

\bibitem[{\citenamefont{Takahashi and Yin}(2021)}]{takahashi2021kilobyte}
\bibinfo{author}{\bibfnamefont{F.}~\bibnamefont{Takahashi}} \bibnamefont{and}
  \bibinfo{author}{\bibfnamefont{W.}~\bibnamefont{Yin}},
  \bibinfo{journal}{Journal of Cosmology and Astroparticle Physics}
  \textbf{\bibinfo{volume}{2021}}, \bibinfo{pages}{007} (\bibinfo{year}{2021}).

\bibitem[{\citenamefont{Marsh}(2016)}]{marsh2016axion}
\bibinfo{author}{\bibfnamefont{D.~J.} \bibnamefont{Marsh}},
  \bibinfo{journal}{Physics Reports} \textbf{\bibinfo{volume}{643}},
  \bibinfo{pages}{1} (\bibinfo{year}{2016}).

\bibitem[{\citenamefont{Kim et~al.}(2021)\citenamefont{Kim, Kim, Semertzidis,
  Shin, and Yin}}]{kim2021cosmic}
\bibinfo{author}{\bibfnamefont{D.}~\bibnamefont{Kim}},
  \bibinfo{author}{\bibfnamefont{Y.}~\bibnamefont{Kim}},
  \bibinfo{author}{\bibfnamefont{Y.~K.} \bibnamefont{Semertzidis}},
  \bibinfo{author}{\bibfnamefont{Y.~C.} \bibnamefont{Shin}}, \bibnamefont{and}
  \bibinfo{author}{\bibfnamefont{W.}~\bibnamefont{Yin}},
  \bibinfo{journal}{Physical Review D} \textbf{\bibinfo{volume}{104}},
  \bibinfo{pages}{095010} (\bibinfo{year}{2021}).

\bibitem[{\citenamefont{Liu and Ng}(2017)}]{liu2017axion}
\bibinfo{author}{\bibfnamefont{G.}~\bibnamefont{Liu}} \bibnamefont{and}
  \bibinfo{author}{\bibfnamefont{K.}~\bibnamefont{Ng}},
  \bibinfo{journal}{Physics of the dark universe}
  \textbf{\bibinfo{volume}{16}}, \bibinfo{pages}{22} (\bibinfo{year}{2017}).

\bibitem[{\citenamefont{Capparelli et~al.}(2020)\citenamefont{Capparelli,
  Caldwell, and Melchiorri}}]{capparelli2020cosmic}
\bibinfo{author}{\bibfnamefont{L.~M.} \bibnamefont{Capparelli}},
  \bibinfo{author}{\bibfnamefont{R.~R.} \bibnamefont{Caldwell}},
  \bibnamefont{and}
  \bibinfo{author}{\bibfnamefont{A.}~\bibnamefont{Melchiorri}},
  \bibinfo{journal}{Physical Review D} \textbf{\bibinfo{volume}{101}},
  \bibinfo{pages}{123529} (\bibinfo{year}{2020}).

\bibitem[{\citenamefont{Lue et~al.}(1999)\citenamefont{Lue, Wang, and
  Kamionkowski}}]{lue1999cosmological}
\bibinfo{author}{\bibfnamefont{A.}~\bibnamefont{Lue}},
  \bibinfo{author}{\bibfnamefont{L.}~\bibnamefont{Wang}}, \bibnamefont{and}
  \bibinfo{author}{\bibfnamefont{M.}~\bibnamefont{Kamionkowski}},
  \bibinfo{journal}{Physical Review Letters} \textbf{\bibinfo{volume}{83}},
  \bibinfo{pages}{1506} (\bibinfo{year}{1999}).

\bibitem[{\citenamefont{Liu et~al.}(2006)\citenamefont{Liu, Lee, and
  Ng}}]{liu2006effect}
\bibinfo{author}{\bibfnamefont{G.-C.} \bibnamefont{Liu}},
  \bibinfo{author}{\bibfnamefont{S.}~\bibnamefont{Lee}}, \bibnamefont{and}
  \bibinfo{author}{\bibfnamefont{K.-W.} \bibnamefont{Ng}},
  \bibinfo{journal}{Physical Review Letters} \textbf{\bibinfo{volume}{97}},
  \bibinfo{pages}{161303} (\bibinfo{year}{2006}).

\bibitem[{\citenamefont{Feng et~al.}(2006)\citenamefont{Feng, Li, Xia, Chen,
  and Zhang}}]{feng2006searching}
\bibinfo{author}{\bibfnamefont{B.}~\bibnamefont{Feng}},
  \bibinfo{author}{\bibfnamefont{M.}~\bibnamefont{Li}},
  \bibinfo{author}{\bibfnamefont{J.-Q.} \bibnamefont{Xia}},
  \bibinfo{author}{\bibfnamefont{X.}~\bibnamefont{Chen}}, \bibnamefont{and}
  \bibinfo{author}{\bibfnamefont{X.}~\bibnamefont{Zhang}},
  \bibinfo{journal}{Physical review letters} \textbf{\bibinfo{volume}{96}},
  \bibinfo{pages}{221302} (\bibinfo{year}{2006}).

\bibitem[{\citenamefont{Aghanim et~al.}(2016)\citenamefont{Aghanim, Ashdown,
  Aumont, Baccigalupi, Ballardini, Banday, Barreiro, Bartolo, Basak, Benabed
  et~al.}}]{aghanim2016planck}
\bibinfo{author}{\bibfnamefont{N.}~\bibnamefont{Aghanim}},
  \bibinfo{author}{\bibfnamefont{M.}~\bibnamefont{Ashdown}},
  \bibinfo{author}{\bibfnamefont{J.}~\bibnamefont{Aumont}},
  \bibinfo{author}{\bibfnamefont{C.}~\bibnamefont{Baccigalupi}},
  \bibinfo{author}{\bibfnamefont{M.}~\bibnamefont{Ballardini}},
  \bibinfo{author}{\bibfnamefont{A.}~\bibnamefont{Banday}},
  \bibinfo{author}{\bibfnamefont{R.}~\bibnamefont{Barreiro}},
  \bibinfo{author}{\bibfnamefont{N.}~\bibnamefont{Bartolo}},
  \bibinfo{author}{\bibfnamefont{S.}~\bibnamefont{Basak}},
  \bibinfo{author}{\bibfnamefont{K.}~\bibnamefont{Benabed}},
  \bibnamefont{et~al.}, \bibinfo{journal}{Astronomy \& Astrophysics}
  \textbf{\bibinfo{volume}{596}}, \bibinfo{pages}{A110} (\bibinfo{year}{2016}).

\bibitem[{\citenamefont{Gruppuso et~al.}(2020)\citenamefont{Gruppuso, Molinari,
  Natoli, and Pagano}}]{gruppuso2020planck}
\bibinfo{author}{\bibfnamefont{A.}~\bibnamefont{Gruppuso}},
  \bibinfo{author}{\bibfnamefont{D.}~\bibnamefont{Molinari}},
  \bibinfo{author}{\bibfnamefont{P.}~\bibnamefont{Natoli}}, \bibnamefont{and}
  \bibinfo{author}{\bibfnamefont{L.}~\bibnamefont{Pagano}},
  \bibinfo{journal}{Journal of Cosmology and Astroparticle Physics}
  \textbf{\bibinfo{volume}{2020}}, \bibinfo{pages}{066} (\bibinfo{year}{2020}).

\bibitem[{\citenamefont{Minami and Komatsu}(2020)}]{minami2020new}
\bibinfo{author}{\bibfnamefont{Y.}~\bibnamefont{Minami}} \bibnamefont{and}
  \bibinfo{author}{\bibfnamefont{E.}~\bibnamefont{Komatsu}},
  \bibinfo{journal}{Physical Review Letters} \textbf{\bibinfo{volume}{125}},
  \bibinfo{pages}{221301} (\bibinfo{year}{2020}).

\bibitem[{\citenamefont{Diego-Palazuelos
  et~al.}(2022)\citenamefont{Diego-Palazuelos, Eskilt, Minami, Tristram,
  Sullivan, Banday, Barreiro, Eriksen, G{\'o}rski, Keskitalo
  et~al.}}]{diego2022cosmic}
\bibinfo{author}{\bibfnamefont{P.}~\bibnamefont{Diego-Palazuelos}},
  \bibinfo{author}{\bibfnamefont{J.~R.} \bibnamefont{Eskilt}},
  \bibinfo{author}{\bibfnamefont{Y.}~\bibnamefont{Minami}},
  \bibinfo{author}{\bibfnamefont{M.}~\bibnamefont{Tristram}},
  \bibinfo{author}{\bibfnamefont{R.~M.} \bibnamefont{Sullivan}},
  \bibinfo{author}{\bibfnamefont{A.~J.} \bibnamefont{Banday}},
  \bibinfo{author}{\bibfnamefont{R.~B.} \bibnamefont{Barreiro}},
  \bibinfo{author}{\bibfnamefont{H.~K.} \bibnamefont{Eriksen}},
  \bibinfo{author}{\bibfnamefont{K.~M.} \bibnamefont{G{\'o}rski}},
  \bibinfo{author}{\bibfnamefont{R.}~\bibnamefont{Keskitalo}},
  \bibnamefont{et~al.}, \bibinfo{journal}{arXiv preprint arXiv:2201.07682}
  (\bibinfo{year}{2022}).

\bibitem[{\citenamefont{Clark et~al.}(2021)\citenamefont{Clark, Kim, Hill, and
  Hensley}}]{clark2021origin}
\bibinfo{author}{\bibfnamefont{S.~E.} \bibnamefont{Clark}},
  \bibinfo{author}{\bibfnamefont{C.-G.} \bibnamefont{Kim}},
  \bibinfo{author}{\bibfnamefont{J.~C.} \bibnamefont{Hill}}, \bibnamefont{and}
  \bibinfo{author}{\bibfnamefont{B.~S.} \bibnamefont{Hensley}},
  \bibinfo{journal}{arXiv preprint arXiv:2105.00120}  (\bibinfo{year}{2021}).

\bibitem[{\citenamefont{Eskilt}(2022)}]{eskilt2022}
\bibinfo{author}{\bibfnamefont{J.~R.} \bibnamefont{Eskilt}},
  \bibinfo{journal}{arXiv preprint arXiv:2201.13347}  (\bibinfo{year}{2022}).

\bibitem[{\citenamefont{Gruppuso et~al.}(2016)\citenamefont{Gruppuso, Maggio,
  Molinari, and Natoli}}]{gruppuso2016note}
\bibinfo{author}{\bibfnamefont{A.}~\bibnamefont{Gruppuso}},
  \bibinfo{author}{\bibfnamefont{G.}~\bibnamefont{Maggio}},
  \bibinfo{author}{\bibfnamefont{D.}~\bibnamefont{Molinari}}, \bibnamefont{and}
  \bibinfo{author}{\bibfnamefont{P.}~\bibnamefont{Natoli}},
  \bibinfo{journal}{Journal of Cosmology and Astroparticle Physics}
  \textbf{\bibinfo{volume}{2016}}, \bibinfo{pages}{020} (\bibinfo{year}{2016}).

\bibitem[{\citenamefont{Zaldarriaga and Seljak}(1997)}]{zaldarriaga55all}
\bibinfo{author}{\bibfnamefont{M.}~\bibnamefont{Zaldarriaga}} \bibnamefont{and}
  \bibinfo{author}{\bibfnamefont{U.}~\bibnamefont{Seljak}},
  \bibinfo{journal}{Phys. Rev. D} \textbf{\bibinfo{volume}{55}}
  (\bibinfo{year}{1997}).

\bibitem[{\citenamefont{Hu and White}(1997)}]{hu1997cmb}
\bibinfo{author}{\bibfnamefont{W.}~\bibnamefont{Hu}} \bibnamefont{and}
  \bibinfo{author}{\bibfnamefont{M.}~\bibnamefont{White}},
  \bibinfo{journal}{Physical Review D} \textbf{\bibinfo{volume}{56}},
  \bibinfo{pages}{596} (\bibinfo{year}{1997}).

\bibitem[{\citenamefont{Isserlis}(1918)}]{isserlis1918formula}
\bibinfo{author}{\bibfnamefont{L.}~\bibnamefont{Isserlis}},
  \bibinfo{journal}{Biometrika} \textbf{\bibinfo{volume}{12}},
  \bibinfo{pages}{134} (\bibinfo{year}{1918}).

\bibitem[{\citenamefont{Newman and Penrose}(1966)}]{newman1966note}
\bibinfo{author}{\bibfnamefont{E.~T.} \bibnamefont{Newman}} \bibnamefont{and}
  \bibinfo{author}{\bibfnamefont{R.}~\bibnamefont{Penrose}},
  \bibinfo{journal}{Journal of Mathematical Physics}
  \textbf{\bibinfo{volume}{7}}, \bibinfo{pages}{863} (\bibinfo{year}{1966}).

\bibitem[{\citenamefont{Varshalovich}(1988)}]{varshalovich1988quantum}
\bibinfo{author}{\bibfnamefont{D.}~\bibnamefont{Varshalovich}},
  \emph{\bibinfo{title}{Quantum theory of angular momentum}}
  (\bibinfo{publisher}{World Scientific}, \bibinfo{year}{1988}).

\bibitem[{\citenamefont{Bucher et~al.}(2016)\citenamefont{Bucher, Racine, and
  van Tent}}]{bucher2016binned}
\bibinfo{author}{\bibfnamefont{M.}~\bibnamefont{Bucher}},
  \bibinfo{author}{\bibfnamefont{B.}~\bibnamefont{Racine}}, \bibnamefont{and}
  \bibinfo{author}{\bibfnamefont{B.}~\bibnamefont{van Tent}},
  \bibinfo{journal}{Journal of Cosmology and Astroparticle Physics}
  \textbf{\bibinfo{volume}{2016}}, \bibinfo{pages}{055} (\bibinfo{year}{2016}).

\bibitem[{\citenamefont{Yadav et~al.}(2007)\citenamefont{Yadav, Komatsu, and
  Wandelt}}]{yadav2007fast}
\bibinfo{author}{\bibfnamefont{A.~P.} \bibnamefont{Yadav}},
  \bibinfo{author}{\bibfnamefont{E.}~\bibnamefont{Komatsu}}, \bibnamefont{and}
  \bibinfo{author}{\bibfnamefont{B.~D.} \bibnamefont{Wandelt}},
  \bibinfo{journal}{The Astrophysical Journal} \textbf{\bibinfo{volume}{664}},
  \bibinfo{pages}{680} (\bibinfo{year}{2007}).

\bibitem[{\citenamefont{Babich and Zaldarriaga}(2004)}]{babich2004primordial}
\bibinfo{author}{\bibfnamefont{D.}~\bibnamefont{Babich}} \bibnamefont{and}
  \bibinfo{author}{\bibfnamefont{M.}~\bibnamefont{Zaldarriaga}},
  \bibinfo{journal}{Physical Review D} \textbf{\bibinfo{volume}{70}},
  \bibinfo{pages}{083005} (\bibinfo{year}{2004}).

\bibitem[{\citenamefont{Shiraishi et~al.}(2019)\citenamefont{Shiraishi,
  Liguori, Fergusson, and Shellard}}]{shiraishi2019general}
\bibinfo{author}{\bibfnamefont{M.}~\bibnamefont{Shiraishi}},
  \bibinfo{author}{\bibfnamefont{M.}~\bibnamefont{Liguori}},
  \bibinfo{author}{\bibfnamefont{J.~R.} \bibnamefont{Fergusson}},
  \bibnamefont{and} \bibinfo{author}{\bibfnamefont{E.}~\bibnamefont{Shellard}},
  \bibinfo{journal}{Journal of Cosmology and Astroparticle Physics}
  \textbf{\bibinfo{volume}{2019}}, \bibinfo{pages}{046} (\bibinfo{year}{2019}).

\bibitem[{\citenamefont{Komatsu and Spergel}(2001)}]{2001acoustic}
\bibinfo{author}{\bibfnamefont{E.}~\bibnamefont{Komatsu}} \bibnamefont{and}
  \bibinfo{author}{\bibfnamefont{D.~N.} \bibnamefont{Spergel}},
  \bibinfo{journal}{Physical Review D} \textbf{\bibinfo{volume}{63}},
  \bibinfo{pages}{063002} (\bibinfo{year}{2001}).

\bibitem[{\citenamefont{Kamionkowski and
  Souradeep}(2011)}]{kamionkowski2011odd}
\bibinfo{author}{\bibfnamefont{M.}~\bibnamefont{Kamionkowski}}
  \bibnamefont{and}
  \bibinfo{author}{\bibfnamefont{T.}~\bibnamefont{Souradeep}},
  \bibinfo{journal}{Physical Review D} \textbf{\bibinfo{volume}{83}},
  \bibinfo{pages}{027301} (\bibinfo{year}{2011}).

\bibitem[{\citenamefont{Komatsu}(2010)}]{komatsu2010hunting}
\bibinfo{author}{\bibfnamefont{E.}~\bibnamefont{Komatsu}},
  \bibinfo{journal}{Classical and Quantum Gravity}
  \textbf{\bibinfo{volume}{27}}, \bibinfo{pages}{124010}
  (\bibinfo{year}{2010}).

\bibitem[{\citenamefont{Lesgourgues}(2011)}]{lesgourgues2011cosmic}
\bibinfo{author}{\bibfnamefont{J.}~\bibnamefont{Lesgourgues}},
  \bibinfo{journal}{arXiv preprint arXiv:1104.2932}  (\bibinfo{year}{2011}).

\bibitem[{\citenamefont{Akrami et~al.}(2020{\natexlab{a}})}]{Planck:2018jri}
\bibinfo{author}{\bibfnamefont{Y.}~\bibnamefont{Akrami}} \bibnamefont{et~al.}
  (\bibinfo{collaboration}{Planck}), \bibinfo{journal}{Astron. Astrophys.}
  \textbf{\bibinfo{volume}{641}}, \bibinfo{pages}{A10}
  (\bibinfo{year}{2020}{\natexlab{a}}), \eprint{1807.06211}.

\bibitem[{\citenamefont{Ade et~al.}(2021)}]{BICEP:2021xfz}
\bibinfo{author}{\bibfnamefont{P.~A.~R.} \bibnamefont{Ade}}
  \bibnamefont{et~al.} (\bibinfo{collaboration}{BICEP, Keck}),
  \bibinfo{journal}{Phys. Rev. Lett.} \textbf{\bibinfo{volume}{127}},
  \bibinfo{pages}{151301} (\bibinfo{year}{2021}), \eprint{2110.00483}.

\bibitem[{\citenamefont{Tristram et~al.}(2021)}]{Tristram:2021tvh}
\bibinfo{author}{\bibfnamefont{M.}~\bibnamefont{Tristram}} \bibnamefont{et~al.}
  (\bibinfo{year}{2021}), \eprint{2112.07961}.

\bibitem[{\citenamefont{Aghanim et~al.}(2020)\citenamefont{Aghanim, Akrami,
  Ashdown, Aumont, Baccigalupi, Ballardini, Banday, Barreiro, Bartolo, Basak
  et~al.}}]{aghanim2020planck}
\bibinfo{author}{\bibfnamefont{N.}~\bibnamefont{Aghanim}},
  \bibinfo{author}{\bibfnamefont{Y.}~\bibnamefont{Akrami}},
  \bibinfo{author}{\bibfnamefont{M.}~\bibnamefont{Ashdown}},
  \bibinfo{author}{\bibfnamefont{J.}~\bibnamefont{Aumont}},
  \bibinfo{author}{\bibfnamefont{C.}~\bibnamefont{Baccigalupi}},
  \bibinfo{author}{\bibfnamefont{M.}~\bibnamefont{Ballardini}},
  \bibinfo{author}{\bibfnamefont{A.~J.} \bibnamefont{Banday}},
  \bibinfo{author}{\bibfnamefont{R.}~\bibnamefont{Barreiro}},
  \bibinfo{author}{\bibfnamefont{N.}~\bibnamefont{Bartolo}},
  \bibinfo{author}{\bibfnamefont{S.}~\bibnamefont{Basak}},
  \bibnamefont{et~al.}, \bibinfo{journal}{Astronomy \& Astrophysics}
  \textbf{\bibinfo{volume}{641}}, \bibinfo{pages}{A5} (\bibinfo{year}{2020}).

\bibitem[{\citenamefont{Kendall and Stuart}(1969)}]{kendall1969advanced}
\bibinfo{author}{\bibfnamefont{M.~G.} \bibnamefont{Kendall}} \bibnamefont{and}
  \bibinfo{author}{\bibfnamefont{A.}~\bibnamefont{Stuart}},
  \emph{\bibinfo{title}{The advanced theory of statistics (vol 2)}}
  (\bibinfo{year}{1969}).

\bibitem[{\citenamefont{Komatsu}(2002)}]{komatsu2002pursuit}
\bibinfo{author}{\bibfnamefont{E.}~\bibnamefont{Komatsu}},
  \bibinfo{journal}{arXiv preprint astro-ph/0206039}  (\bibinfo{year}{2002}).

\bibitem[{\citenamefont{Bartolo et~al.}(2004)\citenamefont{Bartolo, Komatsu,
  Matarrese, and Riotto}}]{bartolo2004non}
\bibinfo{author}{\bibfnamefont{N.}~\bibnamefont{Bartolo}},
  \bibinfo{author}{\bibfnamefont{E.}~\bibnamefont{Komatsu}},
  \bibinfo{author}{\bibfnamefont{S.}~\bibnamefont{Matarrese}},
  \bibnamefont{and} \bibinfo{author}{\bibfnamefont{A.}~\bibnamefont{Riotto}},
  \bibinfo{journal}{Physics Reports} \textbf{\bibinfo{volume}{402}},
  \bibinfo{pages}{103} (\bibinfo{year}{2004}).

\bibitem[{\citenamefont{Akrami et~al.}(2020{\natexlab{b}})\citenamefont{Akrami,
  Arroja, Ashdown, Aumont, Baccigalupi, Ballardini, Banday, Barreiro, Bartolo,
  Basak et~al.}}]{akrami2020planckNG}
\bibinfo{author}{\bibfnamefont{Y.}~\bibnamefont{Akrami}},
  \bibinfo{author}{\bibfnamefont{F.}~\bibnamefont{Arroja}},
  \bibinfo{author}{\bibfnamefont{M.}~\bibnamefont{Ashdown}},
  \bibinfo{author}{\bibfnamefont{J.}~\bibnamefont{Aumont}},
  \bibinfo{author}{\bibfnamefont{C.}~\bibnamefont{Baccigalupi}},
  \bibinfo{author}{\bibfnamefont{M.}~\bibnamefont{Ballardini}},
  \bibinfo{author}{\bibfnamefont{A.}~\bibnamefont{Banday}},
  \bibinfo{author}{\bibfnamefont{R.}~\bibnamefont{Barreiro}},
  \bibinfo{author}{\bibfnamefont{N.}~\bibnamefont{Bartolo}},
  \bibinfo{author}{\bibfnamefont{S.}~\bibnamefont{Basak}},
  \bibnamefont{et~al.}, \bibinfo{journal}{Astronomy \& Astrophysics}
  \textbf{\bibinfo{volume}{641}}, \bibinfo{pages}{A9}
  (\bibinfo{year}{2020}{\natexlab{b}}).

\bibitem[{\citenamefont{Liguori et~al.}(2010)\citenamefont{Liguori, Sefusatti,
  Fergusson, and Shellard}}]{liguori2010primordial}
\bibinfo{author}{\bibfnamefont{M.}~\bibnamefont{Liguori}},
  \bibinfo{author}{\bibfnamefont{E.}~\bibnamefont{Sefusatti}},
  \bibinfo{author}{\bibfnamefont{J.~R.} \bibnamefont{Fergusson}},
  \bibnamefont{and} \bibinfo{author}{\bibfnamefont{E.}~\bibnamefont{Shellard}},
  \bibinfo{journal}{Advances in Astronomy} \textbf{\bibinfo{volume}{2010}}
  (\bibinfo{year}{2010}).

\bibitem[{\citenamefont{Cooray and Hu}(2000)}]{cooray2000imprint}
\bibinfo{author}{\bibfnamefont{A.}~\bibnamefont{Cooray}} \bibnamefont{and}
  \bibinfo{author}{\bibfnamefont{W.}~\bibnamefont{Hu}}, \bibinfo{journal}{The
  Astrophysical Journal} \textbf{\bibinfo{volume}{534}}, \bibinfo{pages}{533}
  (\bibinfo{year}{2000}).

\bibitem[{\citenamefont{Gangui and Martin}(2000)}]{gangui2000cosmic}
\bibinfo{author}{\bibfnamefont{A.}~\bibnamefont{Gangui}} \bibnamefont{and}
  \bibinfo{author}{\bibfnamefont{J.}~\bibnamefont{Martin}},
  \bibinfo{journal}{Monthly Notices of the Royal Astronomical Society}
  \textbf{\bibinfo{volume}{313}}, \bibinfo{pages}{323} (\bibinfo{year}{2000}).

\bibitem[{\citenamefont{Hu}(2000)}]{hu2000weak}
\bibinfo{author}{\bibfnamefont{W.}~\bibnamefont{Hu}},
  \bibinfo{journal}{Physical Review D} \textbf{\bibinfo{volume}{62}},
  \bibinfo{pages}{043007} (\bibinfo{year}{2000}).

\bibitem[{\citenamefont{Rotenberg}(1959)}]{rotenberg19593}
\bibinfo{author}{\bibfnamefont{M.}~\bibnamefont{Rotenberg}},
  \emph{\bibinfo{title}{The 3-j and 6-j Symbols}}
  (\bibinfo{publisher}{Technology Press, Massachusetts Institute of
  Technology}, \bibinfo{year}{1959}).

\bibitem[{\citenamefont{Namikawa et~al.}(2020)\citenamefont{Namikawa, Guan,
  Darwish, Sherwin, Aiola, Battaglia, Beall, Becker, Bond, Calabrese
  et~al.}}]{namikawa2020atacama}
\bibinfo{author}{\bibfnamefont{T.}~\bibnamefont{Namikawa}},
  \bibinfo{author}{\bibfnamefont{Y.}~\bibnamefont{Guan}},
  \bibinfo{author}{\bibfnamefont{O.}~\bibnamefont{Darwish}},
  \bibinfo{author}{\bibfnamefont{B.~D.} \bibnamefont{Sherwin}},
  \bibinfo{author}{\bibfnamefont{S.}~\bibnamefont{Aiola}},
  \bibinfo{author}{\bibfnamefont{N.}~\bibnamefont{Battaglia}},
  \bibinfo{author}{\bibfnamefont{J.~A.} \bibnamefont{Beall}},
  \bibinfo{author}{\bibfnamefont{D.~T.} \bibnamefont{Becker}},
  \bibinfo{author}{\bibfnamefont{J.~R.} \bibnamefont{Bond}},
  \bibinfo{author}{\bibfnamefont{E.}~\bibnamefont{Calabrese}},
  \bibnamefont{et~al.}, \bibinfo{journal}{Physical Review D}
  \textbf{\bibinfo{volume}{101}}, \bibinfo{pages}{083527}
  (\bibinfo{year}{2020}).

\bibitem[{\citenamefont{Bianchini et~al.}(2020)\citenamefont{Bianchini, Wu,
  Ade, Anderson, Austermann, Avva, Balkenhol, Baxter, Beall, Bender
  et~al.}}]{bianchini2020searching}
\bibinfo{author}{\bibfnamefont{F.}~\bibnamefont{Bianchini}},
  \bibinfo{author}{\bibfnamefont{W.}~\bibnamefont{Wu}},
  \bibinfo{author}{\bibfnamefont{P.}~\bibnamefont{Ade}},
  \bibinfo{author}{\bibfnamefont{A.}~\bibnamefont{Anderson}},
  \bibinfo{author}{\bibfnamefont{J.}~\bibnamefont{Austermann}},
  \bibinfo{author}{\bibfnamefont{J.}~\bibnamefont{Avva}},
  \bibinfo{author}{\bibfnamefont{L.}~\bibnamefont{Balkenhol}},
  \bibinfo{author}{\bibfnamefont{E.}~\bibnamefont{Baxter}},
  \bibinfo{author}{\bibfnamefont{J.}~\bibnamefont{Beall}},
  \bibinfo{author}{\bibfnamefont{A.}~\bibnamefont{Bender}},
  \bibnamefont{et~al.}, \bibinfo{journal}{Physical Review D}
  \textbf{\bibinfo{volume}{102}}, \bibinfo{pages}{083504}
  (\bibinfo{year}{2020}).

\bibitem[{\citenamefont{Contreras et~al.}(2017)\citenamefont{Contreras, Boubel,
  and Scott}}]{contreras2017constraints}
\bibinfo{author}{\bibfnamefont{D.}~\bibnamefont{Contreras}},
  \bibinfo{author}{\bibfnamefont{P.}~\bibnamefont{Boubel}}, \bibnamefont{and}
  \bibinfo{author}{\bibfnamefont{D.}~\bibnamefont{Scott}},
  \bibinfo{journal}{Journal of Cosmology and Astroparticle Physics}
  \textbf{\bibinfo{volume}{2017}}, \bibinfo{pages}{046} (\bibinfo{year}{2017}).

\bibitem[{\citenamefont{Ade et~al.}(2017)\citenamefont{Ade, Aikin, Bock,
  Brevik, Filippini, Ghosh, Hildebrandt, Hui, Kefeli, Moncelsi
  et~al.}}]{ade2017bicep2}
\bibinfo{author}{\bibfnamefont{P.}~\bibnamefont{Ade}},
  \bibinfo{author}{\bibfnamefont{R.}~\bibnamefont{Aikin}},
  \bibinfo{author}{\bibfnamefont{J.}~\bibnamefont{Bock}},
  \bibinfo{author}{\bibfnamefont{J.}~\bibnamefont{Brevik}},
  \bibinfo{author}{\bibfnamefont{J.}~\bibnamefont{Filippini}},
  \bibinfo{author}{\bibfnamefont{T.}~\bibnamefont{Ghosh}},
  \bibinfo{author}{\bibfnamefont{S.}~\bibnamefont{Hildebrandt}},
  \bibinfo{author}{\bibfnamefont{H.}~\bibnamefont{Hui}},
  \bibinfo{author}{\bibfnamefont{S.}~\bibnamefont{Kefeli}},
  \bibinfo{author}{\bibfnamefont{L.}~\bibnamefont{Moncelsi}},
  \bibnamefont{et~al.}, \bibinfo{journal}{Physical Review D}
  \textbf{\bibinfo{volume}{96}}, \bibinfo{pages}{Art} (\bibinfo{year}{2017}).

\bibitem[{\citenamefont{Ade et~al.}(2015)\citenamefont{Ade, Arnold, Atlas,
  Baccigalupi, Barron, Boettger, Borrill, Chapman, Chinone, Cukierman
  et~al.}}]{ade2015polarbear}
\bibinfo{author}{\bibfnamefont{P.~A.} \bibnamefont{Ade}},
  \bibinfo{author}{\bibfnamefont{K.}~\bibnamefont{Arnold}},
  \bibinfo{author}{\bibfnamefont{M.}~\bibnamefont{Atlas}},
  \bibinfo{author}{\bibfnamefont{C.}~\bibnamefont{Baccigalupi}},
  \bibinfo{author}{\bibfnamefont{D.}~\bibnamefont{Barron}},
  \bibinfo{author}{\bibfnamefont{D.}~\bibnamefont{Boettger}},
  \bibinfo{author}{\bibfnamefont{J.}~\bibnamefont{Borrill}},
  \bibinfo{author}{\bibfnamefont{S.}~\bibnamefont{Chapman}},
  \bibinfo{author}{\bibfnamefont{Y.}~\bibnamefont{Chinone}},
  \bibinfo{author}{\bibfnamefont{A.}~\bibnamefont{Cukierman}},
  \bibnamefont{et~al.}, \bibinfo{journal}{Physical Review D}
  \textbf{\bibinfo{volume}{92}}, \bibinfo{pages}{123509}
  (\bibinfo{year}{2015}).

\bibitem[{\citenamefont{Gluscevic et~al.}(2012)\citenamefont{Gluscevic, Hanson,
  Kamionkowski, and Hirata}}]{gluscevic2012first}
\bibinfo{author}{\bibfnamefont{V.}~\bibnamefont{Gluscevic}},
  \bibinfo{author}{\bibfnamefont{D.}~\bibnamefont{Hanson}},
  \bibinfo{author}{\bibfnamefont{M.}~\bibnamefont{Kamionkowski}},
  \bibnamefont{and} \bibinfo{author}{\bibfnamefont{C.~M.}
  \bibnamefont{Hirata}}, \bibinfo{journal}{Physical Review D}
  \textbf{\bibinfo{volume}{86}}, \bibinfo{pages}{103529}
  (\bibinfo{year}{2012}).

\bibitem[{\citenamefont{Pogosian et~al.}(2019)\citenamefont{Pogosian, Shimon,
  Mewes, and Keating}}]{pogosian2019future}
\bibinfo{author}{\bibfnamefont{L.}~\bibnamefont{Pogosian}},
  \bibinfo{author}{\bibfnamefont{M.}~\bibnamefont{Shimon}},
  \bibinfo{author}{\bibfnamefont{M.}~\bibnamefont{Mewes}}, \bibnamefont{and}
  \bibinfo{author}{\bibfnamefont{B.}~\bibnamefont{Keating}},
  \bibinfo{journal}{Physical Review D} \textbf{\bibinfo{volume}{100}},
  \bibinfo{pages}{023507} (\bibinfo{year}{2019}).

\bibitem[{\citenamefont{Matsumura et~al.}(2014)\citenamefont{Matsumura, Akiba,
  Borrill, Chinone, Dobbs, Fuke, Ghribi, Hasegawa, Hattori, Hattori
  et~al.}}]{matsumura2014mission}
\bibinfo{author}{\bibfnamefont{T.}~\bibnamefont{Matsumura}},
  \bibinfo{author}{\bibfnamefont{Y.}~\bibnamefont{Akiba}},
  \bibinfo{author}{\bibfnamefont{J.}~\bibnamefont{Borrill}},
  \bibinfo{author}{\bibfnamefont{Y.}~\bibnamefont{Chinone}},
  \bibinfo{author}{\bibfnamefont{M.}~\bibnamefont{Dobbs}},
  \bibinfo{author}{\bibfnamefont{H.}~\bibnamefont{Fuke}},
  \bibinfo{author}{\bibfnamefont{A.}~\bibnamefont{Ghribi}},
  \bibinfo{author}{\bibfnamefont{M.}~\bibnamefont{Hasegawa}},
  \bibinfo{author}{\bibfnamefont{K.}~\bibnamefont{Hattori}},
  \bibinfo{author}{\bibfnamefont{M.}~\bibnamefont{Hattori}},
  \bibnamefont{et~al.}, \bibinfo{journal}{Journal of Low Temperature Physics}
  \textbf{\bibinfo{volume}{176}}, \bibinfo{pages}{733} (\bibinfo{year}{2014}).

\bibitem[{\citenamefont{Knox}(1995)}]{knox1995determination}
\bibinfo{author}{\bibfnamefont{L.}~\bibnamefont{Knox}},
  \bibinfo{journal}{Physical Review D} \textbf{\bibinfo{volume}{52}},
  \bibinfo{pages}{4307} (\bibinfo{year}{1995}).

\bibitem[{\citenamefont{Ng and Liu}(1999)}]{ng1999correlation}
\bibinfo{author}{\bibfnamefont{K.-W.} \bibnamefont{Ng}} \bibnamefont{and}
  \bibinfo{author}{\bibfnamefont{G.-C.} \bibnamefont{Liu}},
  \bibinfo{journal}{International Journal of Modern Physics D}
  \textbf{\bibinfo{volume}{8}}, \bibinfo{pages}{61} (\bibinfo{year}{1999}).

\bibitem[{\citenamefont{Hazumi et~al.}(2020)\citenamefont{Hazumi, Ade, Adler,
  Allys, Alonso, Arnold, Auguste, Aumont, Aurlien, Austermann
  et~al.}}]{hazumi2020litebird}
\bibinfo{author}{\bibfnamefont{M.}~\bibnamefont{Hazumi}},
  \bibinfo{author}{\bibfnamefont{P.~A.} \bibnamefont{Ade}},
  \bibinfo{author}{\bibfnamefont{A.}~\bibnamefont{Adler}},
  \bibinfo{author}{\bibfnamefont{E.}~\bibnamefont{Allys}},
  \bibinfo{author}{\bibfnamefont{D.}~\bibnamefont{Alonso}},
  \bibinfo{author}{\bibfnamefont{K.}~\bibnamefont{Arnold}},
  \bibinfo{author}{\bibfnamefont{D.}~\bibnamefont{Auguste}},
  \bibinfo{author}{\bibfnamefont{J.}~\bibnamefont{Aumont}},
  \bibinfo{author}{\bibfnamefont{R.}~\bibnamefont{Aurlien}},
  \bibinfo{author}{\bibfnamefont{J.}~\bibnamefont{Austermann}},
  \bibnamefont{et~al.}, in \emph{\bibinfo{booktitle}{Space Telescopes and
  Instrumentation 2020: Optical, Infrared, and Millimeter Wave}}
  (\bibinfo{organization}{International Society for Optics and Photonics},
  \bibinfo{year}{2020}), vol. \bibinfo{volume}{11443}, p.
  \bibinfo{pages}{114432F}.

\bibitem[{\citenamefont{Knox and Song}(2002)}]{Knox:2002pe}
\bibinfo{author}{\bibfnamefont{L.}~\bibnamefont{Knox}} \bibnamefont{and}
  \bibinfo{author}{\bibfnamefont{Y.-S.} \bibnamefont{Song}},
  \bibinfo{journal}{Phys. Rev. Lett.} \textbf{\bibinfo{volume}{89}},
  \bibinfo{pages}{011303} (\bibinfo{year}{2002}), \eprint{astro-ph/0202286}.

\bibitem[{\citenamefont{Kesden et~al.}(2002)\citenamefont{Kesden, Cooray, and
  Kamionkowski}}]{Kesden:2002ku}
\bibinfo{author}{\bibfnamefont{M.}~\bibnamefont{Kesden}},
  \bibinfo{author}{\bibfnamefont{A.}~\bibnamefont{Cooray}}, \bibnamefont{and}
  \bibinfo{author}{\bibfnamefont{M.}~\bibnamefont{Kamionkowski}},
  \bibinfo{journal}{Phys. Rev. Lett.} \textbf{\bibinfo{volume}{89}},
  \bibinfo{pages}{011304} (\bibinfo{year}{2002}), \eprint{astro-ph/0202434}.

\bibitem[{\citenamefont{Hirata and Seljak}(2003)}]{Hirata:2002jy}
\bibinfo{author}{\bibfnamefont{C.~M.} \bibnamefont{Hirata}} \bibnamefont{and}
  \bibinfo{author}{\bibfnamefont{U.}~\bibnamefont{Seljak}},
  \bibinfo{journal}{Phys. Rev. D} \textbf{\bibinfo{volume}{67}},
  \bibinfo{pages}{043001} (\bibinfo{year}{2003}), \eprint{astro-ph/0209489}.

\bibitem[{\citenamefont{Seljak and Hirata}(2004)}]{Seljak:2003pn}
\bibinfo{author}{\bibfnamefont{U.}~\bibnamefont{Seljak}} \bibnamefont{and}
  \bibinfo{author}{\bibfnamefont{C.~M.} \bibnamefont{Hirata}},
  \bibinfo{journal}{Phys. Rev. D} \textbf{\bibinfo{volume}{69}},
  \bibinfo{pages}{043005} (\bibinfo{year}{2004}), \eprint{astro-ph/0310163}.

\bibitem[{\citenamefont{Kalaja et~al.}(2021)\citenamefont{Kalaja, Meerburg,
  Pimentel, and Coulton}}]{Kalaja:2020mkq}
\bibinfo{author}{\bibfnamefont{A.}~\bibnamefont{Kalaja}},
  \bibinfo{author}{\bibfnamefont{P.~D.} \bibnamefont{Meerburg}},
  \bibinfo{author}{\bibfnamefont{G.~L.} \bibnamefont{Pimentel}},
  \bibnamefont{and} \bibinfo{author}{\bibfnamefont{W.~R.}
  \bibnamefont{Coulton}}, \bibinfo{journal}{JCAP}
  \textbf{\bibinfo{volume}{04}}, \bibinfo{pages}{050} (\bibinfo{year}{2021}),
  \eprint{2011.09461}.

\bibitem[{\citenamefont{Coulton et~al.}(2020)\citenamefont{Coulton, Meerburg,
  Baker, Hotinli, Duivenvoorden, and van Engelen}}]{Coulton:2019odk}
\bibinfo{author}{\bibfnamefont{W.~R.} \bibnamefont{Coulton}},
  \bibinfo{author}{\bibfnamefont{P.~D.} \bibnamefont{Meerburg}},
  \bibinfo{author}{\bibfnamefont{D.~G.} \bibnamefont{Baker}},
  \bibinfo{author}{\bibfnamefont{S.}~\bibnamefont{Hotinli}},
  \bibinfo{author}{\bibfnamefont{A.~J.} \bibnamefont{Duivenvoorden}},
  \bibnamefont{and} \bibinfo{author}{\bibfnamefont{A.}~\bibnamefont{van
  Engelen}}, \bibinfo{journal}{Phys. Rev. D} \textbf{\bibinfo{volume}{101}},
  \bibinfo{pages}{123504} (\bibinfo{year}{2020}), \eprint{1912.07619}.

\bibitem[{\citenamefont{Bartolo et~al.}(in preparation)\citenamefont{Bartolo,
  Greco, and Gruppuso}}]{workinprogress}
\bibinfo{author}{\bibfnamefont{N.}~\bibnamefont{Bartolo}},
  \bibinfo{author}{\bibfnamefont{A.}~\bibnamefont{Greco}}, \bibnamefont{and}
  \bibinfo{author}{\bibfnamefont{A.}~\bibnamefont{Gruppuso}} (\bibinfo{year}{in
  preparation}).

\bibitem[{\citenamefont{Duivenvoorden et~al.}(2020)\citenamefont{Duivenvoorden,
  Meerburg, and Freese}}]{duivenvoorden2020cmb}
\bibinfo{author}{\bibfnamefont{A.~J.} \bibnamefont{Duivenvoorden}},
  \bibinfo{author}{\bibfnamefont{P.~D.} \bibnamefont{Meerburg}},
  \bibnamefont{and} \bibinfo{author}{\bibfnamefont{K.}~\bibnamefont{Freese}},
  \bibinfo{journal}{Physical Review D} \textbf{\bibinfo{volume}{102}},
  \bibinfo{pages}{023521} (\bibinfo{year}{2020}).

\bibitem[{\citenamefont{Meerburg et~al.}(2016)\citenamefont{Meerburg, Meyers,
  Van~Engelen, and Ali-Ha{\"\i}moud}}]{meerburg2016cmb}
\bibinfo{author}{\bibfnamefont{P.~D.} \bibnamefont{Meerburg}},
  \bibinfo{author}{\bibfnamefont{J.}~\bibnamefont{Meyers}},
  \bibinfo{author}{\bibfnamefont{A.}~\bibnamefont{Van~Engelen}},
  \bibnamefont{and}
  \bibinfo{author}{\bibfnamefont{Y.}~\bibnamefont{Ali-Ha{\"\i}moud}},
  \bibinfo{journal}{Physical Review D} \textbf{\bibinfo{volume}{93}},
  \bibinfo{pages}{123511} (\bibinfo{year}{2016}).

\bibitem[{\citenamefont{Bloomfield et~al.}(2017)\citenamefont{Bloomfield, Face,
  and Moss}}]{bloomfield2017indefinite}
\bibinfo{author}{\bibfnamefont{J.~K.} \bibnamefont{Bloomfield}},
  \bibinfo{author}{\bibfnamefont{S.~H.} \bibnamefont{Face}}, \bibnamefont{and}
  \bibinfo{author}{\bibfnamefont{Z.}~\bibnamefont{Moss}},
  \bibinfo{journal}{arXiv preprint arXiv:1703.06428}  (\bibinfo{year}{2017}).

\bibitem[{\citenamefont{Akrami et~al.}(2020{\natexlab{c}})\citenamefont{Akrami,
  Arroja, Ashdown, Aumont, Baccigalupi, Ballardini, Banday, Barreiro, Bartolo,
  Basak et~al.}}]{akrami2020planck}
\bibinfo{author}{\bibfnamefont{Y.}~\bibnamefont{Akrami}},
  \bibinfo{author}{\bibfnamefont{F.}~\bibnamefont{Arroja}},
  \bibinfo{author}{\bibfnamefont{M.}~\bibnamefont{Ashdown}},
  \bibinfo{author}{\bibfnamefont{J.}~\bibnamefont{Aumont}},
  \bibinfo{author}{\bibfnamefont{C.}~\bibnamefont{Baccigalupi}},
  \bibinfo{author}{\bibfnamefont{M.}~\bibnamefont{Ballardini}},
  \bibinfo{author}{\bibfnamefont{A.~J.} \bibnamefont{Banday}},
  \bibinfo{author}{\bibfnamefont{R.}~\bibnamefont{Barreiro}},
  \bibinfo{author}{\bibfnamefont{N.}~\bibnamefont{Bartolo}},
  \bibinfo{author}{\bibfnamefont{S.}~\bibnamefont{Basak}},
  \bibnamefont{et~al.}, \bibinfo{journal}{Astronomy \& Astrophysics}
  \textbf{\bibinfo{volume}{641}}, \bibinfo{pages}{A10}
  (\bibinfo{year}{2020}{\natexlab{c}}).

\end{thebibliography}
		
\end{document}